\documentclass{aa}  
\usepackage{graphicx}
\usepackage{changepage}
\usepackage{tablefootnote}
\usepackage{txfonts}

\usepackage{chemformula}
\let\ce\ch
\usepackage{comment}
\usepackage{dblfloatfix}
\usepackage[font=small,labelfont=bf]{caption}
\usepackage{subcaption}
\usepackage{hyperref}
\usepackage{siunitx}
\usepackage{pdflscape}
\usepackage{cellspace}

\begin{document}

% \titlerunning
   \title{An extensively validated C/H/O/N chemical network for hot exoplanet disequilibrium chemistry}

%   \subtitle{Rework of the carbon and nitrogen chemistry based on recent experimentally validated combustion models}

% Ajout Pierre-Alexandre, Ahmed

% \authorrunning
   \author{R. Veillet
          \inst{1},
          O. Venot
          \inst{1},
          B. Sirjean
          \inst{2},
          R. Bounaceur
          \inst{2},
          P-A. Glaude
          \inst{2},
          A. Al-Refaie
          \inst{3},
          \and
          E. Hébrard
          \inst{4}
          % Rajouter Ahmed ?
          %\and
          %C. Ptolemy\inst{2}\fnmsep\thanks{Just to show the usage
          %of the elements in the author field}
          }

% commented line 1124 in aa.cls

   %\institute{Laboratoire Interuniversitaire des Systèmes Atmosphériques (LISA), UMR CNRS 7583, %Université Paris-Est-Créteil, Université de Paris, Institut Pierre Simon Laplace, Créteil, France\\
   \institute{Universit\'e Paris Cit\'e and Univ Paris Est Cr\'eteil, CNRS, LISA, F-75013 Paris, France\\
   %\institute{Laboratoire Interuniversitaire des Systèmes Atmosphériques (LISA), UMR CNRS 7583, Universit\'e de Paris Cit\'e and Univ Paris Est Creteil, CNRS, LISA, F-75013 Paris, France\\
              \email{romeo.veillet@lisa.ipsl.fr}%\\
              %\email{olivia.venot@lisa.ipsl.fr}
         \and
             Université de Lorraine, CNRS, LRGP, F-54000 Nancy, France
             %Laboratoire Réactions et Génie des Procédés, LRGP UMP 7274 CNRS, Université de Lorraine, 1 rue Grandville, BP 20401, 54001 Nancy, France%\\
             %\email{baptiste.sirjean@univ-lorraine.fr}\\
             %\email{roda.bounaceur@univ-lorraine.fr}
        \and
        Department of Physics and Astronomy, University College London, Gower Street, London, WC1E 6BT, UK
        \and
            Astrophysics Group, University of Exeter, EX4 4QL Exeter, UK
             %\thanks{The university of heaven temporarily does not
             %        accept e-mails}
             }

   %\date{Received September 15, 1996; accepted March 16, 1997}

% \abstract{}{}{}{}{} 
% 5 {} token are mandatory
\titlerunning{Extensively validated C/H/O/N chemical network}
\authorrunning{R. Veillet et al.}
  \abstract
   {The reliability of one-dimensional disequilibrium chemistry models in hot exoplanet atmospheres depends on the chemical network used. To develop robust networks, we can rely on combustion studies that provide \ce{C/H/O/N} chemical networks validated by vast amount of experimental data generated by the extensive research that has been done on hydrocarbon combustion and \ce{NO_x} formation in the last decades.
   }
  {
  We aimed to build a new and updated C$_0$-C$_2$ chemical network to study the C/H/O/N disequilibrium chemistry of warm and hot exoplanet atmospheres that relies on extensively validated and recent state-of-the-art combustion networks.
  The reliability range of this network was aimed for conditions between 500 - 2500 K and 100 - $10^{-6}$ bar, with cautious extrapolation at lower temperature values.
  }
   {
   We compared the predictions of seven networks over a large set of experiments, covering a wide range of conditions (pressures, temperatures, and initial compositions). To examine the consequences of this new chemical network on exoplanets atmospheric studies, we generated abundances profiles for GJ 436 b, GJ 1214 b, HD 189733 b, and HD 209458 b, using the 1D kinetic model FRECKLL and calculated the corresponding transmission spectra using TauREx 3.1. These spectra and abundance profiles have been compared with results obtained with our previous chemical network.
   }
   {
   Our new kinetic network is composed of 174 species and 1293 reactions mostly reversible. %most of them being reversible.
   This network proves to be more accurate than our previous one for the tested experimental conditions. %, with huge gains in accuracy especially for nitrogen species.
   The nitrogen chemistry update is found to be very impactful on the abundance profiles, particularly for \ce{HCN}, with differences up to four orders of magnitude. The \ce{CO2} profiles are also significantly affected, with important repercussions on the transmission spectrum of GJ 436 b.
   }
   {
   %These important effects on 1D abundance profiles of species that contribute to the transmission spectrum of exoplanets highlights the importance of using extensively validated chemical networks to gain confidence in our models predictions.
   These effects highlight the importance of using extensively validated chemical networks to gain confidence in our models predictions.
   %However, as the experimental validation range will never be as large as the variety of exoplanetary conditions, some care should be taken when adapting these chemical networks, particularly due to missing radical reactions that later appears to be crucial for the upper atmosphere, and results in globalised reactions that can be ill-suited at low pressures with high concentrations of radicals.
   As shown with \ce{CH2NH}, the coupling between carbon and nitrogen chemistry combined with radicals produced by photolysis can have huge effects impacting the transmission spectra. % on the main nitrogen-bearing and carbon-bearing species in these atmospheres impacting the transmission spectra.
   %, which end up impacting the transmission spectra.
   This should be kept in mind when adding new elements like sulfur, as only adding a sub-mechanism neglects these coupling effects.
   }


   \maketitle

%-------------------------------------------------------------------

\section{Introduction}

    Over recent decades, and still remaining relevant today, the characterization of the atmospheric composition of exoplanets has only been possible for massive hydrogen-dominated exoplanets close to their star.
    Because of the detection biases of the transit method and its technical difficulty for exoplanets with a shallow transit depth, the range of masses and semi-major axis that can be probed by spectrometric means remains blind to colder, Earth-like exoplanets.
    The proximity of these observable exoplanets to their star results in highly irradiated atmospheres \citep{linsky2014}, which implies both a high temperature profile that activates endothermic reactions and an intense UV flux that photodissociates the majority of species in the upper atmosphere, resulting in the creation of a high quantity of radicals \citep{heays2017}.
    This proximity also causes huge tidal forces that probably results in these exoplanets to be tidally-locked, which further intensifies the horizontal and vertical temperature gradients in the atmosphere, causing intense advection and strong steady winds \citep{menou2022, charnay2015}.
    We also know that this advection coupled to photolysis in the upper atmosphere should maintain the chemical species abundance profiles in a steady-state out of equilibrium \citep{moses2011, roudier2021, stevenson2010}.
    To take into account the dynamical timescale, it is therefore necessary to accurately describe both the atmospheric advection and the chemical kinetics of the reactions taking place in the atmosphere \citep{drummond2020, zamyatina2023}.
    Accurately reproducing the chemistry in these conditions requires a detailed kinetic network, which describes chemistry in sets of elementary reversible reactions that form, consume, and propagate radicals.
    These reactions then form a parameterized chemical network that can be used to model the chemical kinetics in exoplanet atmospheres. However, the parameters that characterize the kinetic properties of each reaction can be difficult to estimate and their determination is subject to an entire field of research in combustion kinetics \citep{wang2015, curran2019}.
    In the combustion domain, the detailed kinetic networks are validated against experimental data measured in 0D or 1D reactors close to ideal reactors and designed to characterize only the chemical kinetics.
    Such data can include the evolution of combustion products and intermediates as a function of time or temperature, auto-ignition delay times, or laminar flame studies \citep{gcr2011}.
    
    For  atmospheric studies of exoplanets, various detailed kinetic networks have already been developed \citep{moses2011, tsai2017,tsai2021, venot2012,venot2015, venot2020,rimmer2016}.
    Most of these chemical networks were built by grouping reactions with available parameters from databases and/or computed with quantum mechanics calculations. \cite{venot2012, venot2015,venot2020} networks are the only ones based on networks validated by experiments. \cite{venot2012} was the first one to be developed, and was extended to species bearing up to six carbon atoms in \cite{venot2015}. Additional corrections to the methanol chemistry were later introduced (\citealt{venot2020}; hereafter Venot 2020).
    These networks usually describe only the kinetics of carbon-, \mbox{hydrogen-,} oxygen-, and nitrogen-bearing species, and are such labeled C/H/O/N chemical networks.
    In this present work, we aim to develop a new C/H/O/N network for exoplanet atmospheric chemistry based on extensive validations against experimental data, totally revisiting the C/H/O/N chemistry and basing it on two new state-of-the-art combustion networks for C/H/O and N chemistry, respectively, from \cite{curranmodel} and \cite{glarborgmodel}.
        
    % Quelles sont les conditions d'utilisation visées par le modèle ?
    To accurately reproduce very different conditions from warm sub-Neptunes to very hot Jupiters, with potential applications to warm super-Earths, the new chemical network is a detailed network suitable for a wide range of pressures and temperatures.
    The validity domain of the network must therefore be, in principle, from 500 to 2500 K and from 100 to $10^{-6}$ bar.
    The network is also required to accurately describe the kinetics of all C/H/O/N species with fewer than two atoms of carbon in order to correctly model the overall chemistry of every major species observed and potentially visible in exoplanet spectra (\ce{H2O}, \ce{CH4}, \ce{NH3}, \ce{CO}, \ce{CO2}, \ce{HCN}, \ce{C2H2}, \ce{C2H4}, \ce{C2H6} \dots). Although the chemical network is aimed at studying hydrogen dominated atmospheres, it should remain valid at even very high metallicity and for every possible C/H/O/N atomic abundance.
    This implies that it should accurately describe all the reaction kinetics ranging from oxygen-poor, carbon-, and hydrogen-dominated atmospheres for pyrolysis, up to oxygen-rich environments, more favorable to oxidation reactions. Due to limitations in the available computational resources, it is mainly intended for 1D simulations.
    
    Section \ref{section:mechanism_selection}, discusses how we selected the combustion networks with which we developed our new chemical network, the extensive validation that came along with it and the additions and modifications made to the original networks.
    Then, in Sect. \ref{section:model_application}, we apply this network to the study of exoplanet atmospheres. We studied four planets: GJ 436 b, GJ 1214 b, HD 189733 b, and HD 209458 b, and we compared our results with those obtained with the chemical network Venot 2020.
    We also investigated the differences between the two networks to highlight new chemical pathways, in addition to discussing potential repercussions on the transmission spectrum and their implications on the observability and reliability of current models to interpret JWST observations.
    Finally, we conclude in Sect. \ref{section:conclusion} and discuss potential future improvements on this work.

\section{Detailed combustion network selection}
\label{section:mechanism_selection}

    \subsection{Considered combustion networks}

   Seven networks validated on combustion experiments have been compared: NUIGMech1.1, AramcoMech3.0, Burke 2016, Exgas 2014, Konnov 2005, Glarborg 2018, and Venot 2020.
    The first three networks, NUIGMech1.1, AramcoMech3.0, and Burke 2016, have been developed by Curran et al. at the National University of Ireland in Galway, which led the improvements of combustion kinetics in the last years.
    \begin{description}
    \item[\textbf{NUIGMech1.1:}] Currently, NUIGMech1.1 \citep{nuigmodel} is the state-of-the-art kinetic network for C/H/O combustion. This network, that has been extensively validated against experimental data, describes the combustion kinetics of species up to molecules containing seven carbon atoms (\ce{C7}). It also contains nitrogen reactions for the chemistry of \ce{NO_x}, which are regulated pollutants in combustion processes. This level of details to capture the chemistry of \ce{C0}-\ce{C7} species is achieved at the cost of a very large network size (2746 species and 11279 reactions).
    \end{description}
    Because of the large size of NUIGMech1.1, which makes it impractical for 1D calculations, two smaller C/H/O networks from the same team were also considered: AramcoMech3.0 and Burke 2016.
    \begin{description}
    \item[\textbf{AramcoMech3.0:}] AramcoMech3.0 \citep{aramcomodel} is a C/H/O \ce{C4} network of 581 species and 3037 reactions that focuses on improving the simulations of Polycyclic Aromatic Hydrocarbon formation. \\
    \item[\textbf{Burke 2016:}] Burke 2016 is a C/H/O \ce{C3} network of 173 species and 1011 reactions \citep{curranmodel}, which aimed to better reproduce the combustion of methanol, involved in the combustion of biofuels.% since it lately became a chemical of high interest due to the intense research on biofuels.
    \end{description}
    To verify the performances of these networks, we included another network of the literature for C/H/O chemistry in our comparisons: Exgas 2014.
    \begin{description}
    \item[\textbf{Exgas 2014:}] 
    Exgas 2014 \citep{exgasmodel} is a C/H/O \ce{C3} network of 209 species and 1472 reactions generated with Exgas %, a software developed at the Laboratoire Réactions et Génie des Procédés and published in
    \citep{exgasgenerator}, a software that automatically generates combustion detailed kinetic networks. It was used to predict auto-ignition temperatures and delays for gas turbine applications.  %  the appropriate chemical network from the user's requirements on the imposed validity conditions and other parameters. This model was requested by General Electric to study natural gas conditions of ignition in the pipes of gas turbines.
    \end{description}
    Because all these networks besides NUIGMech1.1 lacked nitrogen chemistry, we included three other networks of the literature on C/H/O/N chemistry : Konnov 2005, Venot 2020, and Glarborg 2018.
    \begin{description}
    \item[\textbf{Konnov 2005:}] Konnov 2005 is a C/H/O/N \ce{C2} network of 127 species and 1213 reactions \citep{konnovmodel} designed to study the oxidation of \ce{NO} into \ce{NO2} in a medium containing ethane and was part of the research effort to reduce \ce{NO_x} emissions from car engines due to toxicity and pollution concerns. \\
    \item[\textbf{Venot 2020:}] Venot 2020 is the \cite{venot2020} chemical network, which is an updated version of the \cite{venot2012} network from which the methanol chemistry was reevaluated. It was especially designed for the study of exoplanet disequilibrium chemistry. It is a C/H/O/N \ce{C2} network of 112 species and 944 reactions, also derived from four experimentally validated combustion networks \citep{bounaceur2010, konnov2009, dagaut2008a, curranmodel}. \\
    \item[\textbf{Glarborg 2018:}] Glarborg 2018 is a C/H/O/N \ce{C3} network \citep{glarborgmodel} that aimed at improving the precision of nitrogen chemistry, especially \ce{NO_x} formation. It is a very comprehensive and widely used network for the modelling of nitrogen chemistry in combustion.
    \end{description}
    
    \noindent For clarity, all the networks used for comparison are listed in \mbox{Table \ref{tab:models}}.

    \begin{table}[htbp]
    \caption{Characteristics of the chemical networks considered and compared in this study. Size corresponds to the heavier reactant included in the network.}
    \begin{tabular}{|c|c|c|c|c|}
    \hline
    \textbf{Name} & \textbf{Species} & \textbf{Reactions} & \textbf{Size} & \textbf{Atoms} \\ \hline
    NUIGMech1.1 & 2746 & 11279 & \ce{C7} & C/H/O/N \\
    AramcoMech3.0 & 581 & 3037 & \ce{C4} & C/H/O \\
    Exgas 2014 & 209 & 1472 & \ce{C3} & C/H/O \\
    Burke 2016 & 173 & 1011 & \ce{C3} & C/H/O \\
    Glarborg 2018 & 151 & 1397 & \ce{C3} & C/H/O/N \\
    Konnov 2005 & 127 & 1213 & \ce{C2} & C/H/O/N \\
    Venot 2020 & 112 & 944 & \ce{C2} & C/H/O/N \\
    \hline
    \end{tabular}
    %\caption{Table of the different considered mechanisms for the C/H/O/N base and their characteristics.}
    \label{tab:models}
    \end{table}
    
        % Conditions Stats
        \begin{figure*}[t!]
             \begin{subfigure}{0.5\textwidth}
                 \centering
                 \includegraphics[width=1.0\textwidth]{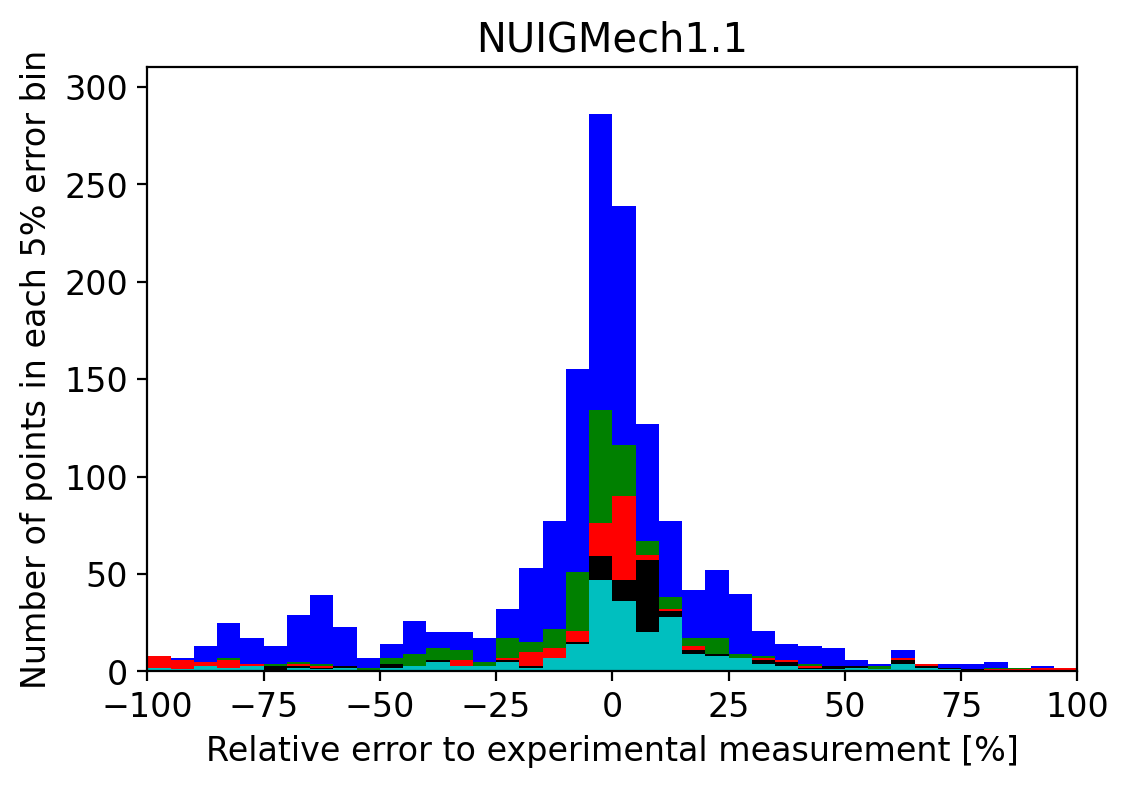}
             \end{subfigure}
             \hfill
             \begin{subfigure}{0.5\textwidth}
                 \centering
                 \includegraphics[width=1.0\textwidth]{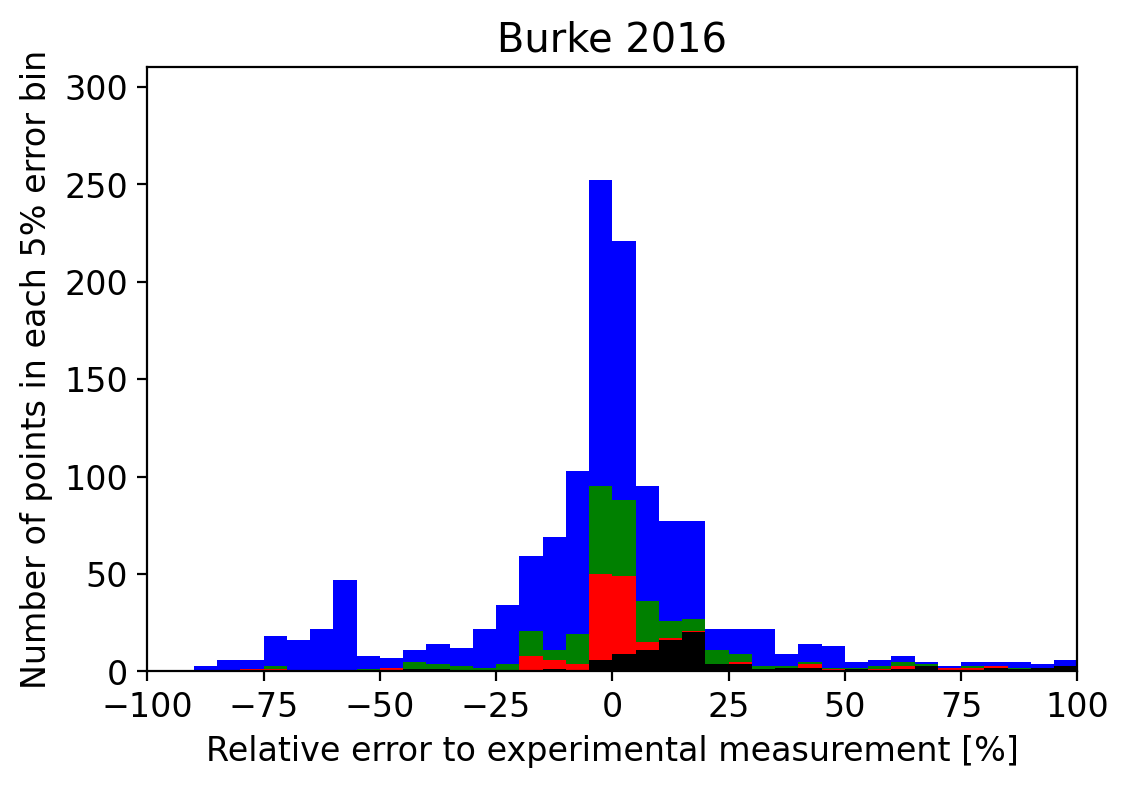}
             \end{subfigure}
             \begin{subfigure}{0.5\textwidth}
                 \centering
                 \includegraphics[width=1.0\textwidth]{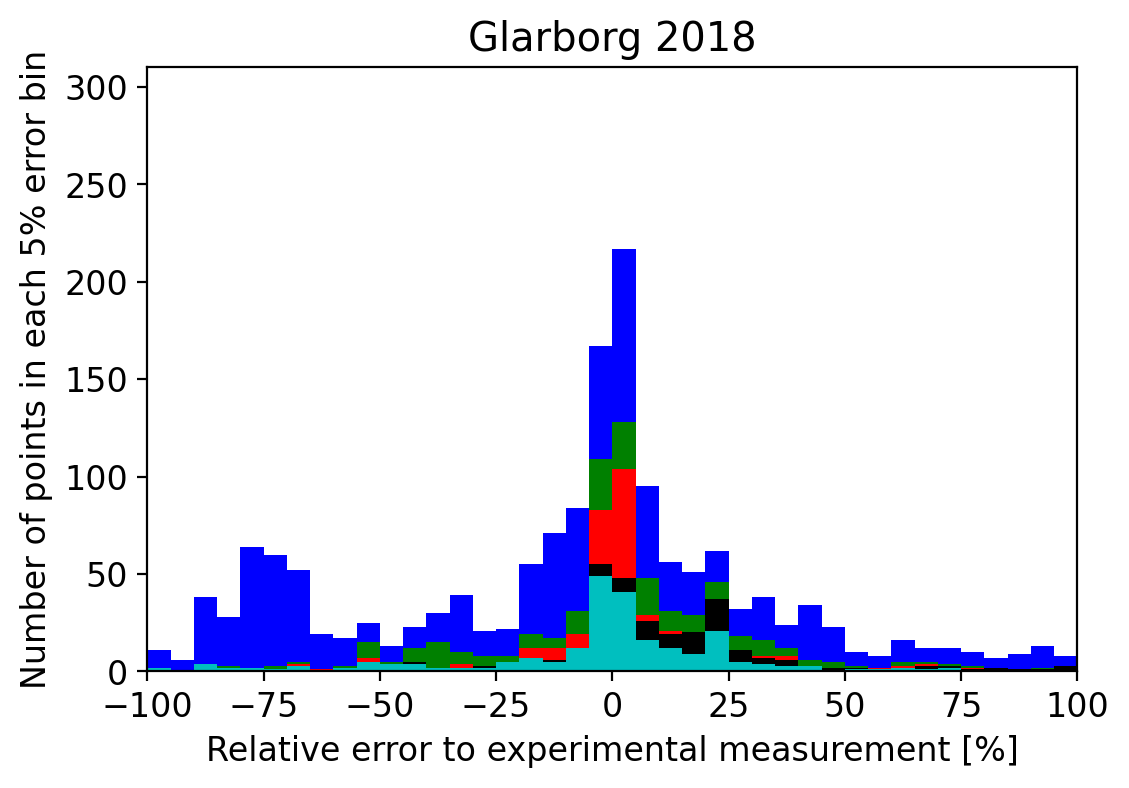}
             \end{subfigure}
             \hfill
             \begin{subfigure}{0.5\textwidth}
                 \centering
                 \includegraphics[width=1.0\textwidth]{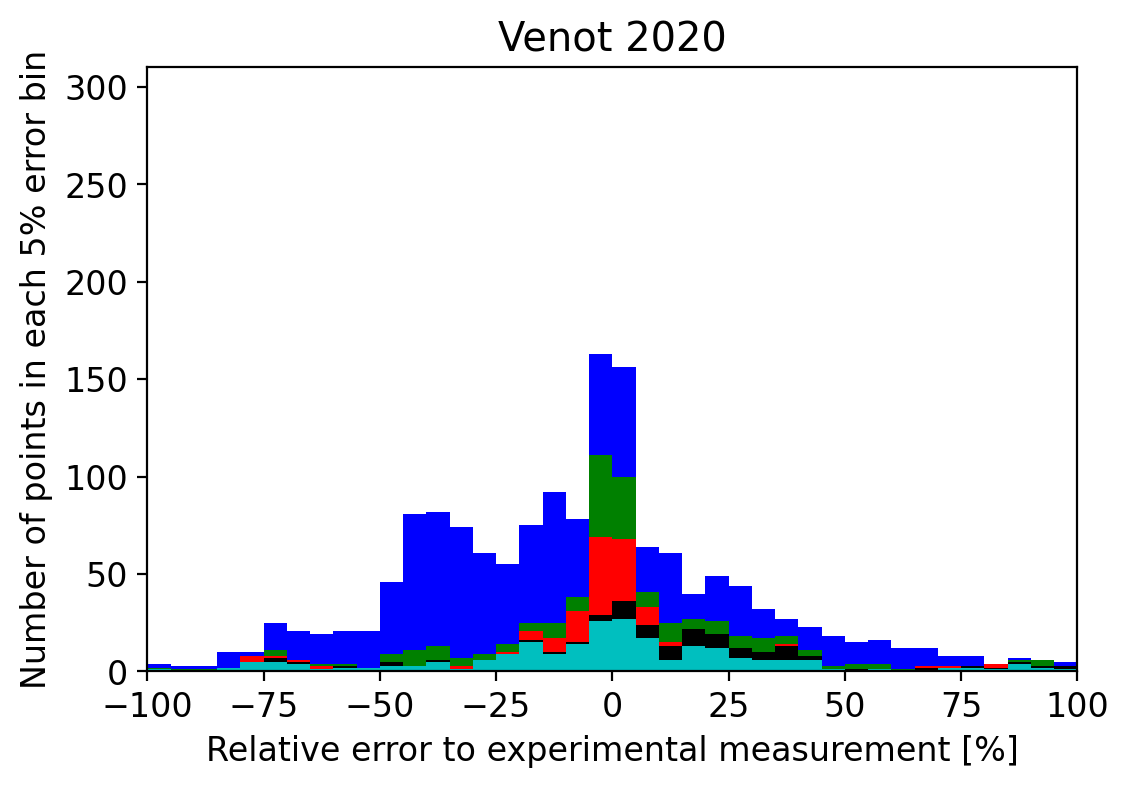}
             \end{subfigure}
             \begin{subfigure}{0.5\textwidth}
                 \centering
                 \includegraphics[width=1.0\textwidth]{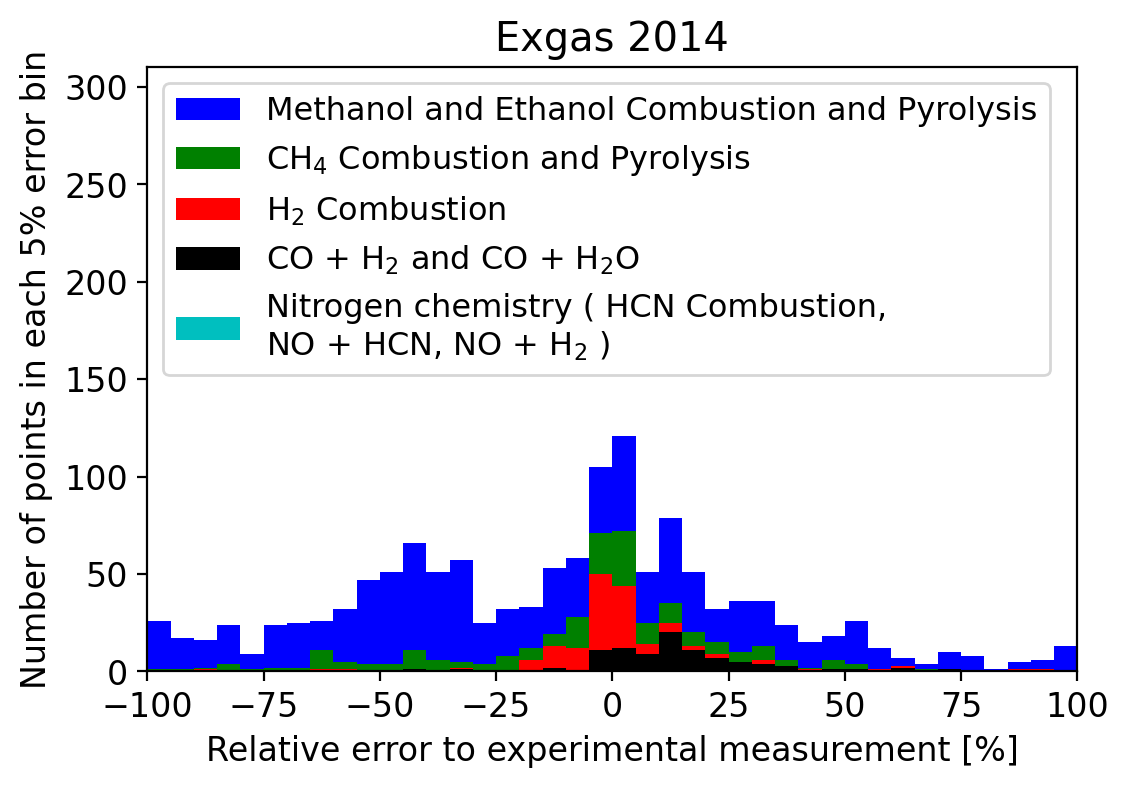}
             \end{subfigure}
             \hfill
             \begin{subfigure}{0.5\textwidth}
                 \centering
                 \includegraphics[width=1.0\textwidth]{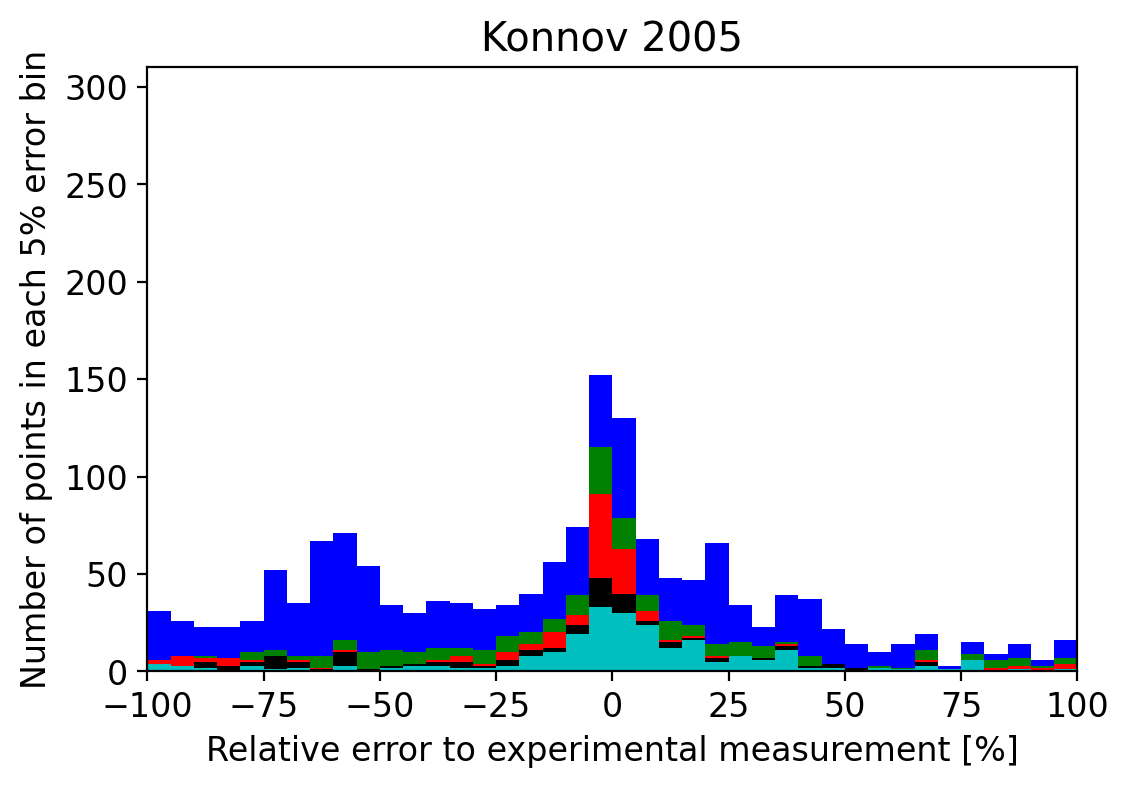}
             \end{subfigure}
                \caption{Statistical distribution of the relative error over every experimental point in the 1618 points data set for each studied chemical network. Points are grouped in colors corresponding to a different type of initial conditions. AramcoMech3.0 is not shown here as it is almost identical to Burke 2016.}
                \label{fig:cond_stats}
        \end{figure*}
        \begin{figure*}[!b]
             \begin{subfigure}{0.5\textwidth}
                 \centering
                 \includegraphics[width=1.0\textwidth]{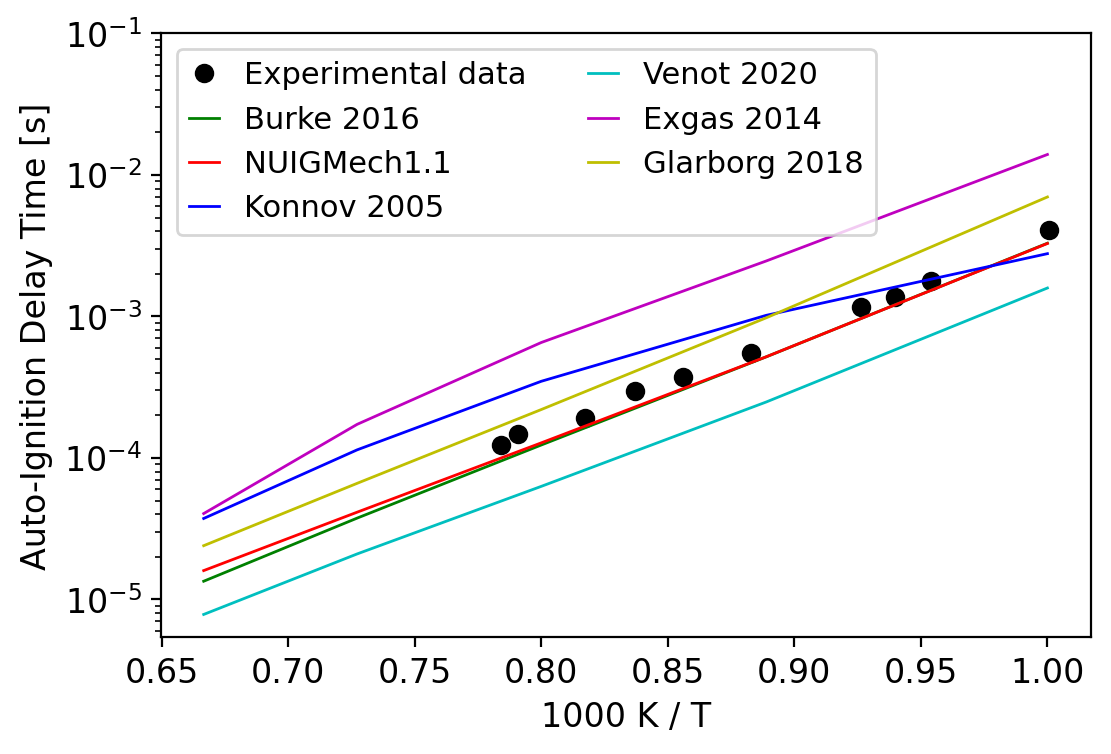}
                 \caption{10 bar}
                 \label{fig:cond1_compare_a}
             \end{subfigure}
             \hfill
             \begin{subfigure}{0.5\textwidth}
                 \centering
                 \includegraphics[width=1.0\textwidth]{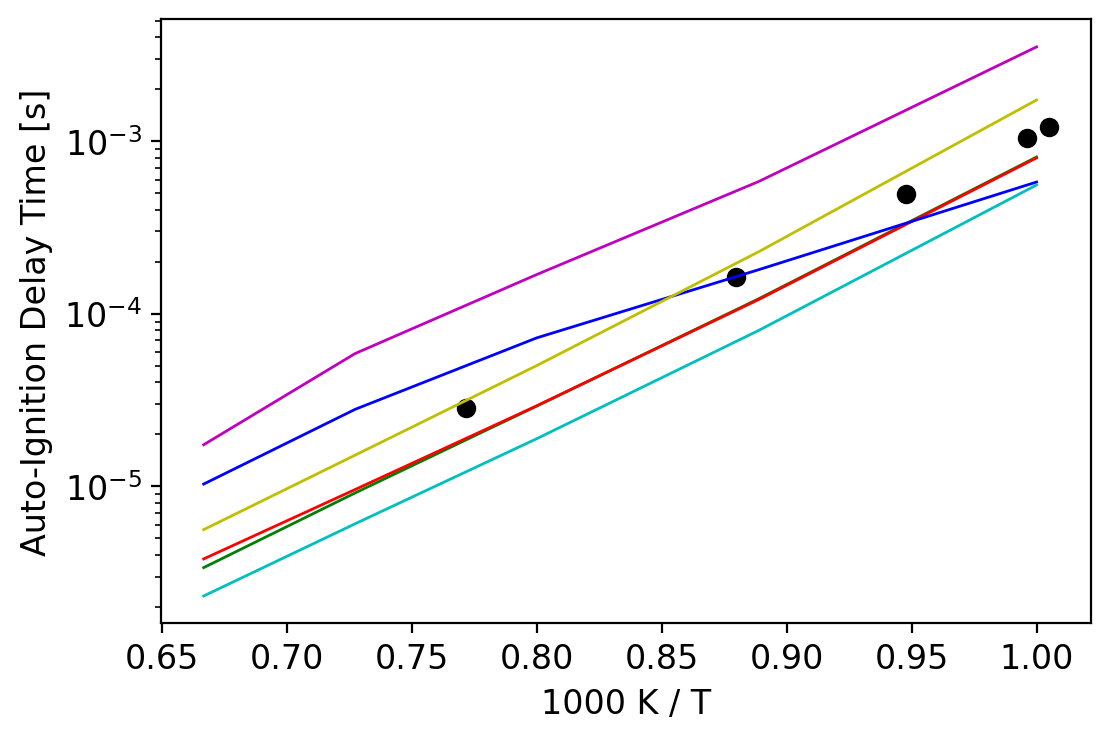}
                 \caption{50 bar}
                 \label{fig:cond1_compare_b}
             \end{subfigure}
             \begin{subfigure}{0.5\textwidth}
                 \centering
                 \includegraphics[width=1.0\textwidth]{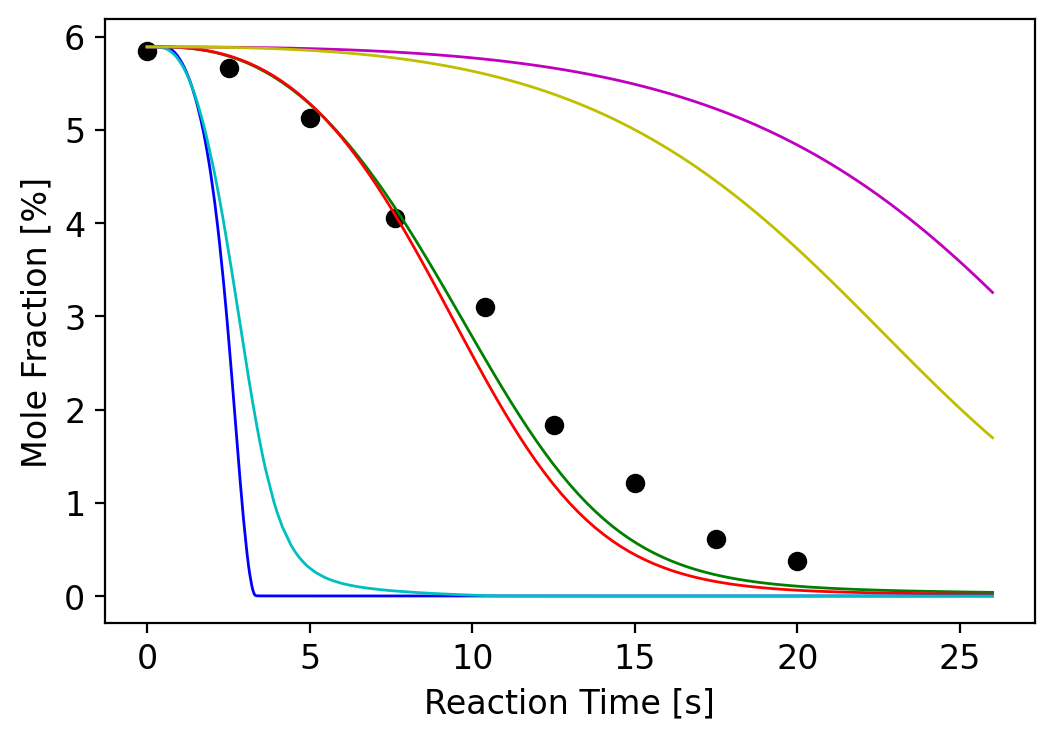}
                 \caption{\ce{CH3OH}}
                 \label{fig:cond2_compare_a}
             \end{subfigure}
             \hfill
             \begin{subfigure}{0.5\textwidth}
                 \centering
                 \includegraphics[width=1.0\textwidth]{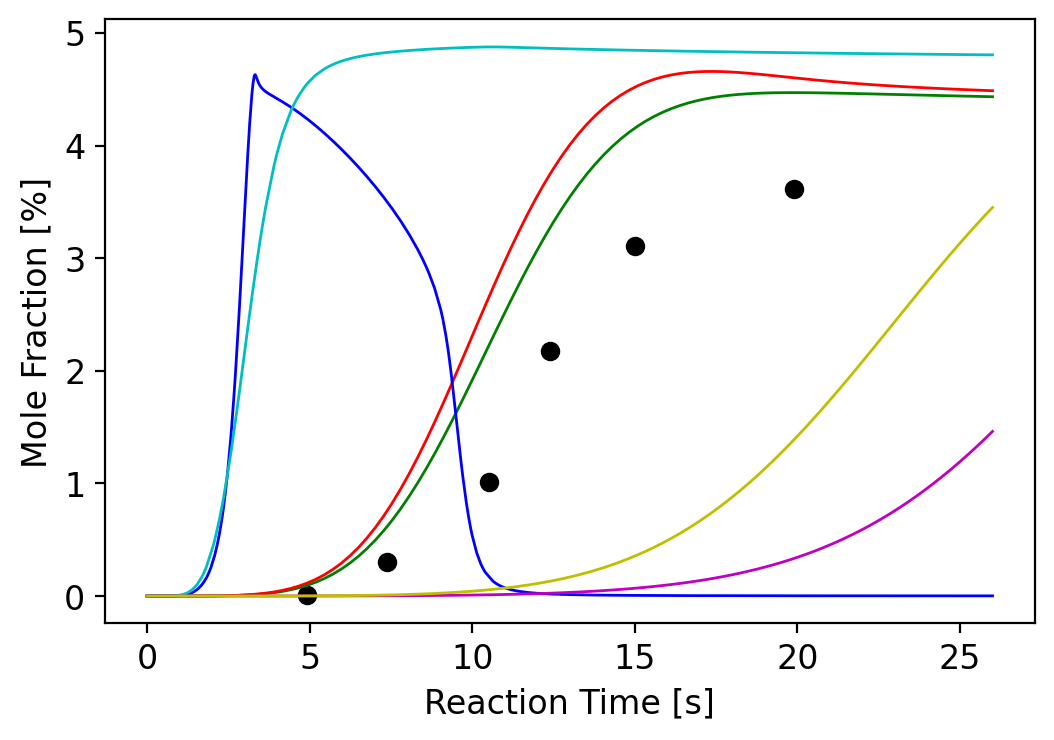}
                 \caption{CO}
                 \label{fig:cond2_compare_b}
             \end{subfigure}
             \caption{Ignition delay time of methanol at 10 bar (a) and 50 bar (b) in condition 1 and mole fraction of \ce{CH3OH} (c) and CO (d) over time in condition 2 of Table \ref{tab:conditions} for all tested chemical networks. AramcoMech3.0 is not shown here as it is almost identical to Burke 2016.}
             %\label{fig:cond2_compare}
             \label{fig:cond1and2_compare}
        \end{figure*}
        
        \begin{figure*}[!b] % Species4 Stats: alcools species
             \begin{subfigure}{0.5\textwidth}
                 \centering
                 \includegraphics[width=1.0\textwidth]{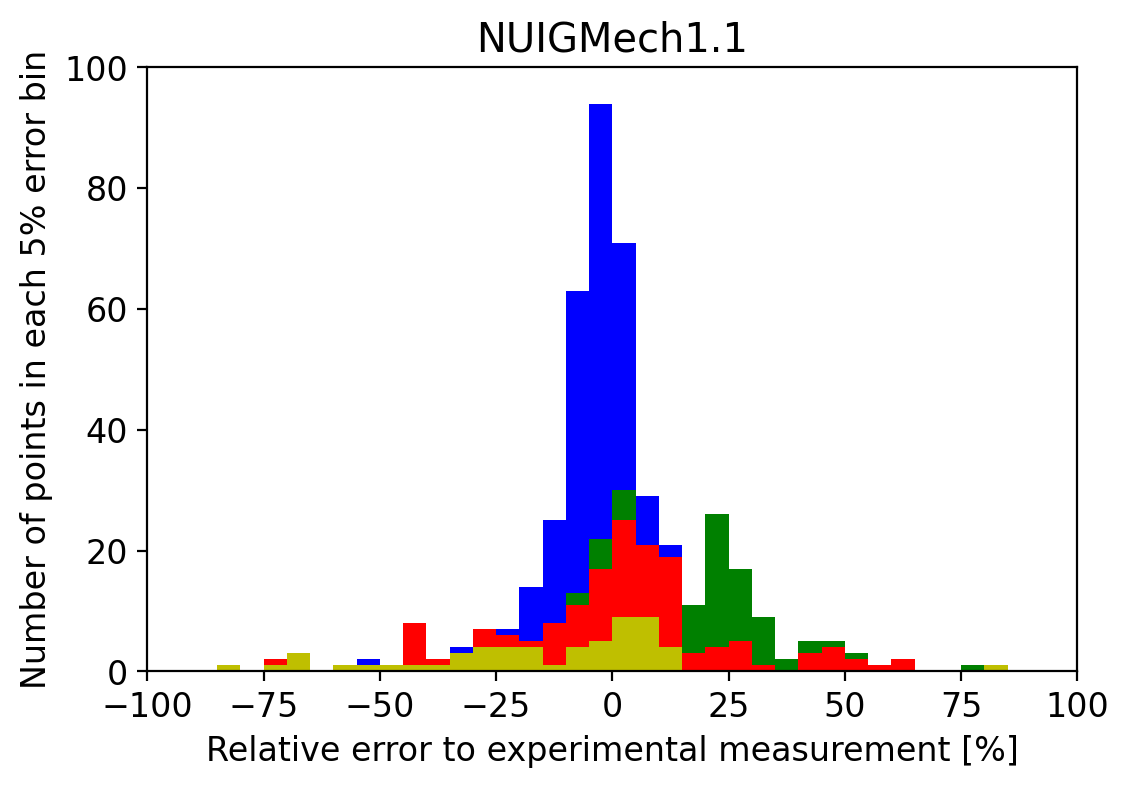}
             \end{subfigure}
             \hfill
             \begin{subfigure}{0.5\textwidth}
                 \centering
                 \includegraphics[width=1.0\textwidth]{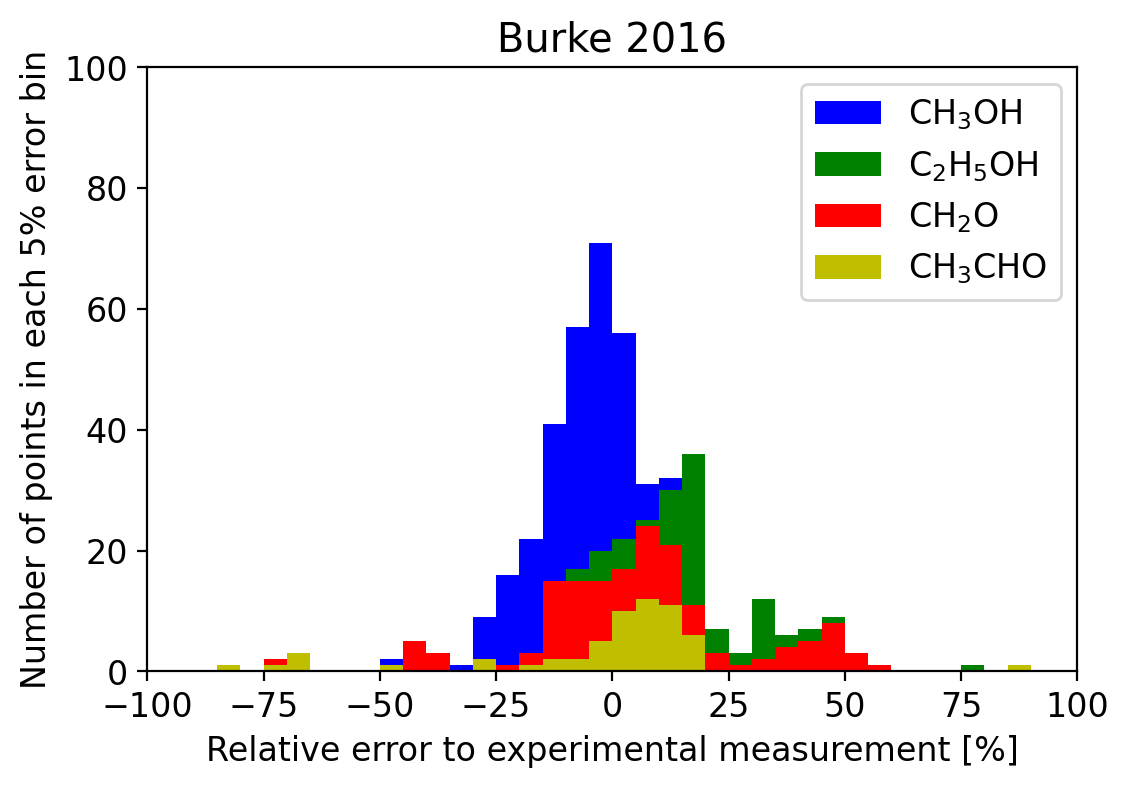}
             \end{subfigure}
             \begin{subfigure}{0.5\textwidth}
                 \centering
                 \includegraphics[width=1.0\textwidth]{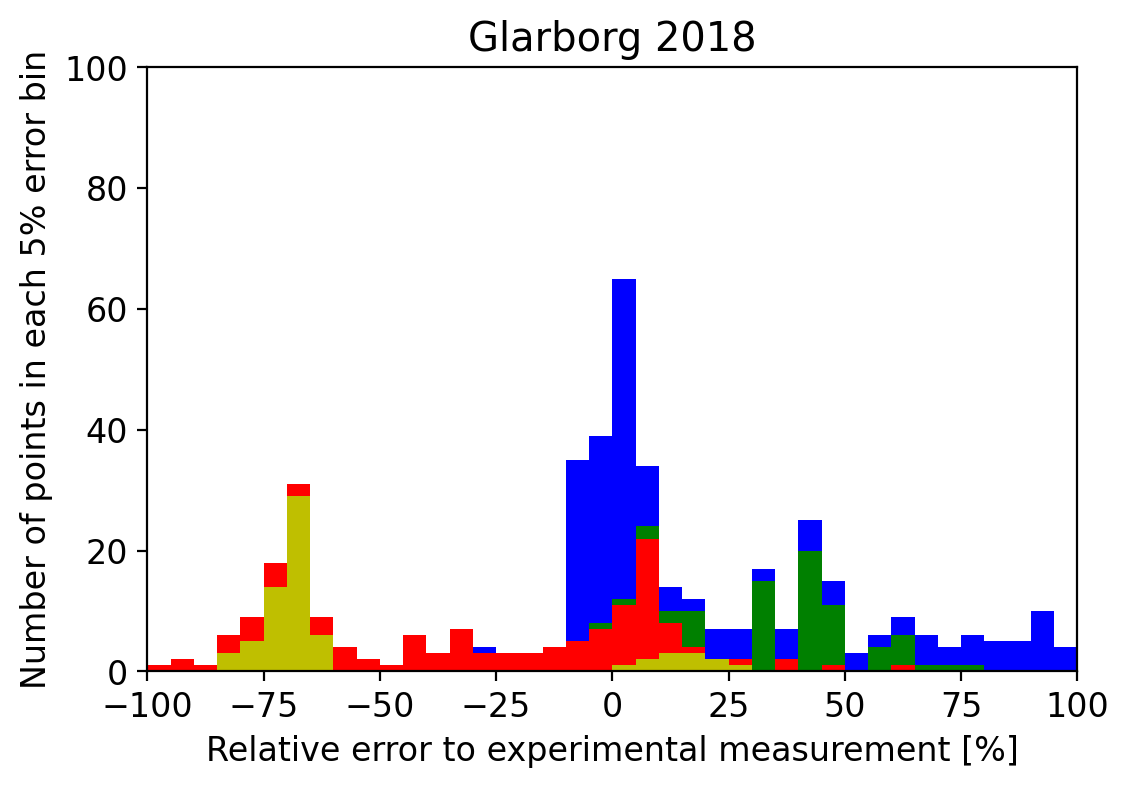}
             \end{subfigure}
             \hfill
             \begin{subfigure}{0.5\textwidth}
                 \centering
                 \includegraphics[width=1.0\textwidth]{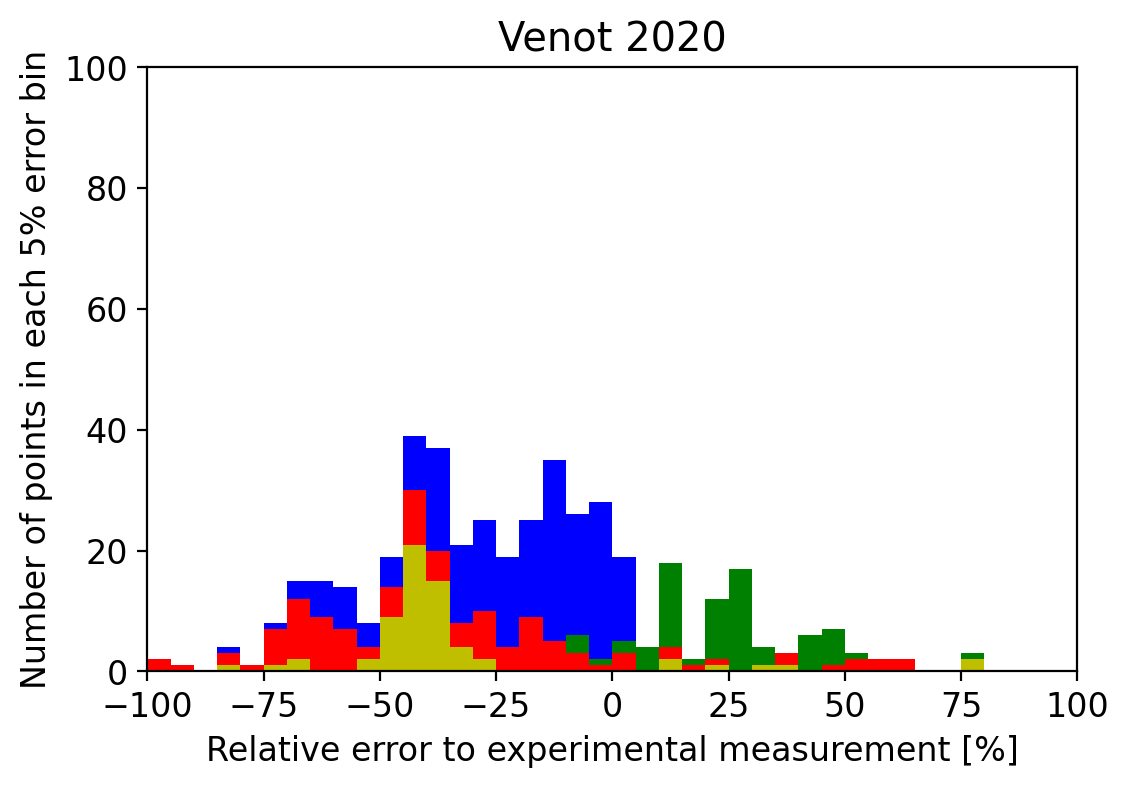}
             \end{subfigure}
                \caption{Statistical distribution of the relative error over the 449 experimental data points for intermediate products (\ce{CH3CHO}, \ce{CH2O}) in combustion and pyrolysis of \ce{C2H5OH} and \ce{CH3OH}. Each color gathers all molar fraction measurements of the corresponding species. Contribution from combustion and pyrolysis data are shown separately in Figs. \ref{fig:combustion_alcools} and \ref{fig:pyrolysis_alcools}.}
                \label{fig:species4_stats}
        \end{figure*}
        \begin{figure*}[htbp] % Species1 Stats: general species
             \begin{subfigure}[b]{0.5\textwidth}
                 \centering
                 \includegraphics[width=1.0\textwidth]{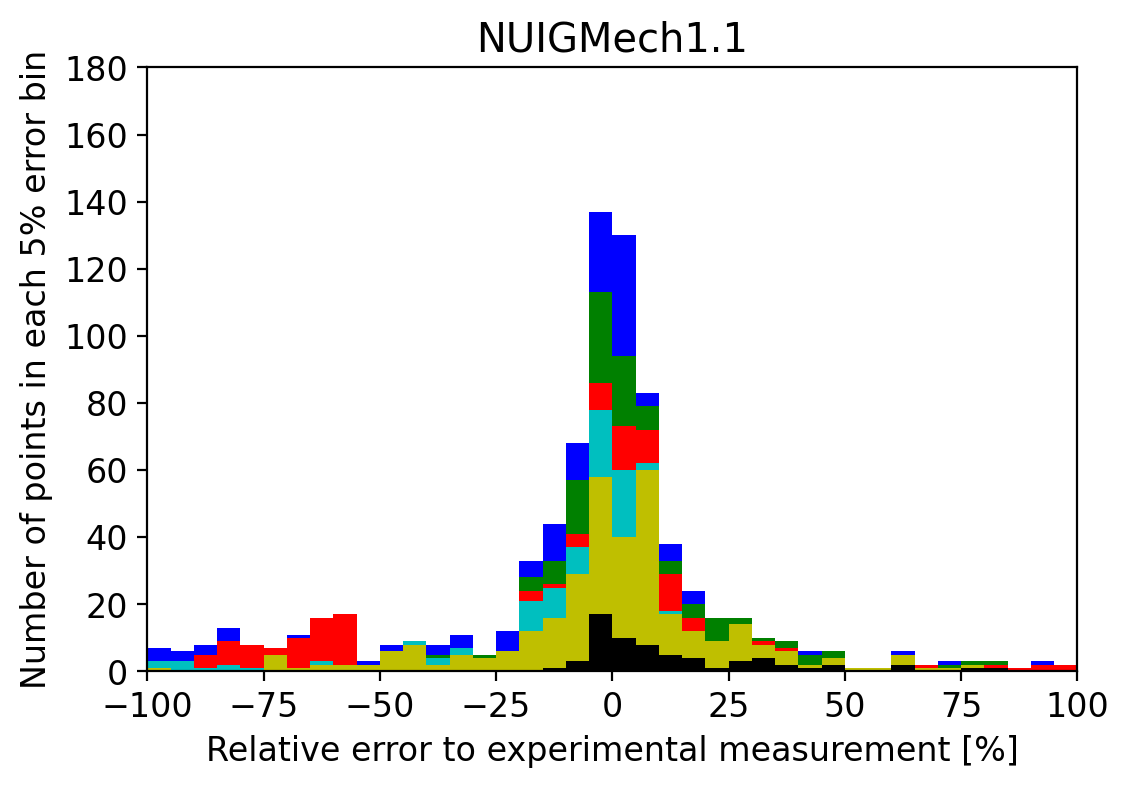}
             \end{subfigure}
             \hfill
             \begin{subfigure}[b]{0.5\textwidth}
                 \centering
                 \includegraphics[width=1.0\textwidth]{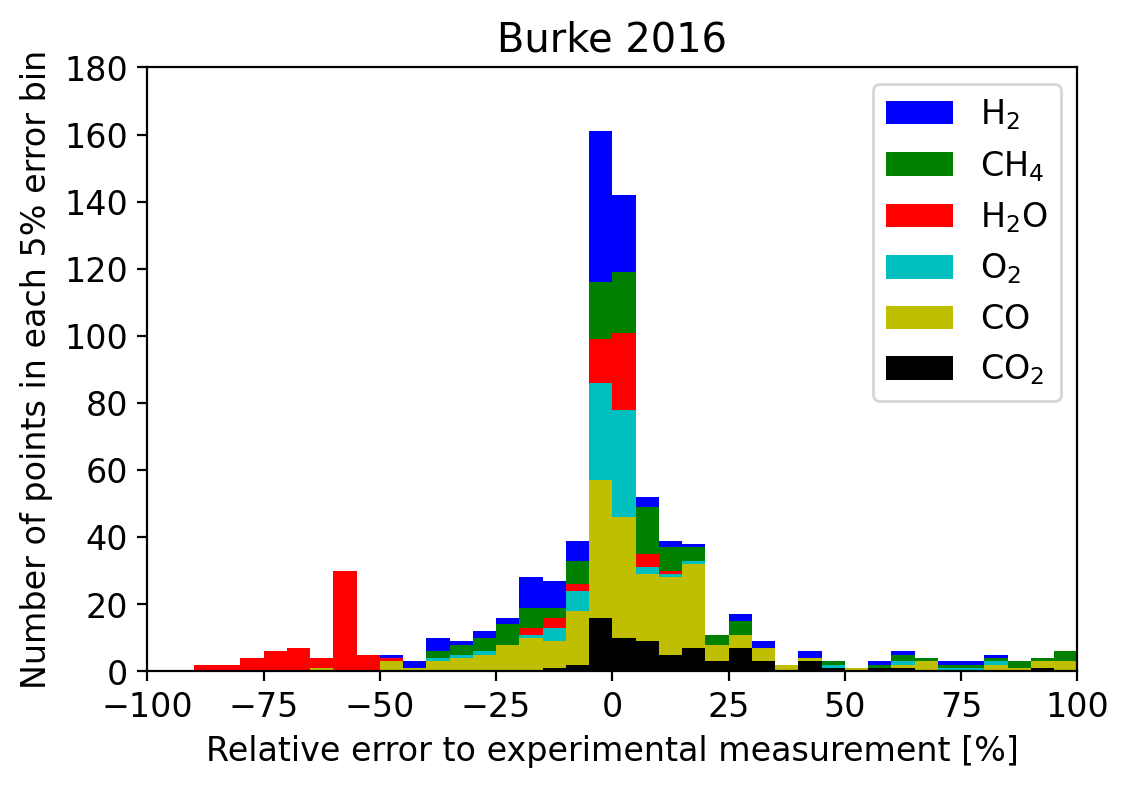}
             \end{subfigure}
             \begin{subfigure}[b]{0.5\textwidth}
                 \centering
                 \includegraphics[width=1.0\textwidth]{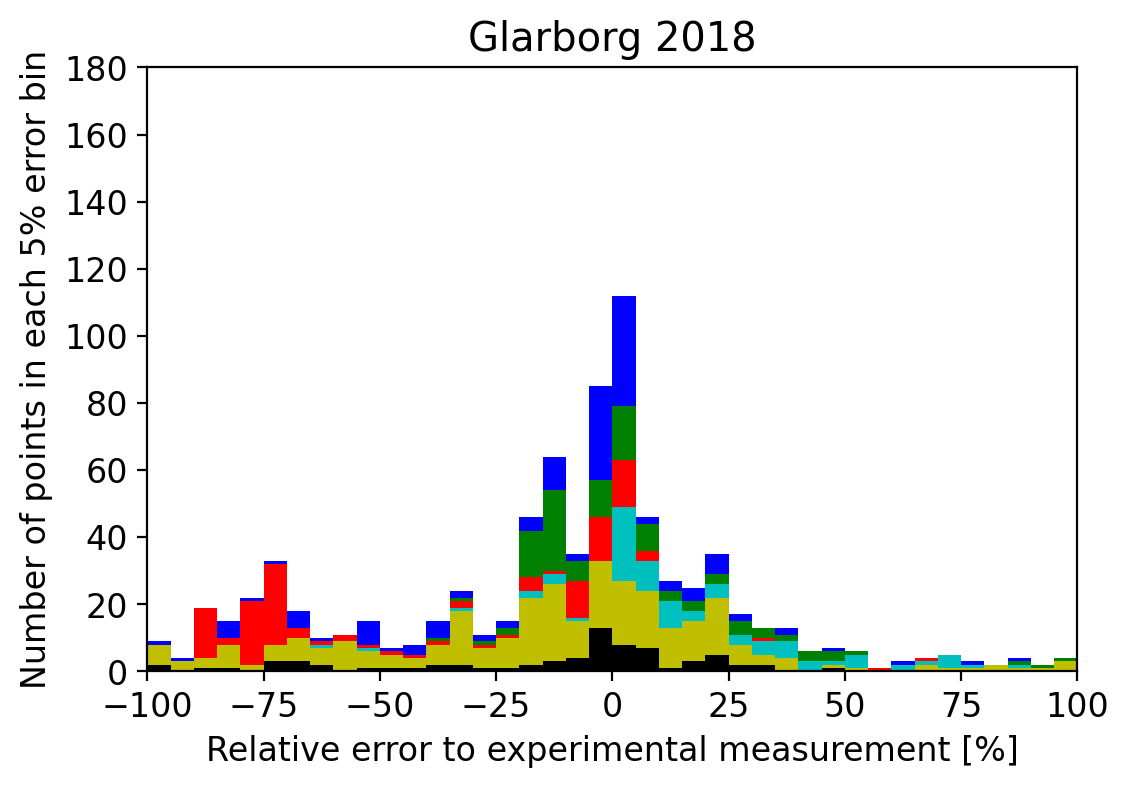}
             \end{subfigure}
             \hfill
             \begin{subfigure}[b]{0.5\textwidth}
                 \centering
                 \includegraphics[width=1.0\textwidth]{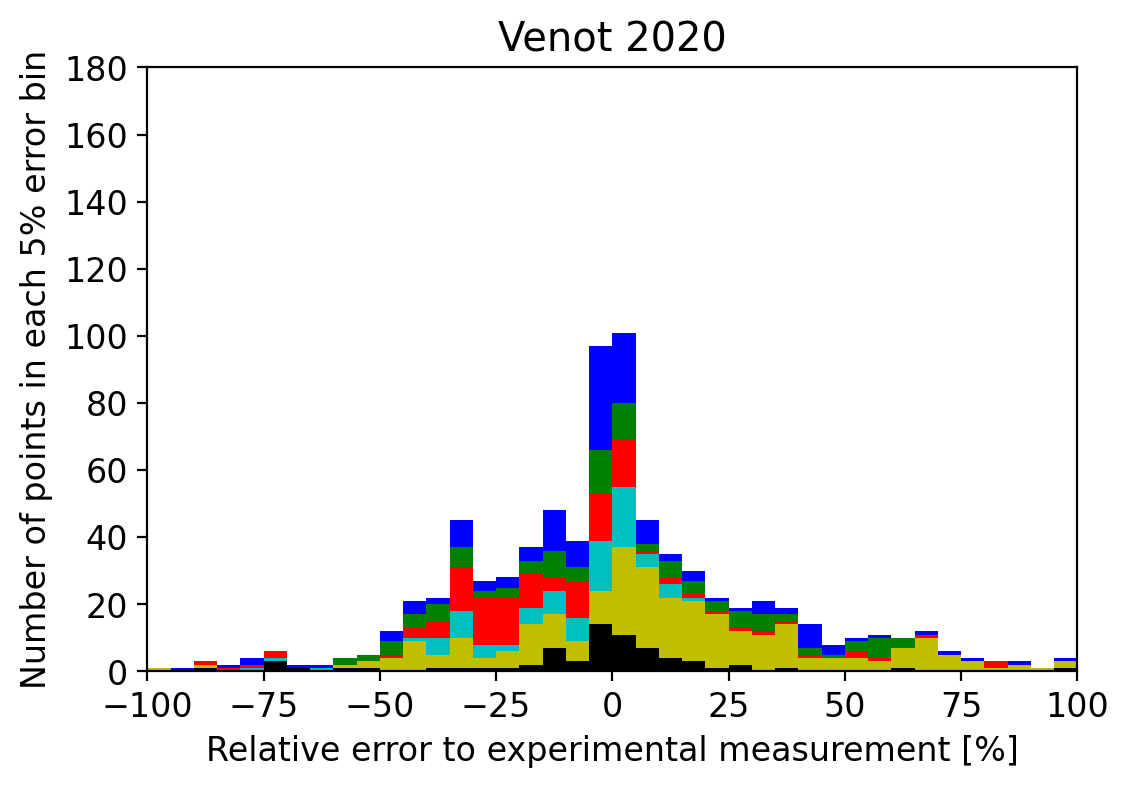}
             \end{subfigure}
                \caption{Statistical distribution of the relative error over the 823 experimental data points for the main pyrolysis products (\ce{H2}, \ce{CH4}), combustion products (\ce{H2O}, CO, \ce{CO2}), and reactants (\ce{O2}) in combustion and pyrolysis of \ce{CH3OH}, \ce{CH4}, \ce{H2}, HCN, and \ce{C2H5OH}. For some points, CO and \ce{H2O} are reactants (condition 13 of Table \ref{tab:conditions}). Each color gathers all molar fraction measurements of the corresponding species. Contribution from combustion and pyrolysis data are shown separately in Figs. \ref{fig:combustion_species} and \ref{fig:pyrolysis_species}.}
                \label{fig:species1_stats}
        \end{figure*}

    \subsection{Experimental data}

    % Data description
    In order to select the best chemical network for our requirements, we gathered 1618 combustion experimental data points, tested the %candidate
    seven different networks over conditions detailed in Table \ref{tab:conditions} in the appendix using the Ansys software Chemkin-Pro \citep{chemkin}, and, finally, we compared them to the experimental data.
    For a large majority, these data consisted in molar fraction measurements of different species (reactants and products, 1558 measurements out of 1618), but also in the measurement of auto ignition delay times (IDT, 60 measurements out of 1618).
    %This delay corresponds to the time it takes for a fuel mixture to spontaneously combust at a given temperature and pressure. In experiments, the measured quantity is the delay between the temperature and pressure increase from the experimenter (induced by a shock wave for shock tube reactors) that causes the combustion, and the maximum concentration of the first excited state of the OH radical. Therefore, in simulations, this delay is also chosen to be the time at the maximum concentration of the OH radical.
    This delay corresponds to the time it takes for a fuel mixture to spontaneously ignite at a given temperature and pressure. In experiments, it is measured as the time between the moment when the gas is brought to temperature and pressure conditions and the moment when the ignition is detected. It is most often detected by a pressure or concentration peak of excited OH or CH radicals. In our simulations, the ignition delay time is chosen to be the time at the maximum concentration of the OH radical.
    These experimental data were collected over 21 different publications in total, from a wide range of conditions fully described in the appendix.
    The first eight experimental conditions considered were taken from those used in \cite{venot2020}, to determine how the other chemical networks compare to it on the original data used for its validation.
    %Among these conditions, 7 were combustion conditions and 1 was a pyrolysis one.
    The data collected for the first six experimental conditions consisted in the temporal evolution of the abundances of major species at play at the start and end of reaction (\(\ce{CH3OH}\), \(\ce{O2}\), \(\ce{CO}\), \(\ce{CO2}\), \(\ce{H2O}\), \(\ce{HCHO}\), \(\ce{H2}\) \dots) in three different reactor types (closed reactor, plug flow reactor, shock tube). The seventh consisted in the auto-ignition delay time measurement in a shock tube at an initial pressure of 10 and 50 bar, at 10 and 5 different initial temperatures, respectively, over a range of 1000 to 1300 K. For the eighth, it consisted in the evolution with temperature of abundances of major species at the exit of a perfectly-stirred reactor.
    The rest of the experimental conditions (9 to 21 in Table \ref{tab:conditions}) were focused on exploring a wider range of initial species and conditions by varying equivalence ratios, from very oxygen-rich combustion to pyrolysis, in addition to varying the fuel type, which consisted in combustion of \ce{H2}, of \ce{HCN}, pyrolysis of \ce{CH4}, of \ce{C2H5OH}, as well as reactions of nitrogen species like \ce{N2O}, \ce{NO}, or \ce{NH3}.
    Like the eight first ones, these data consisted in auto-ignition delay times, abundances over time or abundances at steady state over temperature, and sometimes with a parameter study on equivalence ratio, pressure, or different initial species.
    The species concerned by these abundance data can be reactants (\(\ce{H2}\), \(\ce{CH4}\), \(\ce{HCN}\), \(\ce{C2H5OH}\), \(\ce{O2}\) \dots), products (\(\ce{H2O}\), \(\ce{CO2}\), \(\ce{CO}\),  \(\ce{C2H2}\),  \(\ce{C2H4}\),  \(\ce{C2H6}\), \(\ce{CH3CHO}\), \(\ce{HCHO}\) \dots) or appear as both in the data set depending on the conditions (\ce{CH4}, \ce{H2}).

    In combustion conditions, the parameter describing the abundance of fuel to oxidizer is the equivalence ratio: $$ \phi = \frac{n_{fuel}/n_{ox}}{[n_{fuel}/n_{ox}]_{sto}},$$ \noindent with $n_{fuel}$ as the fuel quantity, $n_{ox}$ as the oxidizer quantity, and $[n_{fuel}/n_{ox}]_{sto}$ as the ratio of these values in a stoichiometric mixture.
    As the equivalence ratio grows larger, the fuel proportion gets higher and the combustion conditions get closer to pyrolysis conditions.
    Pyrolysis corresponds to high temperature conditions in a reducing medium with no oxygen, while combustion refers to high temperature conditions in an oxidizing medium, usually oxygen.
    Covering a wide range of equivalence ratios in our dataset allows us to test the kinetic networks on different compositions to ensure their ability to accurately model the chemistry occurring on exoplanets with very different elemental abundances.
    %Covering a large span of equivalence ratios in our data set allows for testing the kinetic networks over different compositions to gain confidence in its ability to accurately model the chemistry occurring in exoplanets with very different elemental abundances.
    Pyrolysis and a high equivalence ratio combustion corresponds to low metallicities with high C/O, N/O, and H/O ratios, while a low equivalence ratio combustion corresponds to high metallicities with low C/O, N/O, and H/O ratios.
    In total, the full experimental data set spanned equivalence ratios from 0.05 to 5 and pyrolysis conditions, pressures from 0.2 to 50 bar and temperatures from 800 to 2400 K.

    \subsection{Error calculations}

    % Models/Experiments comparison
    To test the agreement of the different chemical networks with the experimental data, network predictions were plotted against experimental points and compared. This resulted in over 500 plots, which is too much to be shown here. Therefore, we will focus on the distribution of errors for each chemical network, compiled in histograms of Fig. \ref{fig:cond_stats}, and discuss the main tendencies visible in the overall dataset.

    % Method explanation of error calculations
    To sum up these numerous plots into a statistical distribution of errors shown in Fig. \ref{fig:cond_stats}, we chose to compute these errors using the following formula: $$ y_{error} = \frac{y_{mod} - y_{exp}}{y_{max}}, $$\noindent with $y_{exp}$ being the values of each experimental point in our dataset, $y_{mod}$ being the network prediction at that point and $y_{max}$ being the value of the highest experimental point over the experimental range.
    Each experimental point corresponds to a measurement of the molar fraction of a species (either product or reactant, for 1558 measurements out of 1618), but also of the IDT of a mixture (60 measurements out of 1618). Depending on the reactor type, for a given experimental range of measurements, pressure, temperature, or reaction time can change.
    This range depends on the type of data, and corresponds to the temperature range of the original measurements for temperature studies, and to the time range of the original measurements for mole fraction over time studies.
    This choice is done to give a relative error that can be compared between different experiments, while avoiding non-representative errors due to data points close to zero and experimental and pointing noise causing diverging relative errors.
    The assumed network prediction corresponds to the linear interpolation between the two closest computed points. For temperature studies, the computed points were evenly distributed over the experimental range, compromising between the density of the distribution and the computational time. For time studies, the computed points were determined by the software used for the calculations.

    In these histograms, NUIGMech1.1, Burke 2016, and Glarborg 2018, display the models with the best prediction accuracy over the dataset.
    In the following, the focus is more on in-depth descriptions of the causes underlying these results.
    In total, over the 1618 experimental points, about 50 were beyond 100\% calculated error for all models, with maximum values reaching around 2500\%. %These extreme values do not appear in the plotted distribution of errors,
    These high deviations are found with all the chemical networks and were not plotted in error distributions.
    They mainly come from conditions 16, 17, and 18 in Table \ref{tab:conditions} for ethanol and methane pyrolysis, the important discrepancies observed between experiments and simulations appearing for plug flow experiments. For these specific experiments, important shifts in time (or temperature) are observed which dramatically affect the $y_{error}$ calculated.
    % Concerned species are mainly \ce{C2H2} in conditions 16 and 17, and \ce{H2} in conditions 18, especially at high temperatures.
    There is also an abnormal distribution of errors around -70\% seen in Fig. \ref{fig:cond_stats} for all models, mainly coming from conditions 19 of Table \ref{tab:conditions} and concerning \ce{C2H4} and \ce{H2O}. %, as seen in Fig. \ref{fig:species1_stats} and \ref{fig:species3_stats_NUIG}.
    These errors could be due to issues in the experimental points, for example, due to ethanol reacting before entering the reactor.
    %, or due to uncaptured \ce{H2} by the measurement system. It could also be due to common weak points in all the models for the formation of \ce{C2} species at high temperature in pyrolysis conditions.

    \subsection{Auto ignition delay time}
    %\subsection{Error comparisons}

    When first comparing the different plots for each network on methanol combustion (conditions 1-9 in Table \ref{tab:conditions}), the first thing that stood out was that the chemical networks based on the work of Curran and co-authors (AramcoMech3.0, NUIGMech1.1, and Burke 2016), were all way better at describing auto ignition delay times. They agree with the auto ignition data of methanol shown in Fig. \ref{fig:cond1_compare_a} and \ref{fig:cond1_compare_b} with a mean error of 5\% at 10 bar and within 10\% at 50 bar, with almost no visible difference between them.
    On the contrary, the Exgas network severely overestimates this delay, with almost an order of magnitude difference. The Konnov network does not reproduce the temperature dependence: delays are underestimated at low temperatures (under 1050 K) and overestimated at high temperatures (over 1050 K) by around 150\%. For the Glarborg network, the delay is overestimated at both pressures by around 30\%, and for the Venot 2020, this delay is too short by around -40\%.
    The ability of each network to accurately describe the auto ignition delay time in given initial conditions has major consequences on kinetic simulations of mole fraction over time. When a network underestimates the IDT, fuel consumption will tend to be overestimated. This correlation is clearly visible in our dataset. Fig. \ref{fig:cond1_compare_a} and \ref{fig:cond1_compare_b} shows that Venot 2020 underestimates the IDT while Glarborg 2018 overestimates it. This is related to Fig. \ref{fig:cond2_compare_a}, where \ce{CH3OH} consumption is overestimated for Venot 2020 and underestimated for Glarborg 2018. This impact is clearly seen in combustion conditions, as shown in   Fig. \ref{fig:combustion_alcools}, leading to \ce{CH3OH} error distribution of Venot 2020 to be mostly between -100\% and 0\% error, and between 0 and 100\% error for Glarborg 2018. Figure \ref{fig:cond2_compare_b} shows that in consequence, products like CO tend to be overestimated around ignition time for networks with underestimated IDT, as in Venot 2020, and, conversely, for networks with overestimated IDT like Glarborg 2018. As the oxidation of intermediate species like CO is not directly linked to IDT, other parameters may control CO consumption, as visible with Konnov 2005.
    
    %The ability of each network to accurately describe the auto ignition delay time in given initial conditions has major consequences on kinetic simulations of mole fraction over time. If this delay is too small in comparison with experimental data, the abundances of the reactants will be underestimated, whereas products will be overestimated, and the opposite will be true if the delay is too large, as show in Fig. \ref{fig:cond2_compare_a} and \ref{fig:cond2_compare_b}.
    %\textbf{However, for intermediate combustion products like CO, IDT can influence its rate of production but not its rate of consumption, therefore leading to a difference between Venot 2020 and Konnov 2005 after 5 seconds of reaction time for example, despite these models having a similar \ce{CH3OH} IDT at these low temperatures. Venot 2020 overestimates CO after this time, and Konnov 2005 underestimates it, thus indicating that CO consumption pathways are faster in Konnov 2005. The consequences of this impact of IDT on model predictions of molar fractions over reaction time can be seen in the statistical distribution of errors, resulting in a more scattered distribution for networks that are inaccurate on this delay.}

    \subsection{Combustion and pyrolysis of methanol, ethanol, formaldehyde, and acetaldehyde}
    %This importance of ignition delay time is further shown in Fig. \ref{fig:species4_stats}. 
    Focusing on data related to methanol thermal decomposition, we can see in Fig. \ref{fig:species4_stats} that Exgas, Glarborg, and Konnov networks overestimate \ce{CH3OH} abundances profiles, while Venot 2020 tends to underestimate methanol abundances. Eventually, for these networks, their best performances on methanol points are for data in perfectly stirred reactors or plug flow reactors.

    In addition, ethanol results are also displayed in Fig. \ref{fig:species4_stats}. The experimental conditions concerning this species include only ethanol pyrolysis, with temperature- and pressure-dependent species profiles. \ce{CH3CHO} data come exclusively from these conditions (18 and 19 of Table \ref{tab:conditions}), whereas \ce{CH2O} errors also include methanol combustion experiments (2, 6, 7, and 8 of Table \ref{tab:conditions}).
    The Curran-based networks give quite similar results for these species, except for NUIGMech1.1 that is significantly better on methanol. This is %thanks to its better results over the methanol pyrolysis sub-dataset with temperature parameter study,
    probably due to a better representation of the growth mechanism towards heavier molecules occurring under pyrolysis conditions. %creates more complex species that are more likely to be modeled in a heavier mechanism.
    Overall, these networks are similar over these species. The Glarborg network %does surprisingly well
    is accurate for methanol pyrolysis, but is less effective for methanol combustion. For \ce{C2H5OH}, its performances are less accurate than the previous networks, and for \ce{CH3CHO}, experimental abundance is underestimated by about -75\%.
    %Konnov, Exgas and 
    Venot 2020 also reproduce these experimental points quite badly, especially on \ce{CH2O} and \ce{CH3CHO}.

    \subsection{Main products and reactants in combustion and pyrolysis}

    Main species are shown in Fig. \ref{fig:species1_stats}. This histogram gathers all computed errors of model predictions on experimental measurements (mole fraction and IDT) of \ce{H2}, \ce{CH4}, \ce{H2O}, \ce{O2}, CO, and \ce{CO2} regardless of their role in the experiment (either product or reactant). In condition of combustion of \ce{H2}, all the networks were within 5\% or 10\% error.
    For \ce{H2} mole fraction measurements coming from methane pyrolysis experiments however, almost every network underestimated its production compared to the experimental points, although the Curran-based networks were the closest to experiments.
    In ethanol pyrolysis, on the contrary, the \ce{H2} production was severely overestimated by all the networks, especially at high temperatures, with errors up to 200\% at 1300 K with usually reliable networks like NUIGMech1.1. 

    For methane pyrolysis, the Curran-based networks were under 5\% error, while other networks like Exgas or Venot 2020 overestimated \ce{CH4} abundances by around 20\%.
    For methane combustion, almost all networks were under the 5\% error range except for the Konnov network whose temperature dependence was totally off.
    On ethanol pyrolysis conditions, methane production were underestimated by all networks.

    For water, results were good for all the networks, except in the ethanol pyrolysis conditions, where \ce{H2O} production was severely underestimated by all the networks by around 75\%, which is shown as red bars in Fig. \ref{fig:pyrolysis_alcools}.

    For \ce{O2} consumption in methanol or hydrogen burning conditions, the best networks were the Curran-based networks, with the same problems as noted previously for others, which were  due to bad methanol ignition delay time predictions.

    One species that displayed significant  gains in accuracy with the Curran-based networks is \ce{CO}, which has a wide range of errors with such networks as Exgas 2014, Glarborg 2018, Konnov 2005, and Venot 2020. However, for \ce{CO2}, we do not see a significant improvement over our dataset in relation to these experimental data.
        \begin{table*}[b]
        \caption{Different added reactions to the network and their parameters. Values are in \si{mol}, \si{cm^3}, \si{cal}, \si{s}. Also, \textbf{A}, \textbf{n,} and \textbf{E} are the parameters of the modified Arrhenius equation, while $\Delta_r H$ is the enthalpy of formation of the reaction, whose values are taken from NIST. As C + NH and N + H are radical-radical combinations, they are barrierless reactions (\textbf{E} = 0). +M indicates low pressure limit reactions.}
        \renewcommand{\arraystretch}{1.5}
        \scalebox{1}{
        \begin{tabular}{|c|c|c|c|c|c|}
        \hline
        \textbf{Reaction} & \textbf{A} & \textbf{n} & \textbf{E} & \textbf{Analogy for A and n} & \textbf{Source for E} \\ \hline
        \ce[label-offset=-1pt]{NH2 <=>[+M] NH + H} & $5.6 \times 10^{15}$ & 0 & 96600 & \ce[label-offset=-1pt]{CH2 <=>[+M] CH + H} \citep{bauerle1995} & $\Delta_r H$ \\
        \ce{C + NH <=> CN + H} & $5.0 \times 10^{13}$ & 0 & 0 & \ce{C + OH <=> CO + H} \citep{glarborg1985} & N/A \\
        \ce[label-offset=-1pt]{N + H <=>[+M] NH} & $4.7 \times 10^{18}$ & -1 & 0 & \ce[label-offset=-1pt]{O + H <=>[+M] OH} \citep{tsang1986} & N/A \\
        \ce[label-offset=-1pt]{CN <=>[+M] C + N} & $1.5 \times 10^{16}$ & 0 & 180260 & \ce[label-offset=-1pt]{C2 <=>[+M] C + C} \citep{kruse1997} & $\Delta_r H$ \\
        \ce[label-offset=-1pt]{NO <=>[+M] N + O} & $1.5 \times 10^{16}$ & 0 & 150920 & \ce[label-offset=-1pt]{C2 <=>[+M] C + C} \citep{kruse1997} & $\Delta_r H$ \\
        \ce[label-offset=-1pt]{N2 <=>[+M] N + N} & $1.5 \times 10^{16}$ & 0 & 225940 & \ce[label-offset=-1pt]{C2 <=>[+M] C + C} \citep{kruse1997} & $\Delta_r H$ \\
        \hline
        \end{tabular}
        }
        \label{tab:reactions}
        \end{table*}

    \subsection{Network base choice and \ce{C2} reduction}

        To derive our C/H/O/N chemical network from these combustion networks, multiple options were considered.
        The first one was to simply take the Glarborg 2018 network, as it is already a C/H/O/N network.
        However, as seen in the corresponding error distributions in Figs. \ref{fig:cond_stats} and \ref{fig:species4_stats}, this network performance, although better than older networks like Venot 2020 or Konnov 2005, is surpassed by the oxygenated species and alcohol combustion conditions of recent methanol-focused networks, such as Burke 2016, or the generic state-of-the-art network NUIGMech1.1.
        However, with respect to the nitrogen chemistry, it is the most state-of-the-art network, although the difference with NUIGMech1.1 and Konnov 2005 was not shown very clearly in our dataset.
        In the end, as both nitrogen and C/H/O chemistry are equally important for exoplanets, we decided to fuse the Glarborg 2018 network with the Burke 2016 network, keeping both the most state-of-the-art chemistry network with the best performing reasonable sized network on our data, while making sure  that the methanol chemistry is accurate.
        To further reduce our network size, we first removed 81 out of 91 species from the \ce{C3} sub-mechanism of Burke 2016 and their reactions, but kept the last 10 that were necessary to preserve the accuracy on some \ce{C2} species.
        This reduction was made because for exoplanets, \ce{C3} species abundances are usually low and their interest is limited in comparison to the increase in computation time they require due to the higher number of possible isomers.
        In addition, limiting calculation times allows for future additions of other species such as sulfur and its use in retrievals using TauREx with the FRECKLL plugin.

    \subsection{Additional modifications to the network}
    \label{section:network_additions}

        %During the test of the chemistry in exoplanet chemistry conditions at high altitudes
        While applying our chemical network to exoplanet studies, we noted that \ce{NH3} formation at high altitudes was primarily driven by a reversed globalized reaction: \ce{CH + NH3 -> H2CN + H + H}. This reaction was assumed to be the combination of two reactions: \ce{CH + NH3 -> CH2NH + H} and \ce{CH2NH -> H2CN + H}, but was written in this compact form in the original network of Glarborg 2018, implicitly assuming that the latter reaction would always happen only after the former. This simplification (while certainly reasonable in nitrogen combustion chemistry) is not suited to exoplanetary conditions, especially in the upper atmosphere, where photolysis combined with low density conditions maintains a really high concentration of hydrogen radicals that heavily favors the reverse reaction, resulting in an unphysical \ce{NH3} production pathway.
        Hence, we decided to rewrite the reversible reaction \ce{CH + NH3 <=> H2CN + H + H} into two others: \ce{CH + NH3 <=> CH2NH + H}, for which we kept the parameters of the \ce{CH + NH3 <=> H2CN + H + H} reaction, and \ce{H2CN + H <=> CH2NH}.
        For this second reaction, the choice of parameters was based on the reaction \ce{NH2 + H <=> NH3} by analogy.
        Both reactions are indeed the recombination of a nitrogen radical with a hydrogen atom, and therefore occur with no activation energy. They also should share a similar pre-exponential factor, with no temperature dependence.
        This factor was hence estimated at $ 1.6 \times 10^{14} $ \si{cm^3.mol^{-1}.s^{-1}}.
        However, this value is only accurate in the high pressure limit because at low pressures, this reaction needs a third body to stabilize the product, causing a strong pressure dependence of the rate constant. Further work is needed to correctly take into account this pressure dependence, using advanced Variational Reaction Coordinate-VTST and Master Equation methods \citep{klippenstein1992, georgievskii2003a, georgievskii2003b}.
        We discuss the impact of this approximation in Section \ref{section:HCN_discussion}.
        % Consequences of this change ? To be presented later ?

        In addition, we disabled the reversibility of 24 other globalized reactions yielding three products, to prevent similar unexpected chemical pathways from occurring. However, we did not find any conditions resulting in the reverse direction of these reactions to be favored.

        To further improve the reliability of the chemical network in the upper atmosphere, we searched for possibly missing radical reactions in the network that could significantly impact chemistry for these specific conditions.
        We listed all the major species and radicals typically encountered in exoplanets or produced by photolysis and checked for their reactions with N and NH radicals. 
        Many reactions of N and NH are negligible in usual combustion conditions; hence, the coupling between all radicals is not systematic, especially with radical compounds such as NH.
        %The choice to investigate on N and NH radicals was done because usually in combustion mechanisms, some radical reactions are not included because they are negligible in usual combustion conditions.
        In exoplanets, the photochemistry of \ce{NH3} produces a lot of N and NH radicals, in a medium where radicals are especially abundant, which causes them to mainly react between each other through pathways that may be usually neglected.
        While searching for these kinds of reactions, we identified six potentially missing reactions and determined their parameters by analogy with other reactions. These added reactions were also checked to ensure that they do not exceed the theoretical collision limit (see Table \ref{tab:reactions}).
        %Five of them were considered as pressure dependent, because of the lack of degrees of freedom to absorb the bonding energy due to the formed compound being diatomic that results in the need for a third body to take away this energy. As the high pressure limit is never reached for gas state densities in these reactions, their parameters all correspond to the low pressure limit.
        We compared the chemical network results on our combustion data set before and after these modifications, confirming that it did not affect the network performances. This scheme can be downloaded from the KInetic Database for Astrochemistry \citep{wakelam2012}\footnote{https://kida.astrochem-tools.org/} and also from the ANR EXACT website\footnote{https://www.anr-exact.cnrs.fr/fr/chemical-schemes/}.
    \begin{table*}[htbp]
    \centering
    \caption{Table of the exoplanets simulated with V20 and V23 and the input parameters used. Here, \textbf{D} is the distance to the host star, \textbf{R} the planet radius, \textbf{T} the temperature at 1 bar, $\pmb K_{zz}$ the eddy diffusion coefficient, and \textbf{M} the metallicity relative to solar abundances.}
    %\scalebox{1}{
    \begin{tabular}{|c|c|c|c|c|c|c|c|}
    \hline
    \textbf{Planet name} & \textbf{Planet type} & \textbf{Star type} & \textbf{D (UA)} & \textbf{R ($\pmb R_J$)} & \textbf{T (K)} & \textbf{$\pmb K_{zz}$ (cm²/s)} & \textbf{M (solar)} \\ \hline
    GJ 436 b & Warm Neptune & M3V & 0.029 & 0.38 & 1094 & $10^9$ & 1 \\
    GJ 436 b & Warm Neptune & M3V & 0.029 & 0.38 & 1094 & $10^9$ & 100 \\
    GJ 1214 b & Warm Neptune & M4.5V & 0.014 & 0.24 & 1054 & $3 \times 10^7 \times P^{-0.4}$ & 100 \\
    HD 189733 b & Hot Jupiter & K2V & 0.031 & 1.14 & 1470 & profile & 1 \\
    HD 209458 b & Hot Jupiter & F9V & 0.047 & 1.38 & 1671 & profile & 1 \\
%    Neptune & Cold Neptune & G2V & 30 & 0.36 & 71 & $10^8$ & 480 \\
%    Uranus & Cold Neptune & G2V & 19 & 0.37 & 79 & $10^8$ & 160 \\
%    ULAS J1335 & Brown Dwarf & T9 & 0.01 & 1.21 & 640 & $10^6$ & 1 \\
    \hline
    \end{tabular}
    %}
    \label{tab:planets}
    \end{table*}

%\section{Differences with the V20 Model}
\section{Application to exoplanetary atmospheres}
\label{section:model_application}

    %\subsection{}

    \subsection{Models and data sources}

        % Comment caractériser les différences du nouveau modèle par rapport aux autres ?
        Our prime motivation for this extensive work on combustion networks was to develop a very robust scheme for the study of exoplanetary atmospheres. Thus, in this section, we now apply this new scheme to model the atmospheric chemical composition of various exoplanets.
        Our new C/H/O/N scheme was tested against multiple exoplanet cases and compared to the one published in \cite{venot2020}. In the following, we refer to the Venot 2020 chemical network as V20 and to our update as V23.
        
        In order to span different type of hydrogen-dominated atmospheres, we chose to model GJ 436 b and GJ 1214 b (warm Neptunes) as well as HD 209458 b and HD189733 b (hot Jupiters) using the same thermal profiles, initial conditions, and parameters given in \cite{venot2020} (Table \ref{tab:planets}).
        For each planet, we compared the abundances obtained with both V23 and V20.
        We calculated the chemical abundance profiles using FRECKLL \citep{freckll}, which is the Python version of the code used in \cite{venot2020}.
        The results obtained with this code are identical, but the computational time has been greatly improved.
        The thermal profiles are discretized in a 130-layers grid, evenly distributed in pressure log space. We assumed a solar metallicity for HD 189733 b and HD 209458 b, a 100x solar metallicity for GJ 1214 b and both metallicities for GJ 436 b. Elemental abundances were based on \cite{lodders2010}, with 20\% less oxygen to account for sequestration in refractory elements.

        % Quelle a été la démarche utilisée pour la mise à jour des photolyses ?
        We updated the photodissociation data (cross-sections and branching ratios, Table \ref{tab:ref_sections}), compared to that used in V20.
        %For the photolysis data, we first updated versions of the cross-sections and branching ratios used in \cite{venot2020} (add ref to UV cross section data).
        To discriminate the changes due to this update and to chemistry, we first compared the abundance profiles of each exoplanet model for some of the major species (\ce{H2}, \ce{H2O}, \ce{CH4}, \ce{CO}, \ce{N2}, \ce{NH3}, \ce{CO2}, \ce{HCN}, and \ce{H}) between the old photolysis and the new photolysis data for the V20 chemical network with FRECKLL.
        This update turned out to have little impact on photochemistry for most species on hot Jupiters (HD 189733 b and HD 209458 b) and for all species on warm Neptunes (GJ 436 b and GJ 1214 b).
        However, for \ce{HCN}, the addition of two new photodissociation pathways of \ce{NH3} into \ce{NH} (\ce{NH3 -> NH + H + H} and \ce{NH3 -> NH + H2}) creates differences of up to one order of magnitude in the upper atmosphere of HD 189733 b and HD 209458 b between $10^{-6}$ and $10^{-7}$ bar.
        The consequences of this photolysis update are summarized in Fig. \ref{fig:photolysis_changes}.

        In the following, we compare the chemical abundances obtained with V20 and V23 for each planet case, using only this most recent UV cross-section data and branching ratios. We also investigate on the reasons explaining the observed differences and identify the main chemical pathways at play in each network.
        To evaluate the impact on observables, we generated the transmission spectrum of every planet with TauREx 3.1 \citep{taurex}, using a spectral resolution of 50 and opacities data from ExoMol \citep{exomol} for HCN and from \cite{taurexdata} for \ce{CH4}, CO, \ce{CO2}, \ce{H2O}, and \ce{NH3}. Rayleigh diffusion for \ce{CH4}, CO, \ce{CO2}, \ce{H2}, \ce{H2O}, He, \ce{N2}, and \ce{NH3} as well as collision-induced absorption from HITRAN \citep{hitran} for \ce{H2}-\ce{H2} and \ce{H2}-He were also included.

        \subsection{Results for GJ 436 b}
        %\subsubsection{GJ 436 b, $K_{zz} = 10^{8}$ cm²/s, 1x solar metallicity}

\begin{figure}[htbp]
                \centering
                \includegraphics[width=0.5\textwidth]{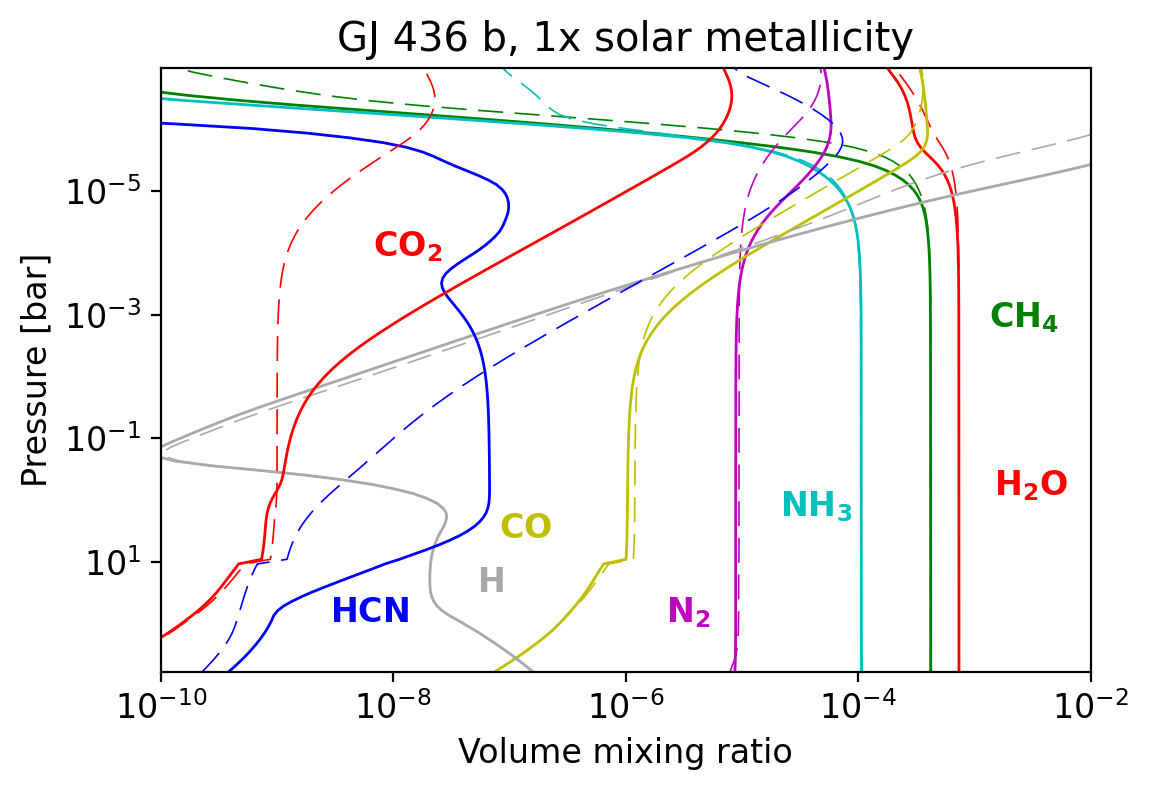}
                \caption{Abundance profiles of GJ 436 b for solar metallicity and a constant eddy diffusion coefficient of $10^{9}$ \si{cm^2.s^{-1}}. Dashed lines are for V20, while solid lines are for V23. \ce{H2} is not shown to focus on other species, but its abundance profile in V23 is almost identical to V20.}
                \label{fig:abundance_GJ436b_solar}
            \end{figure}

            For the warm Neptune GJ 436 b, we first simulated the 1D chemical abundance profiles assuming a solar metallicity and a constant eddy diffusion coefficient of $10^{9}$ \si{cm^2.s^{-1}}.
            While some of the main species (\ce{H2O}, \ce{CH4}, \ce{NH3}, \ce{N2}, CO, and H) are found to have similar abundance profiles (Fig. \ref{fig:abundance_GJ436b_solar}) with both networks, we observe that two species differ by various orders of magnitude: \ce{CO2} below 0.1 bar, and \ce{HCN} on the whole pressure profile (100 to $10^{-7}$ bar).
            In the upper atmosphere, the molar fraction of \ce{CO2} is higher with V23 than in V20, with a difference up to three orders of magnitude around $10^{-6}$ bar.
            For HCN, its molar fraction is lower in V23 than in V20 for pressures under $10^{-2}$ bar, with a difference of up to four orders of magnitude around $10^{-6}$ bar.
            For pressures higher than $10^{-2}$ bar, HCN molar fraction is higher in V23 than in V20, with a difference of up to two orders of magnitude around 10 bar.
            In the following, we discuss the origin of the differences for these two species.

            \subsubsection{\ce{CO2} differences}
            \label{section:CO2_discussion}

\begin{figure}[htbp]
                \centering
                \includegraphics[width=0.5\textwidth]{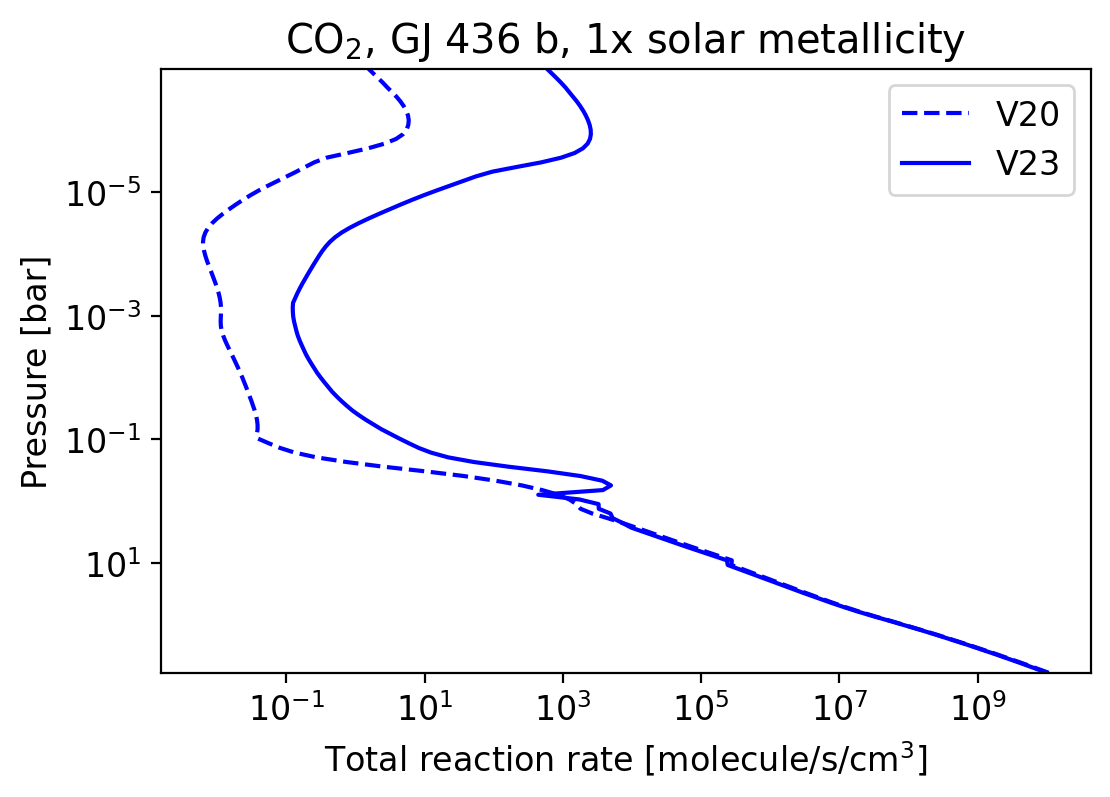}
                                \caption{Total reaction rate profile for \ce{CO2} in GJ 436 b with a solar metallicity with V20 (dashed lines) and V23 (solid lines).}
                \label{fig:reactions_rate_CO2_GJ436b_solar}
            \end{figure}

            % CO2
            Upon investigating the reasons of this discrepancy, we found that between V23 and V20, the total \ce{CO2} reactions rate profile was different.
            Figure \ref{fig:reactions_rate_CO2_GJ436b_solar} shows the total rate of \ce{CO2} formation and destruction in each layer, which are equal when including vertical mixing because the profiles are at steady state. We see that these total reaction rates are larger below 1 bar with V23 than with V20.

\begin{figure}[h!]
                \centering
                \includegraphics[width=0.5\textwidth]{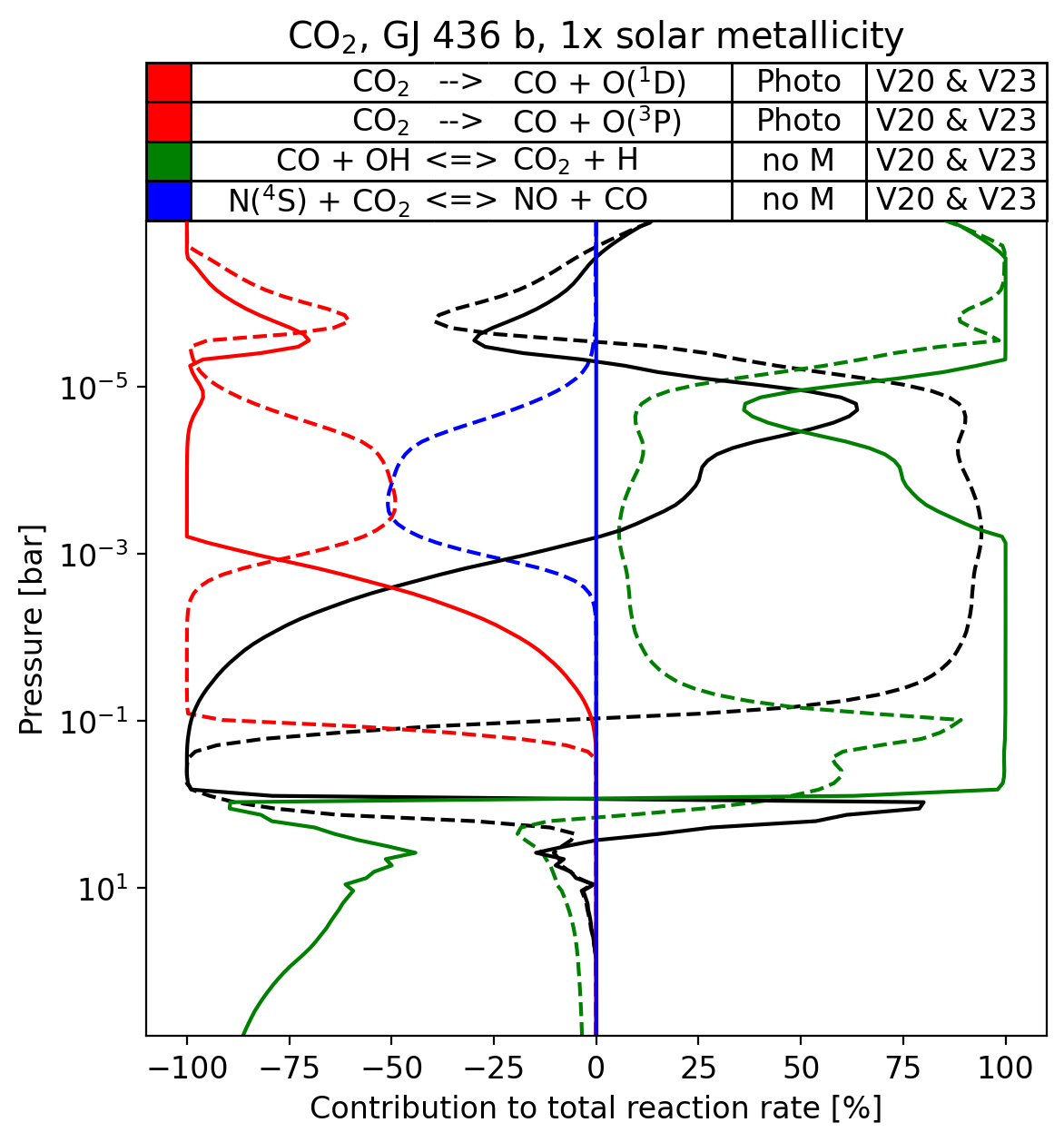}
                \caption{Contribution profile of most major production and loss reactions for \ce{CO2} in GJ 436 b. Positive values are production contributions and negative values are loss contributions. Black lines correspond to vertical mixing compensation, such as the sum in each layer is always zero due to steady state. Dashed lines are for V20, while solid lines are for V23. The contribution of photodissociation pathways to \ce{O(^3P)} and \ce{O(^1D)} are combined, \ce{O(^3P)} being favored above $10^{-5}$ bar and \ce{O(^1D)} being favored for lower pressures. The third column in the legend indicates the reaction type. "Photo" corresponds to photodissociations and "no M" corresponds to reactions without pressure dependence. The last column indicates which model includes this reaction.}
                \label{fig:contrib_CO2_GJ436b_solar}
            \end{figure}

\begin{figure}[htbp]
                \centering
                \includegraphics[width=0.5\textwidth]{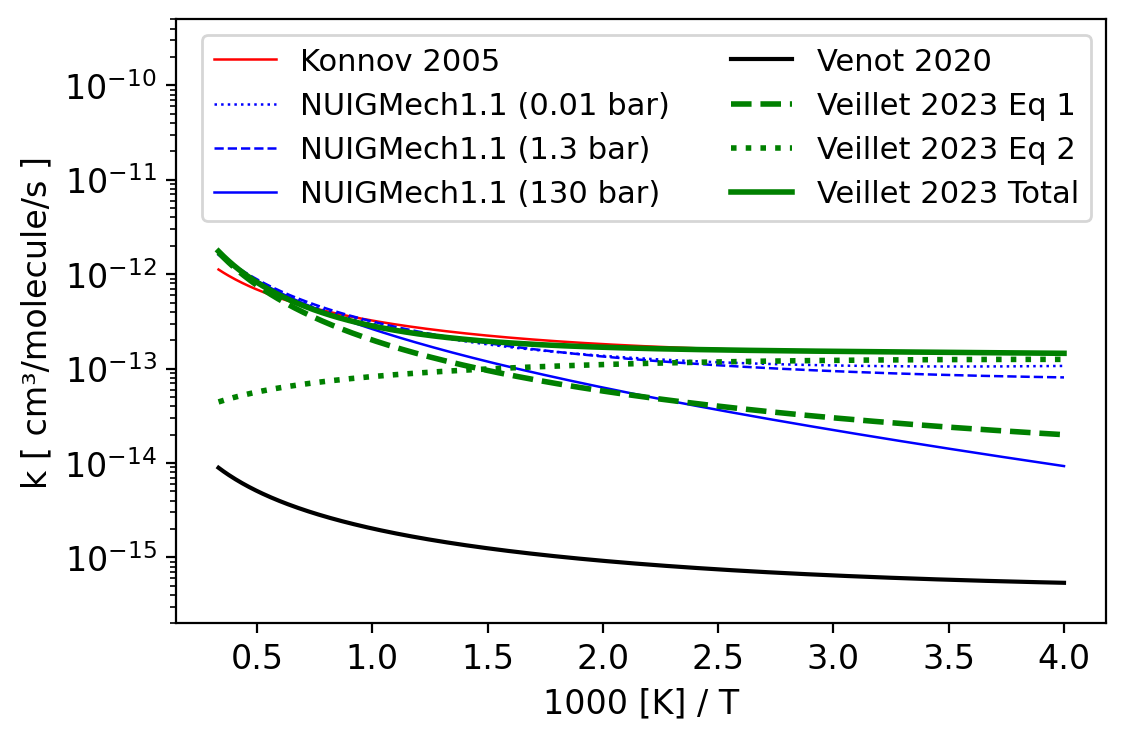}
                \caption{Reaction rates of the reaction \ce{CO + OH -> CO2 + H} in V20 compared to V23, NUIGMech1.1, and Konnov 2005. In V23, this reaction rate is expressed as the sum of two modified Arrhenius equations (Eqs. 1 and  2). In NUIGMech1.1, the rate constant is pressure-dependent.}
            \label{fig:reactionrate}
            \end{figure}

            Figure \ref{fig:contrib_CO2_GJ436b_solar} presents the main contributions of each reaction to this total rate, the sum of positive contributions, and the sum of negative contribution both being equal to the total reactions rate when accounting for vertical mixing because of the steady state.
            For V23, Fig. \ref{fig:contrib_CO2_GJ436b_solar} shows that the reaction \ce{CO + OH -> CO2 + H} is always the main \ce{CO2} production reaction above 1 bar and the main \ce{CO2} destruction reaction below 1 bar.
            Vertical mixing mainly transports the \ce{CO2} produced in the middle atmosphere (1 - $10^{-3}$ bar) towards the lower atmosphere (below 1 bar), where it is destroyed into \ce{CO} and \ce{OH} through the \ce{CO2 + H -> CO + OH} reaction.
            For V20, this reaction is not the main destruction pathway of \ce{CO2} in the lower atmosphere and needs the vertical mixing to compensate for the destruction of \ce{CO2} through the \ce{N(^4S) + CO2 -> NO + CO} reaction, although it also remains the main \ce{CO2} production reaction for pressures lower than 1 bar.
            %This coupling reaction to \ce{N(^4S)} atoms is minor in V23, and is linked to the previously mentioned spike in \ce{CO2} destruction rate, which causes a dip in the \ce{CO2} abundance profile of V20 around $10^{-3}$ bar.
            At the peak of \ce{CO2} abundance around $10^{-6}$ bar, the main production reaction is \ce{CO + OH -> CO2 + H} and the main loss mechanism is by photodissociation into CO through \ce{CO2 -> CO + O(^1D)}, for both V20 and V23.
            Photodissociation rate being proportional to the \ce{CO2} concentration, its rate increase is directly linked to the higher \ce{CO2} levels in V23.
            The UV cross-sections used in V23 and V20 being the same in these simulations, there is no difference in the main loss reaction parameters at this pressure between the two chemical networks, thus, this difference must come from the production reaction \ce{CO + OH -> CO2 + H}.
            In V20, this production reaction is taken from \cite{baulch1994} with the pre-exponential factor divided by 6, resulting in a production rate of around 5 \si{molecule.cm^{-3}.s^{-1}} around $10^{-6}$ bar.
            In V23, these reaction parameters are taken from \cite{joshi2006}, where it is treated as a sum of two modified Arrhenius equations, both having opposite temperature dependence. This results in a total production rate of around 2500 $\si{molecule.cm^{-3}.s^{-1}}$, which is three orders of magnitude greater than V20.
            This difference is directly observed in the rate constant of this reaction for the two networks and is fully attributable to the very different parameters used for modeling this reaction, as shown in Fig. \ref{fig:reactionrate}.
            %A comparison of the \ce{CO2} formation and destruction rate profile between V20 and V23 can be found in Fig. \ref{fig:reactions_rate_CO2_GJ436b_solar}.
            This difference has already been pointed out by \cite{tsai2021} %, which highlights the abundance of references yielding reaction rates between $10^{-12}$ and $10^{-13}$ \si{cm^3.molecule^{-1}.s^{-1}} 
            for temperatures in the 250-2000 K range (their Figure 41).
            As for other combustion networks, such as Konnov 2005 or NUIGMech1.1, they are on a similar order of magnitude as V23 and also three orders of magnitude greater than V20, even using different sources for the reaction parameters.
            NUIGMech1.1 is particularly close to V23 for pressures around 100 bar and temperatures above 500 K, despite using data from \cite{senosiain2005} and a single reaction with pressure dependent parameters.
            For temperatures below this value, some care should be taken as reaction rates differences between chemical networks raise above one order of magnitude around 250K, where the validity of these networks is no longer ensured.

            \subsubsection{HCN differences}
            \label{section:HCN_discussion}

\begin{figure}[htbp]
                \centering
                \includegraphics[width=0.5\textwidth]{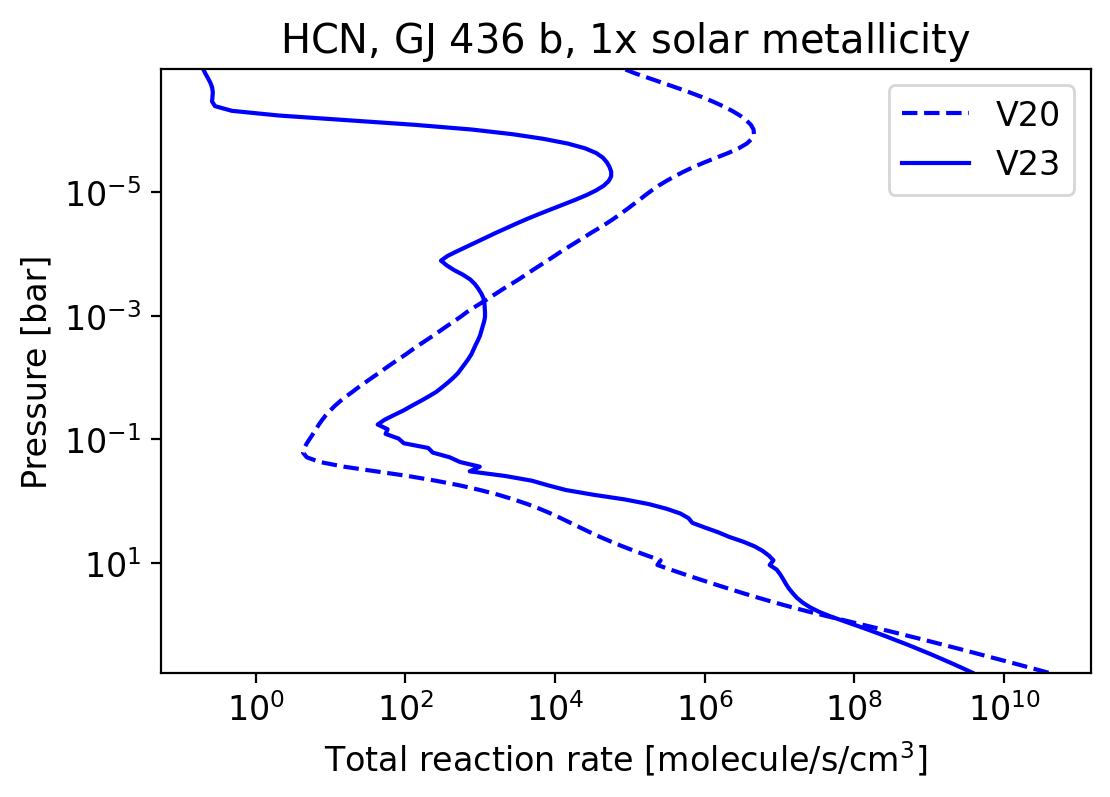}
                                \caption{Total reaction rate profile for \ce{HCN} in GJ 436 b with a solar metallicity with V20 (dashed lines) and V23 (solid lines).}
                \label{fig:reactions_rate_HCN_GJ436b_solar}
            \end{figure}

\begin{figure}[htbp]
                \centering
                \includegraphics[width=0.5\textwidth]{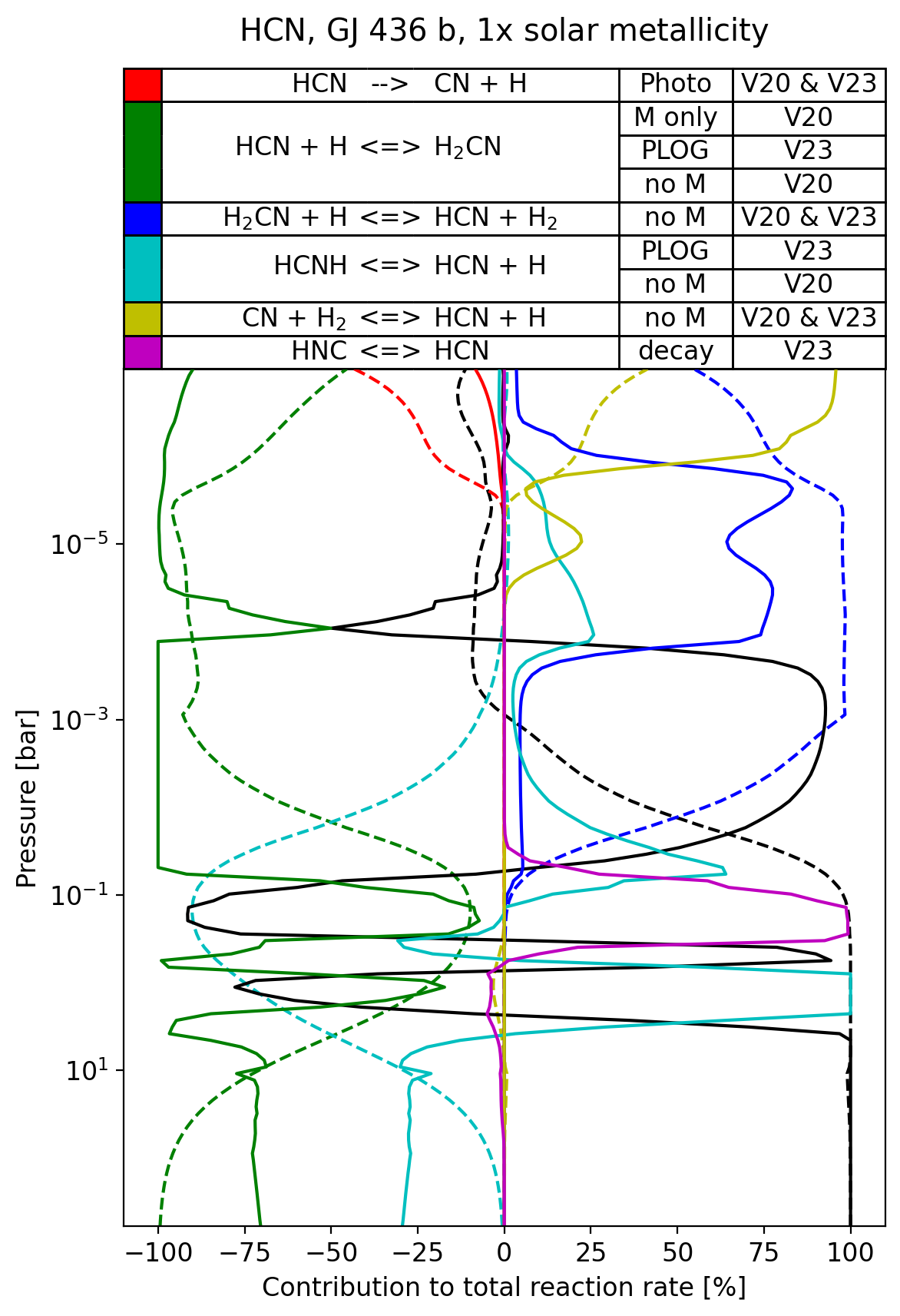}
                \caption{Contribution profile of most major production and loss reactions for \ce{HCN} in GJ 436 b. Positive values are production contributions and negative values are loss contributions. Black lines correspond to vertical mixing compensation, such as the sum in each layer is always zero due to steady state. Dashed lines are for V20, while solid lines are for V23. The third column in the legend indicates the reaction type. "no M" corresponds to reactions without pressure dependence, "PLOG" to full pressure dependence and fall off description with PLOG formalism, "M only" to pressure dependence without fall off nor high pressure limit and "decay" to reversible, pressure dependent unimolecular reactions such as isomerization or electronic decay. The last column indicates which model includes this reaction.}
                \label{fig:contrib_HCN_GJ436b_solar}
            \end{figure}

\begin{figure}[htbp]
                \centering
                \includegraphics[width=0.5\textwidth]{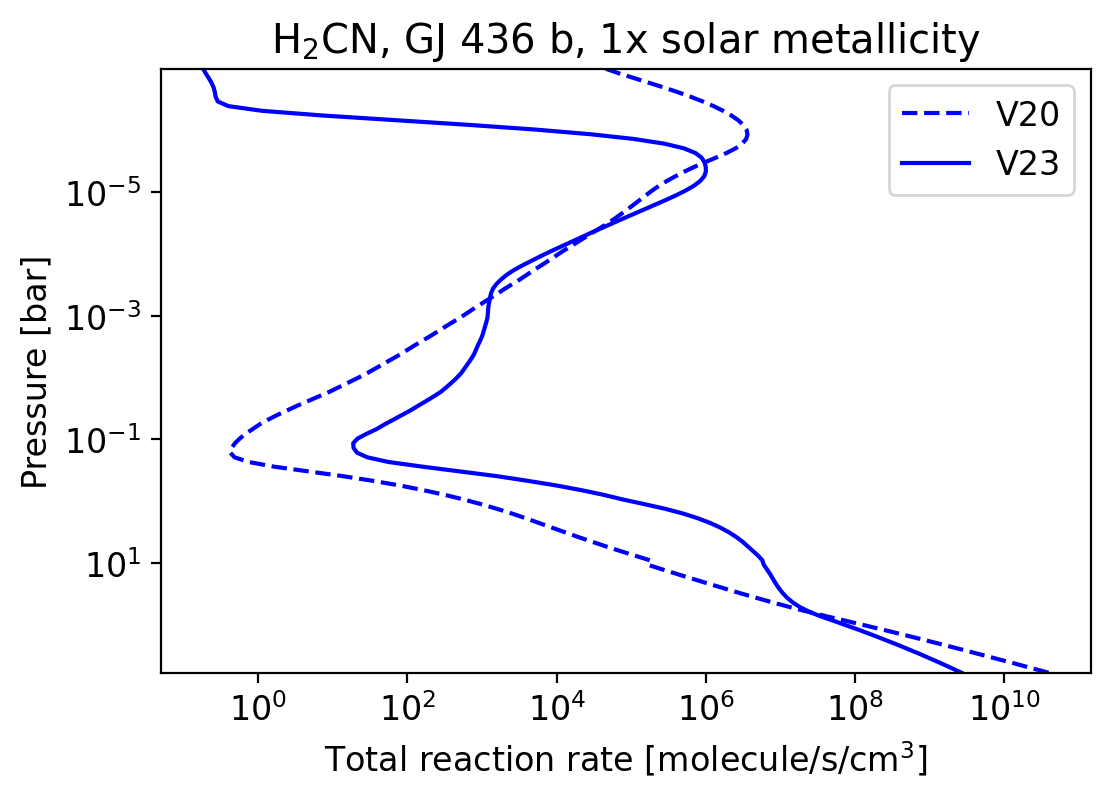}
                                \caption{Total reaction rate profiles for \ce{H2CN} in GJ 436 b with a solar metallicity with V20 (dashed lines) and V23 (solid lines).}
                \label{fig:reactions_rate_H2CN_GJ436b_solar}
            \end{figure}

\begin{figure}[htbp]
                \centering
                \includegraphics[width=0.5\textwidth]{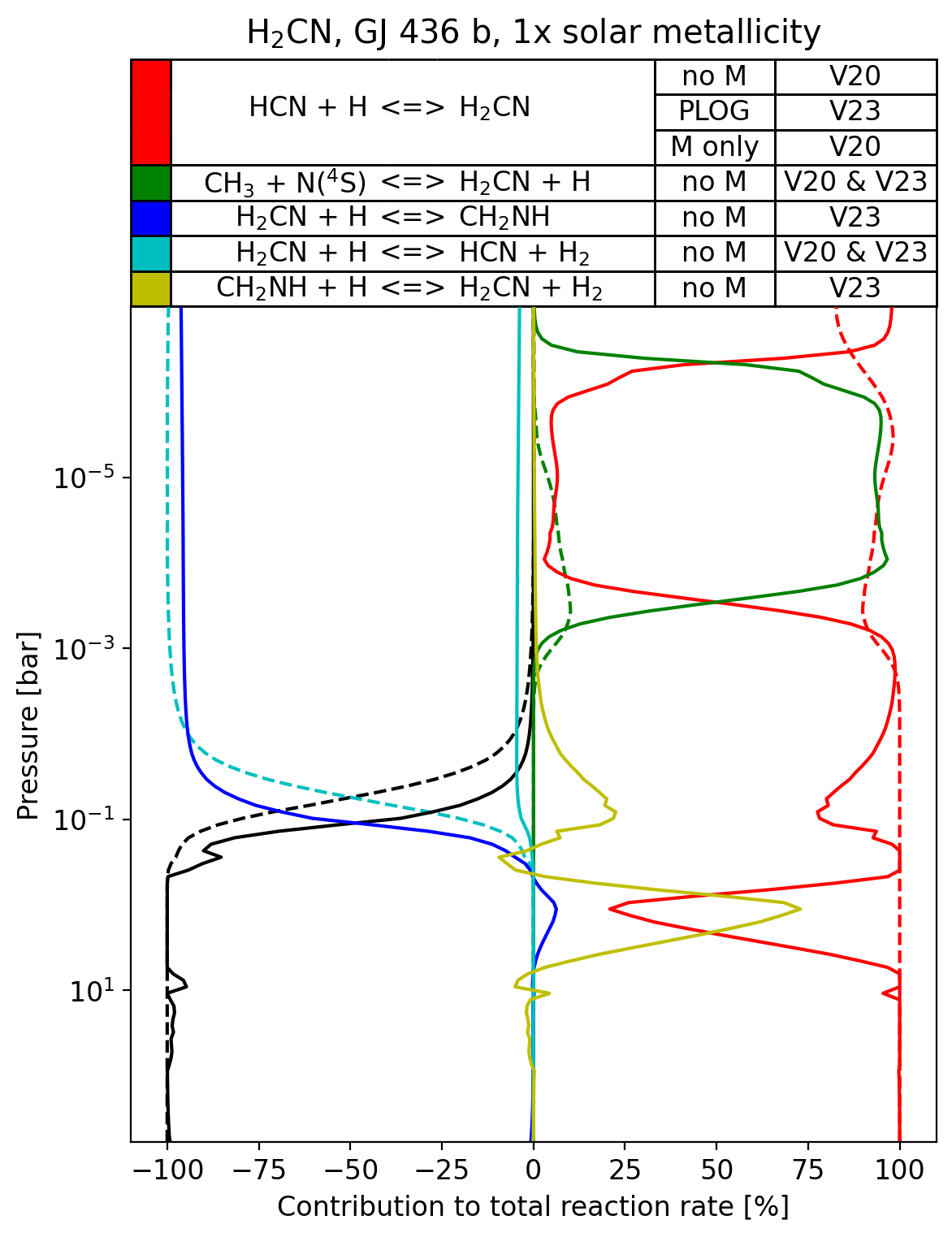}
                \caption{Contribution profile of most major production and loss reactions for \ce{H2CN} in GJ 436 b. Positive values are production contributions and negative values are loss contributions. Black lines correspond to vertical mixing compensation, such as the sum in each layer is always zero due to steady state. Dashed lines are for V20, while solid lines are for V23. The third column in the legend indicates the reaction type. "no M" corresponds to reactions without pressure dependence, "PLOG" to full pressure dependence and fall off description with PLOG formalism and "M only" to pressure dependence without fall off nor high pressure limit. The last column indicates which model includes this reaction.}
                \label{fig:contrib_H2CN_GJ436b_solar}
            \end{figure}

\begin{figure}[htbp]
                \centering
                \includegraphics[width=0.5\textwidth]{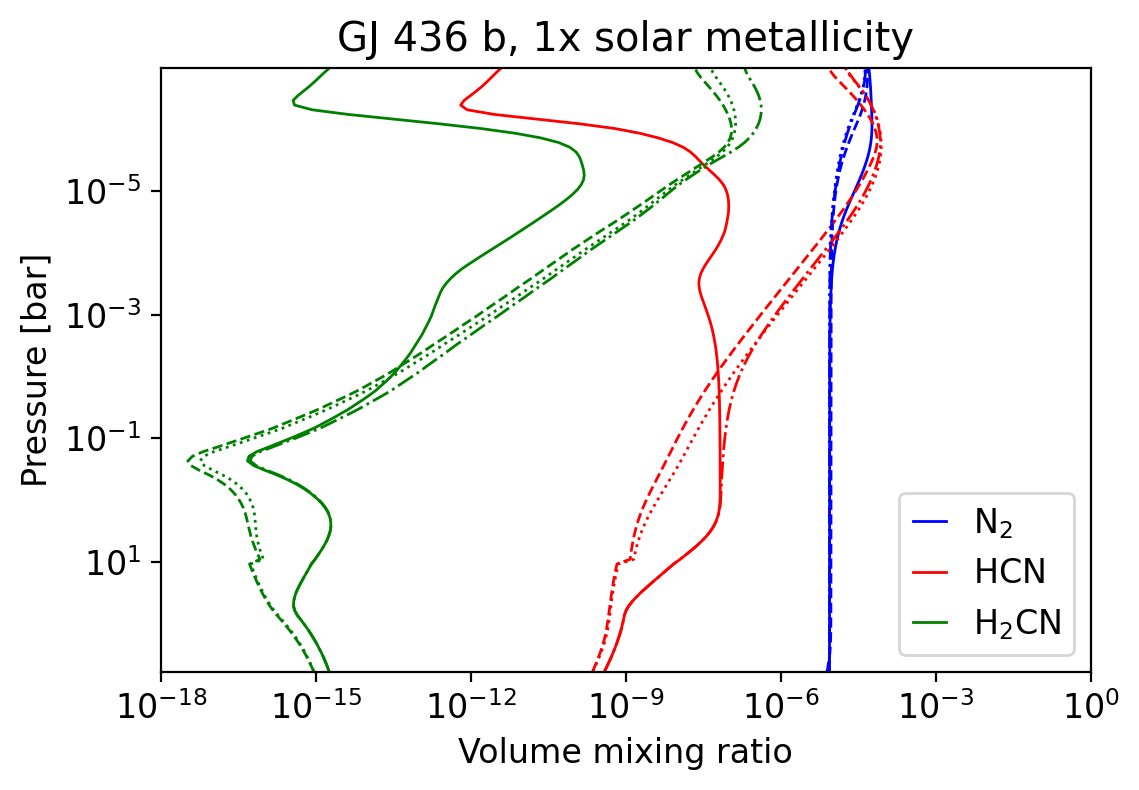}
                \caption{Abundance profile of \ce{N2}, HCN and \ce{H2CN} with V23 (solid lines), V20 (dashed lines), V23 without the reaction \ce{H2CN + H -> CH2NH} (dashdot lines) and V23 without \ce{H2CN + H -> CH2NH}, \ce{CH2NH + H -> H2CN + H}, \ce{CH2NH + CH3 -> H2CN + CH4}, \ce{CH2NH + NH2 -> H2CN + NH3}, \ce{CH2NH + OH -> H2CN + H2O} and with V20 thermochemical data for \ce{H2CN}, HCNH and HCN (dotted lines).}
                \label{fig:V23_reaction_impact}
            \end{figure}

            % NH3 and HCN
            We focus in this part in the differences observed between V23 and V20 for HCN. The total reaction rate profile for HCN production and consumption is shown in Fig. \ref{fig:reactions_rate_HCN_GJ436b_solar}, and the respective contributions of major reactions are shown in Fig. \ref{fig:contrib_HCN_GJ436b_solar}.
            As expected, the HCN total reaction rate profile roughly matches the differences in HCN abundance, with the total reaction rate in V20 being up to six orders of magnitude higher than in V23 around $10^{-7}$ bar, and up to one order of magnitude lower between 100 and $10^{-3}$ bar. In both networks, the main production reaction between $10^{-2}$ and $10^{-6}$ bar is the reaction \ce{HCN + H <=> H2CN}. This reaction is implemented in V23 with a pressure dependent rate, with parameters interpolated in log space (PLOG) between 3 pressure values at 10, 1, and 0.1 bar. In V20, this reaction is described as two separate reactions: one for the high pressure limit, and one for the low pressure limit, requiring a third body. The parameters of the high pressure limit reaction are close to the values in the V23 PLOG reaction for 0.1 bar, and the low pressure limit contribution decreases with altitude, becoming negligible for pressures below 0.1 bar. Therefore, these differences cannot explain those between the two HCN abundance profiles. It is also important to note that this description likely overestimates the \ce{HCN + H -> H2CN} reaction rate for pressures below 0.1 bar, for both networks. A PLOG implementation on the full pressure range (1000 - $10^{-8}$ bar) would be necessary to describe the full pressure dependence.
            In both networks, the main HCN production reaction is \ce{H2CN + H -> HCN + H2} between 0.1 and $10^{-6}$ bar. The parameters for this reaction are identical between the two networks, therefore it is not responsible for the differences between HCN abundance profiles.
            The other major contributing reaction, \ce{HCN + H -> HCNH} is similar to the reaction \ce{HCN + H -> H2CN}, but results in the formation of HCNH, an isomer of \ce{H2CN}. V23 uses similar values to V20 for this reaction, although the pressure dependence is described between 0.1 and 10 bars with the PLOG formalism. Another difference between V20 and V23 is the inclusion of HNC in V23 and its isomerization reaction \ce{HNC <=> HCN}. However, this reaction does not impact HCN profiles, which we confirmed by running the simulation with V23 without this reaction.
            The combination of the reactions \ce{HCN + H -> H2CN} and \ce{H2CN + H -> HCN + H2} results in an equilibrium between the species \ce{HCN} and its radical \ce{H2CN}. Hence, HCN differences between V23 and V20 are directly driven by production and consumption rate differences of \ce{H2CN}, which are shown in Fig. \ref{fig:reactions_rate_H2CN_GJ436b_solar}.
            As expected, we observe a similar reaction rate profile to HCN, with a total reaction rate up to six orders of magnitude above V20 for V23 at $10^{-7}$ bar, and up to two orders between 10 and $10^{-3}$ bar.
            This indicates that HCN abundance is mainly controlled by \ce{H2CN} abundance and its associated consumption and production reactions.
            Figure \ref{fig:contrib_H2CN_GJ436b_solar} shows a few differences on the major reactions of \ce{H2CN}.
            Firstly, in V20, the main consumption reaction on the whole pressure range is the reaction \ce{H2CN + H -> HCN + H2}, previously mentioned as the main HCN production reaction in both networks that controls the HCN/\ce{H2CN} equilibrium.
            For pressures above $10^{-1}$ bar, \ce{H2CN} consumption isn't local anymore, and the vertical mixing advects \ce{H2CN} to be consumed in the upper layers by the reaction \ce{H2CN + H -> HCN + H2}. 
            Conversely, in V23, this reaction is negligible, and \ce{H2CN} consumption is entirely driven by the reaction \ce{H2CN + H -> CH2NH}, discussed in Sect. \ref{section:network_additions}.
            This difference is crucial, because the \ce{CH2NH} species and its linked reactions are absent from V20.
            Secondly, in V20, in the pressure range $10^{-4}$ to $10^{-6}$ bar, the main \ce{H2CN} production reaction is \ce{HCN + H -> H2CN}, which is the second mentioned reaction controlling the HCN/\ce{H2CN} equilibrium.
            In V23, the main reaction in this range is the reaction \ce{CH3 + N(^4S) -> H2CN + H}, that is included in both networks using the same parameters.
            Finally, in V23 at around 1 bar, the reaction \ce{CH2NH + H -> H2CN + H2} becomes dominant for \ce{H2CN} production. This reaction opposes the \ce{H2CN + H -> CH2NH} reaction, and reverts \ce{CH2NH} back to \ce{H2CN}.
            To better understand the role each of each reaction on the differences observed in the HCN abundance profile, we ran these simulations with modified versions of the V23 network presented in Fig. \ref{fig:V23_reaction_impact}.
            This figure shows that disabling the reaction \ce{H2CN + H -> CH2NH} results in a HCN profile almost identical to V20 (dashed and dash-dotted lines) for pressures lower than $10^{-1}$.
            For pressures above this value however, no visible difference with V23 was observed. We disabled the reaction \ce{CH2NH + H -> H2CN + H2}, but other reactions such as \ce{CH2NH + CH3 -> H2CN + CH4} and \ce{CH2NH + NH2 -> H2CN + NH3} would replace its function in the HCN formation pathway, leading to smaller but very significant differences.
            Thus, we disabled all the H-abstraction reactions of \ce{CH2NH} in addition to the reaction \ce{H2CN + H -> CH2NH}, and found almost the same profile as V20.
            As disabling these reactions without disabling the reaction \ce{H2CN + H -> CH2NH} almost does not alter the V23 abundance profile, both of these reactions seem to be required to explain the differences between V23 and V20.
            The remaining difference was mainly located under 100 bar, where HCN and \ce{H2CN} abundances approach chemical equilibrium. Because the thermochemical data for V23 are different from V20 (Fig. \ref{fig:thermo_changes}), we ran V23 with V20 thermochemical data for \ce{H2CN}, HCNH and HCN, and found a perfect match in this pressure range.
            Thus, we conclude that the differences between V23 and V20 observed in HCN abundance profiles for GJ 436 b at 1x solar metallicity are caused by the addition of the species \ce{CH2NH} to the network. The upper atmosphere differences are caused by the reaction \ce{H2CN + H -> CH2NH}, and the lower atmosphere differences are caused mainly by \ce{CH2NH + H -> H2CN + H2}, but also to a lower extent by others H-abstraction reactions of \ce{CH2NH} with \ce{CH3}, \ce{NH2} or OH.
            
\begin{figure}[htbp]
                \centering
                \includegraphics[width=0.5\textwidth]{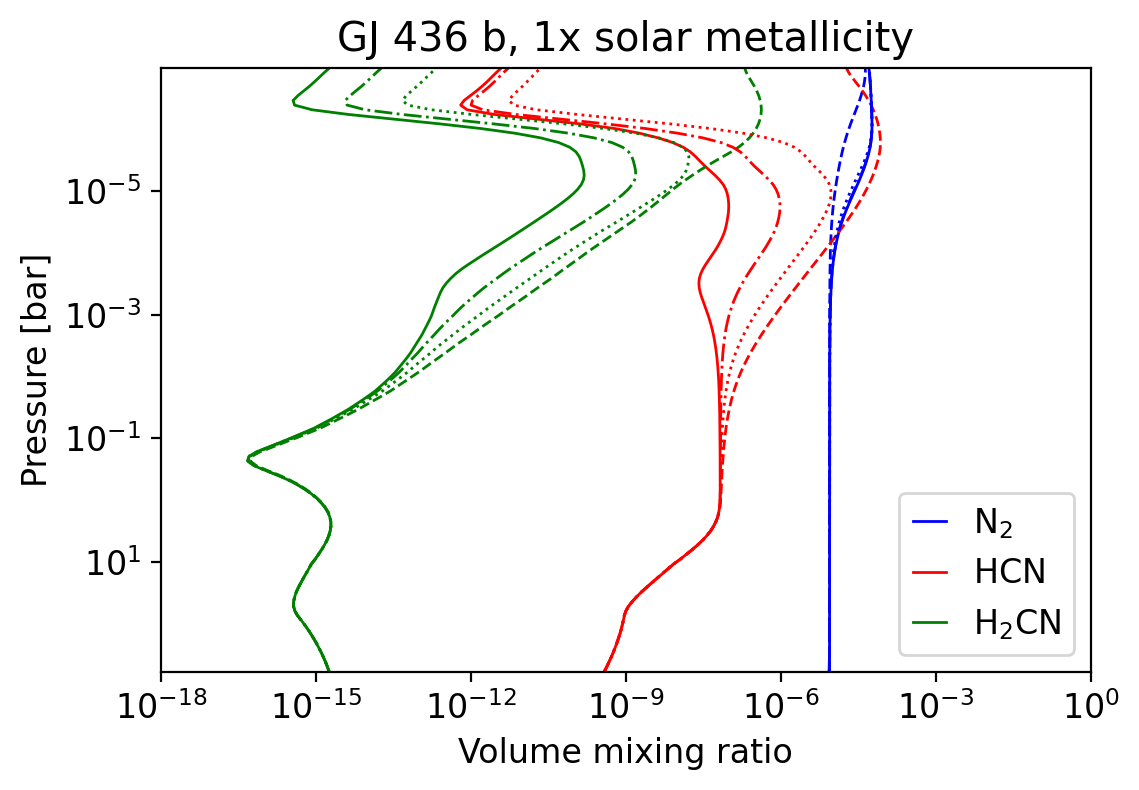}
                \caption{Sensitivity analysis of the pre-exponential factor $A$ of the reaction \ce{H2CN + H -> CH2NH}. Abundances profiles are calculated with the full V23 network (solid lines), with $A$ divided by 10 (dashdot lines), with $A$ divided by 100 (dotted lines) and without this reaction (dashed lines).}
                \label{fig:V23_sensitivity_analysis}
            \end{figure}

            Because of the importance of the reaction \ce{H2CN + H -> CH2NH} in these simulations and the large uncertainty in its reaction parameters (discussed in Sect. \ref{section:network_additions}), we investigated its impact on the simulations with a sensitivity analysis. Fig. \ref{fig:V23_sensitivity_analysis} shows a sensitivity analysis on the pre-exponential factor $A$ of this reaction.
            Multiple simulations were run with different pre-exponential factors for this reaction, up to a division factor of 100. 
            Significant differences in HCN abundance of up to two orders of magnitude are found between values of $A$ and of $A$/100, especially around $10^{-5}$ bar. At lower pressures however, the abundance is shown to be quite insensitive to changes in $A$, showing that the presence of the reaction remains impactful even with low estimates for this pre-exponential factor.
            
\begin{figure}[htbp]
                \centering
                \includegraphics[width=9cm]{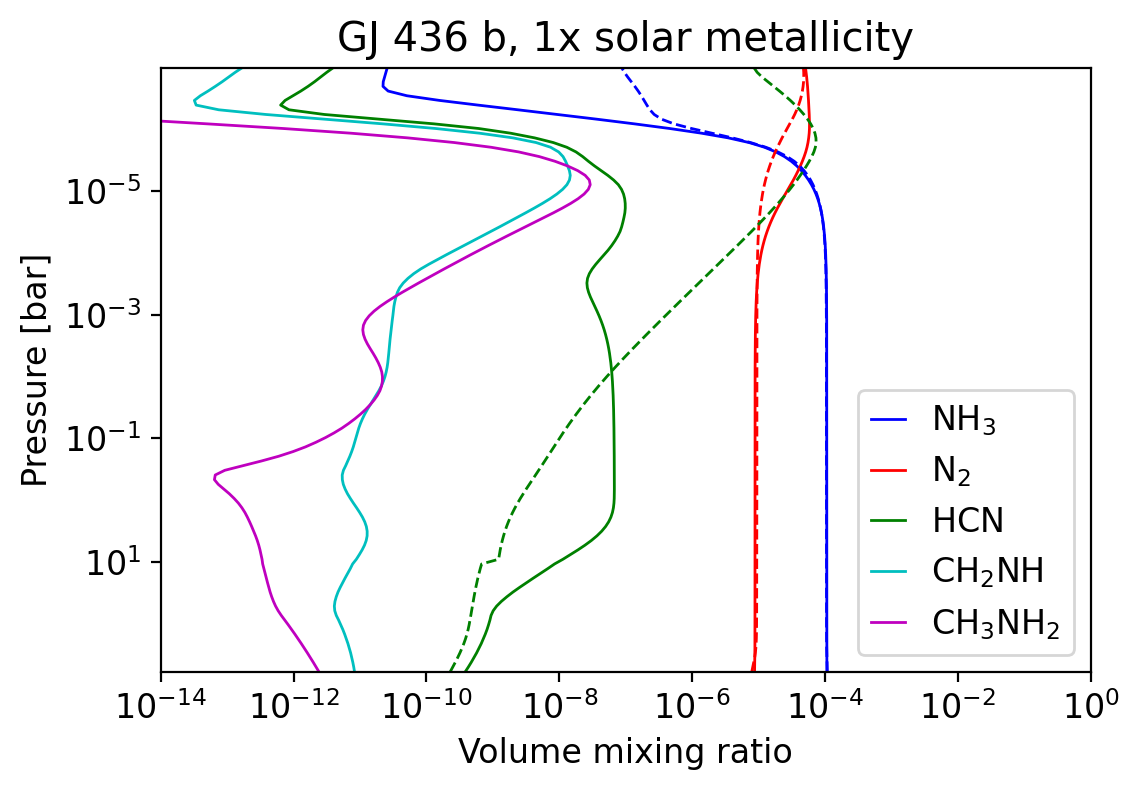}
                \caption{Abundance profiles of \ce{CH2NH} and \ce{CH3NH2} compared with major N-bearing species for GJ 436 b with solar metallicity in V20 (dashed lines) and V23 (solid lines). \ce{CH2NH} and \ce{CH3NH2} are not included in V20.}
                \label{fig:abundances_CH2NH}
            \end{figure}

\begin{figure}[htbp]
                \centering
                \includegraphics[width=9cm]{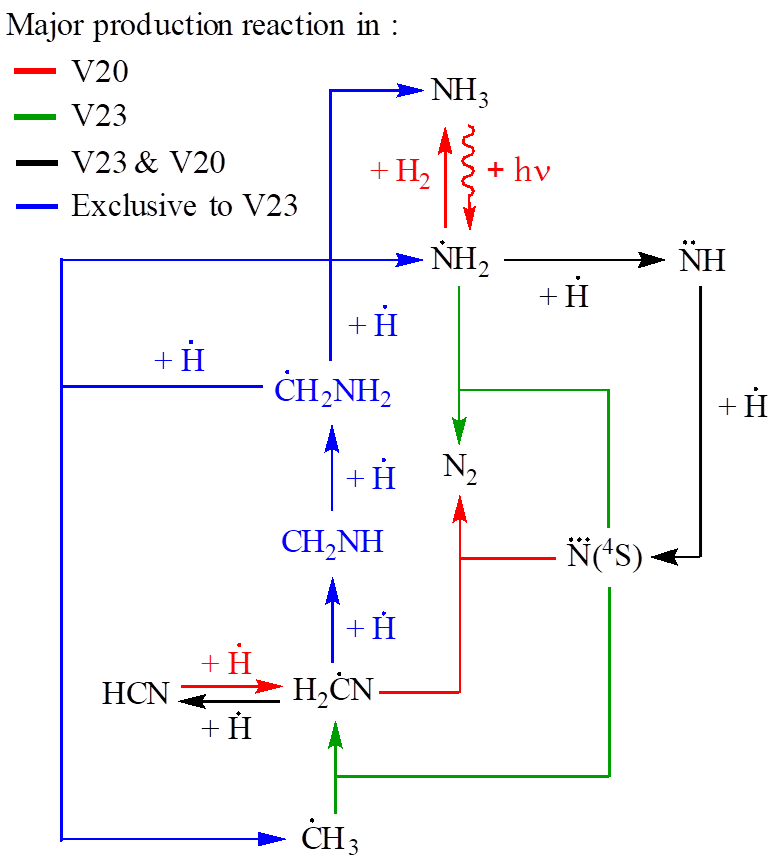}
                \caption{\ce{HCN} formation mechanism in V23 around $10^{-5}$ bar. The blue path is exclusive to V23 and absent from V20. The red reactions are dominant in V20 for the production of the species and minor but included in V23. The green reactions are dominant in V23 and minor but included in V20. The black reactions are dominant in both networks.}
                \label{fig:HCN_diffs}
            \end{figure}

            Given this complete shift in the major N-bearing species above $10^{-5}$ bar, we could expect \ce{CH2NH} abundance to be quite high. However, as shown in Fig. \ref{fig:abundances_CH2NH}, the \ce{CH2NH} abundance profile remains two to four orders of magnitude lower than HCN abundance for pressures higher than $10^{-4}$ bar, and stays similar around $10^{-5}$ bar.
            For the \ce{N2} abundance profile, we can see that its abundance increases with V23 in comparison to V20 around $10^{-5}$ bar, as this species becomes the main N-bearing species instead of HCN.
            The species \ce{CH3NH2} is also plotted, as we could expect the \ce{CH2NH} double bond to be saturated by \ce{H2} and H atoms, but the main reaction producing \ce{CH3NH2} is the reaction \ce{CH3 + NH2 -> CH3NH2}. This reaction uses a PLOG description between 0.1 and 10 bar, hence its contribution for very low pressures is likely to be heavily overestimated. However, a detailed treatment of its pressure dependence raises the same problems as the reaction \ce{H2CN + H -> CH2NH} because it is a barrierless, pressure dependent reaction.
            Therefore, we conclude that \ce{CH2NH} is an intermediate species that links HCN and \ce{N2} abundances. The full mechanism linking these two species around $10^{-5}$ is detailed in Fig. \ref{fig:HCN_diffs}.
            \ce{CH2NH} is mainly hydrogenated through the reaction \ce{CH2NH + H -> CH2NH2}. The \ce{CH2NH2} radical then gets its C-N bond broken by the addition of another H atom through the reaction \ce{CH2NH2 + H -> CH3 + NH2}. These two reactions and the species \ce{CH2NH} and \ce{CH2NH2} are only included in V23 and not in V20. The \ce{NH2} radical formed then gets destroyed by \ce{N(^4S)} atoms through the reaction \ce{NH2 + N(^4S) -> N2 + H + H}. This reaction has the same parameters between V23 and V20, although it is reversible in V20 and not in V23. Despite being reversible, the enthalpy difference between the reactants and the products is too high for it to be significantly reversed in V20. This is important because this reaction is an implicit combination of two other reactions, \ce{NH2 + N(^4S) -> NNH + H} and \ce{NNH -> N2 + H}, and reversing the resulting combination \ce{NH2 + N(^4S) -> N2 + H + H} would be unphysical.
            In addition, \ce{N(^4S)} atoms are produced from \ce{NH2} through the reactions \ce{NH2 + H -> NH + H2} and \ce{NH + H -> N(^4S) + H2}. While the parameters for this second reaction are really similar between V20 and V23, for the first one, they are very different. Both consider this reaction as \ce{NH + H2 -> NH2 + H} and use a temperature exponent of zero, but in V20 the activation energy is close to 20 kcal/mol while it is close to 15 kcal/mol in V23. The pre-exponential factor is also different, with a value of $10^{14}$ \si{cm^3.mol^{-1}.s^{-1}} in V20 and $2.1 \times 10^{13}$ \si{cm^3.mol^{-1}.s^{-1}} in V23, 5 times lower. The reaction \ce{H2CN + N(^4S) -> N2 + ^3CH2}, which is the main formation reaction of \ce{N2} in V20 has similar parameters between V20 and V23, although they differ on the pre-exponential factor by a factor of 3.

            \subsubsection{Consequences on transmission spectra}

\begin{figure}[htbp]
                \centering
                \includegraphics[width=0.5\textwidth]{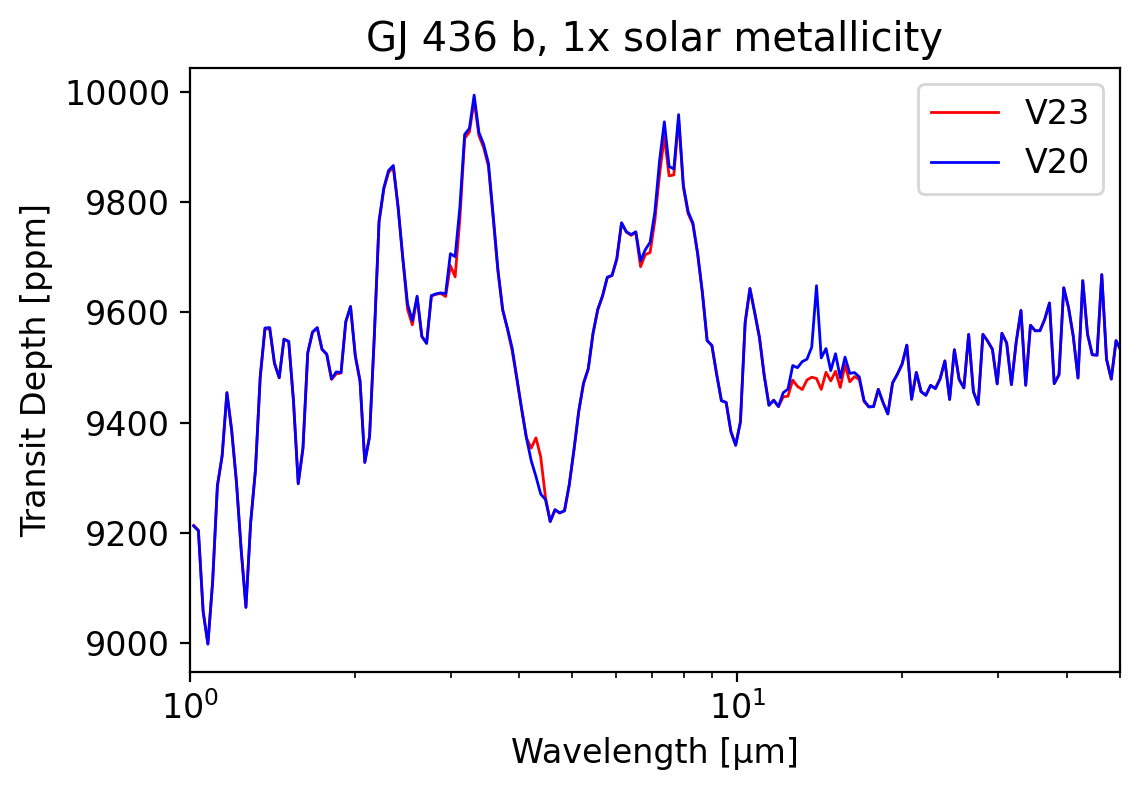}
                \caption{Synthetic transmission spectra of GJ 436 b with a solar metallicity, at a resolution of 50, corresponding to the atmospheric compositions calculated with V23 (in red) and V20 (in blue).}
                \label{fig:spectrum_GJ436b_solar}
            \end{figure}

\begin{figure}[htbp]
                \centering
                \includegraphics[width=0.5\textwidth]{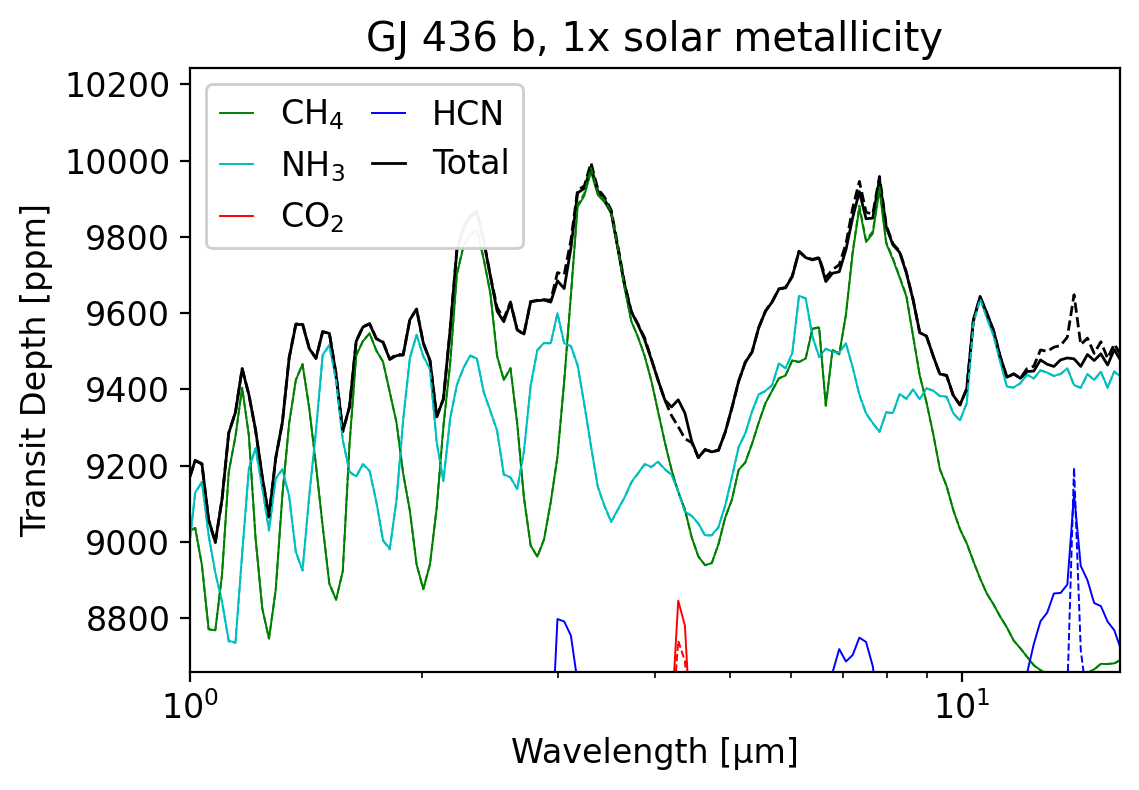}
                \caption{Contributions of major species to the total synthetic transmission spectra of GJ 436 b with a solar metallicity and $K_{zz} = 10^{9}$ \si{cm^2.s^{-1}}. Dashed lines are for V20, while solid lines are for V23. In the middle, we see the \ce{CO2} contribution that leads to a new feature in the spectrum with V23.}
            \label{fig:spectrumcontrib}
            \end{figure}

            These differences in the abundance profiles are also expressed in the synthetic transmission spectra showed in Fig. \ref{fig:spectrum_GJ436b_solar}.
            The higher \ce{CO2} abundance obtained with V23 in the upper atmosphere increases the apparent radius of GJ 436 b around 4.2 \si{\micro\metre}, leading to the apparition of a new \ce{CO2} feature with an amplitude of about 100 ppm. This happens because the transmission spectrum contribution of \ce{CO2} approaches that of \ce{CH4} and \ce{NH3}, which dominate at this wavelength for the abundance predicted with V20 (Fig. \ref{fig:spectrumcontrib}).
            Another major change in the spectrum is the disappearance of the HCN feature around 13 \si{\micro\metre} compared to an amplitude of about 200 ppm with V20, due to the drop in its abundance.
            %The amplitude of these changes is somewhat similar to those detectable by currently available instruments like the JWST. For comparison, \ce{SO2} detection with the JWST in \cite{tsai2022} was based on a peak amplitude of around 400 ppm, which is comparable to the differences observed between V20 and V23 in the HCN feature around 13 \si{\micro\metre} of the synthetic transmission spectrum of GJ 436 b with a solar metallicity.
            %\textbf{The V20 network would then predict an observable HCN feature, while V23 would not for this planet and these modeling parameters.
            %In the present case, these changes in \ce{CO2} and HCN features have a rather small amplitude, but it is likely that other modeling parameters would produce larger features changes in the spectrum for these two species.}

            \subsubsection{Case of 100 times solar metallicity}

\begin{figure}[htbp]
                \centering
                \includegraphics[width=0.5\textwidth]{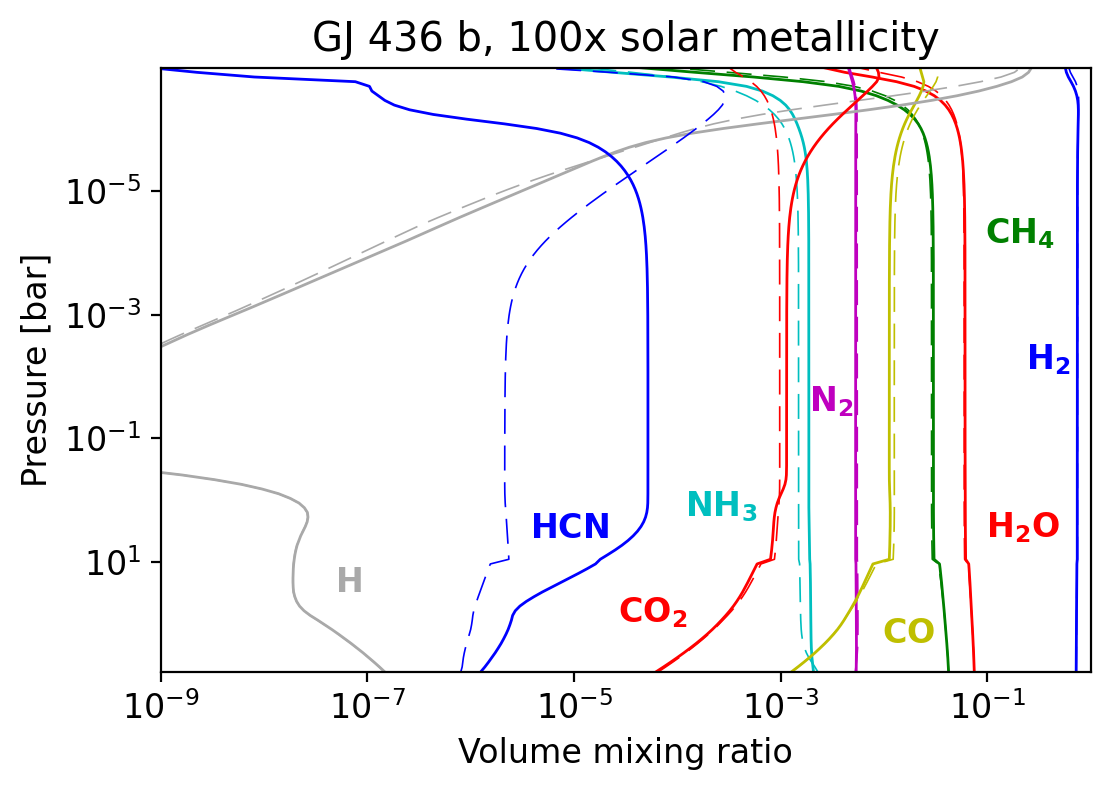}
                \caption{Abundance profiles of GJ 436 b for 100 times solar metallicity and a constant eddy diffusion coefficient of $10^{9}$ \si{cm^2.s^{-1}}. Dashed lines are for V20, while solid lines are for V23.}
                \label{fig:abundance_GJ436b_100xsolar}
            \end{figure}

\begin{figure}[htbp]
                \centering
                \includegraphics[width=0.5\textwidth]{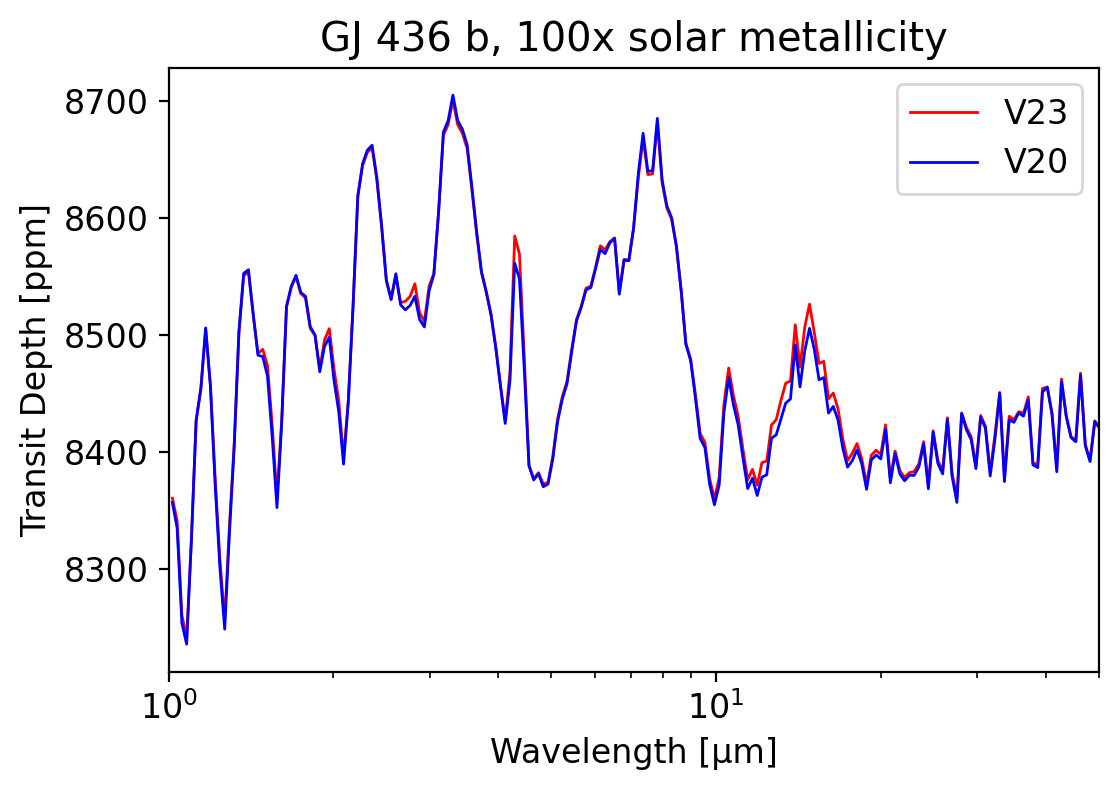}
                \caption{Synthetic transmission spectra of GJ 436 b with 100 times solar metallicity, at a resolution of 50, corresponding to the atmospheric compositions calculated with V23 (in red) and V20 (in blue).}
                \label{fig:spectrum_GJ436b_100xsolar}
            \end{figure}

            We also simulated the atmosphere of GJ 436 b assuming a higher metallicity (100x solar) ($10^9$ \si{cm^2.s^{-1}}), but keeping the same PT profile.
            In this case, variations between V20 and V23 are also observed, but to a lower extent (Fig. \ref{fig:abundance_GJ436b_100xsolar}).
            Compared to the previous case with solar metallicity, the amplitude of \ce{HCN} differences is lower, because the profile is strongly quenched, due to the higher abundance of related species that causes a higher flux of species. However, for pressures from 10 to $10^{-5}$ bar, HCN abundance in V23 is still almost two orders of magnitude above that of V20. For pressures around $10^{-6}$ bar, the HCN abundance in V20 is almost four orders of magnitude greater than in V23. These changes directly relate to the network differences discussed for the solar metallicity case in Sect. \ref{section:HCN_discussion}, and particularly the reaction \ce{H2CN + H -> CH2NH} for the upper atmosphere differences, and the reaction \ce{CH2NH + H -> H2CN + H2} for the lower atmosphere. We also observe differences in the thermochemical equilibrium region, which are due to discussed differences in the thermochemical data. This was verified through the same method described earlier, by disabling these specific reactions to see how they impact HCN abundance profile.
            In addition, \ce{CO2} is more abundant in V23 than in V20 around  $10^{-6}$ bar by almost one order of magnitude, due to differences in the \ce{CO + OH -> CO2 + H} reaction rates (Sect. \ref{section:CO2_discussion}).
            \ce{NH3} abundance profile is also slightly higher in V23 than V20, because the quenching point seems to happen at slightly higher pressures.
            For this metallicity case, these variations in abundances have very little impact on the transmission spectrum (Fig. \ref{fig:spectrum_GJ436b_100xsolar}).
            The slight change in \ce{NH3} abundance profiles barely causes some features to undergo an amplitude change, but HCN clearly does not show any impact on the spectrum, because its contribution is well under the contributions of \ce{CH4} and \ce{NH3} (Fig. \ref{fig:spectra_contrib}).

            \subsection{Case of GJ 1214 b}

\begin{figure}[htbp]
                \centering
                \includegraphics[width=0.5\textwidth]{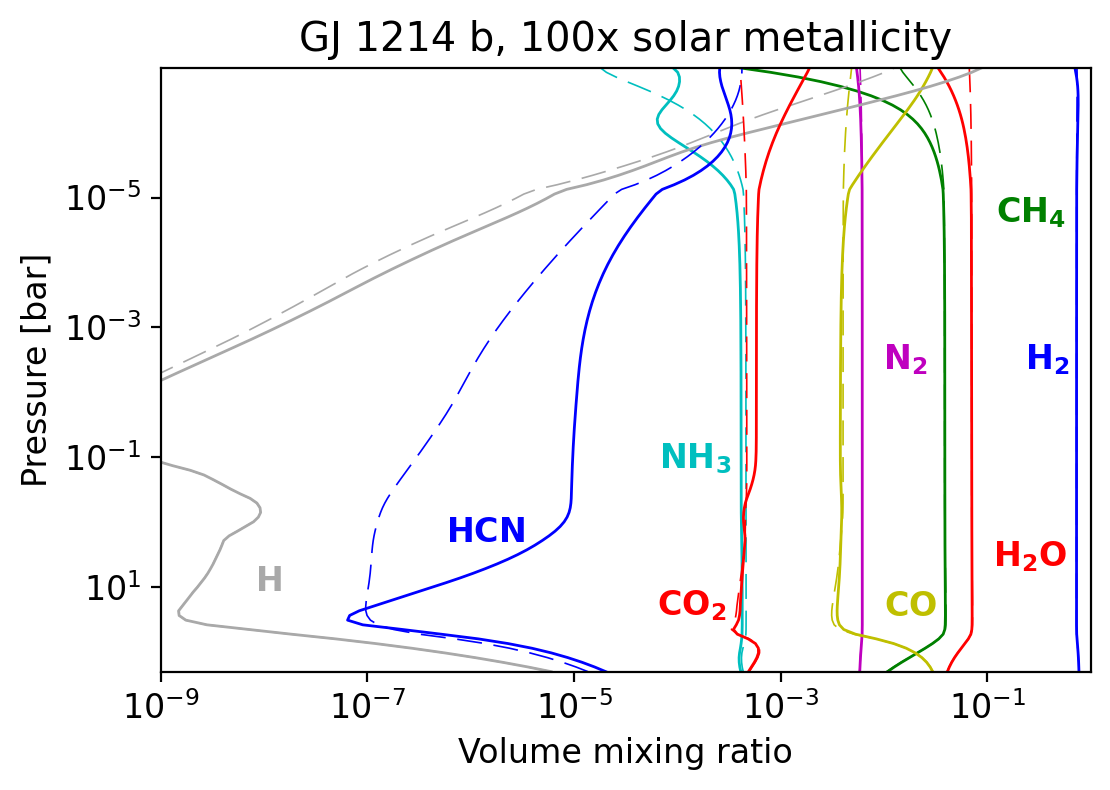}
                \caption{Abundance profiles of GJ 1214 b for 100 times solar metallicity and a pressure-dependent eddy diffusion coefficient. Dashed lines are for V20, while solid lines are for V23.}
                \label{fig:abundance_GJ1214b}
            \end{figure}

            For network comparisons with the warm Neptune GJ 1214 b, we used a pressure dependent eddy diffusion coefficient profile calculated with the formula $K_{zz} = 3 \times 10^{7} \times P^{-0.4}$ \si{cm^2.s^{-1}} given in \cite{charnay2015}. As this planet is expected to have a high metallicity \citep{desert2011, bean2011, gao2023, kempton2023}, we chose to model this planet with 100 times solar metallicity. The PT profile was taken from \cite{venot2020} and the UV flux used was that of GJ 436.
            The resulting abundance profiles (Fig. \ref{fig:abundance_GJ1214b}) do not show a lot of difference between the two chemical networks.
            \ce{HCN} abundance profile is still above the value predicted by V20 up to a factor of 100 around \textbf{1} bar, while other species are mildly affected, except for very low pressure regions around $10^{-6}$ bar, which are one order of magnitude above for \ce{CO} and \ce{CO2}.
            Contrary to the previous cases of GJ 1214 b at 1x and 100x solar metallicity, the HCN abundance profile of V23 is never lower than that of V20, expect at the very limit of the P-T profile, around $10^{-7}$ bar. This means that the reaction \ce{H2CN + H -> CH2NH} has very little impact on HCN in the upper atmosphere of this planet, and it was indeed verified by disabling the reaction. 
            Similarly to previous cases and with the same method, for the remaining differences in the lower atmosphere, we identified the same responsible reactions to be \ce{H2CN + H -> CH2NH}, \ce{CH2NH + H -> H2CN + H2} and the related reactions previously discussed in Sect. \ref{section:HCN_discussion}.
            For the \ce{CO2} abundance profile, the differences stem from the reaction \ce{CO + OH -> CO2 + H}, as discussed in Sect. \ref{section:CO2_discussion}.
            The CO differences come from the reaction \ce{C + OH -> CO + H}, which is the only major production reaction at those pressures in V23. This reaction is also included in V20 with the same parameters. Then,
            \ce{CH4} and \ce{H2O} follow the opposite trend, resulting (respectively) from a higher loss contribution of the reaction \ce{CH4 + CH -> C2H4 + H} and the reaction \ce{H2O + CH -> H2CO + H} which is exclusive to V23.

\begin{figure}[htbp]
                \centering
                \includegraphics[width=0.5\textwidth]{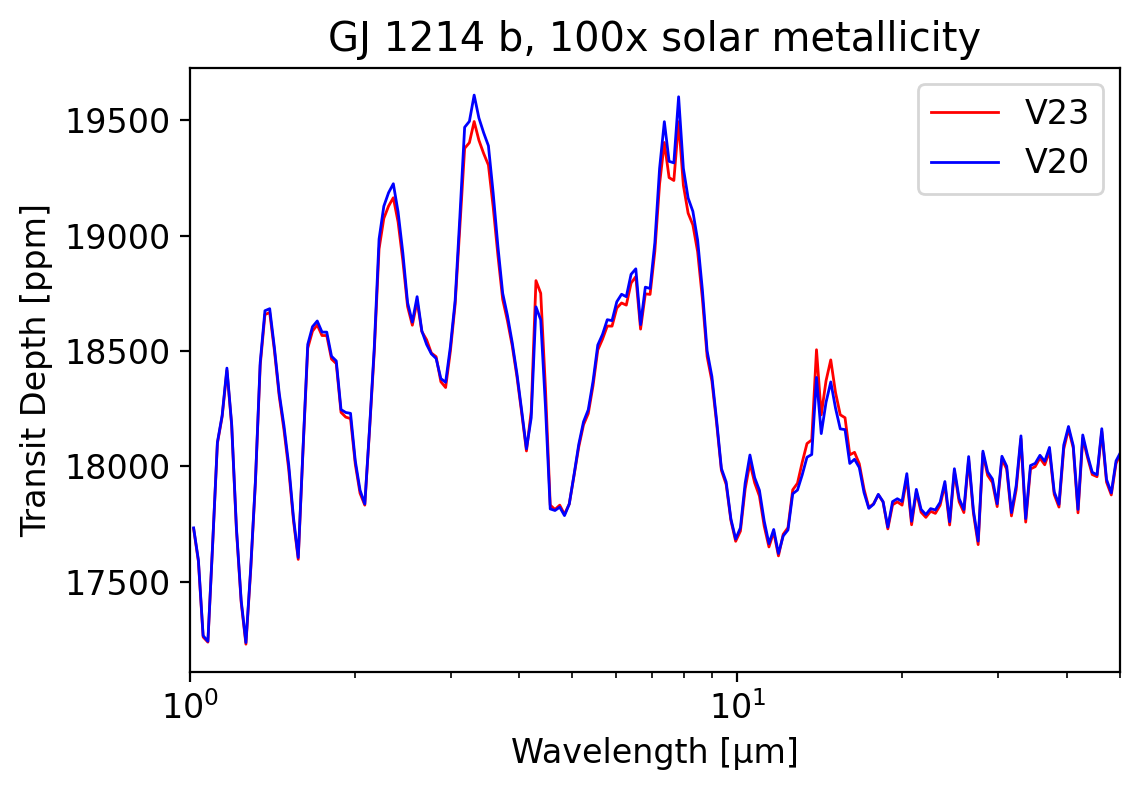}
                \caption{Synthetic transmission spectra of GJ 1214 b, at a resolution of 50, corresponding to the atmospheric compositions calculated with V23 (in red) and V20 (in blue).}
                \label{fig:spectrum_GJ1214b}
            \end{figure}

            Due to the low amplitude of these changes, the corresponding synthetic spectra (Fig. \ref{fig:spectrum_GJ1214b}) shows no new features, with only a few minor variations in the amplitude of existing features, largely under observable values.

            \subsection{Case of HD 189733 b and HD 209458 b}
            
\begin{figure}[htbp]
                \centering
                \includegraphics[width=0.5\textwidth]{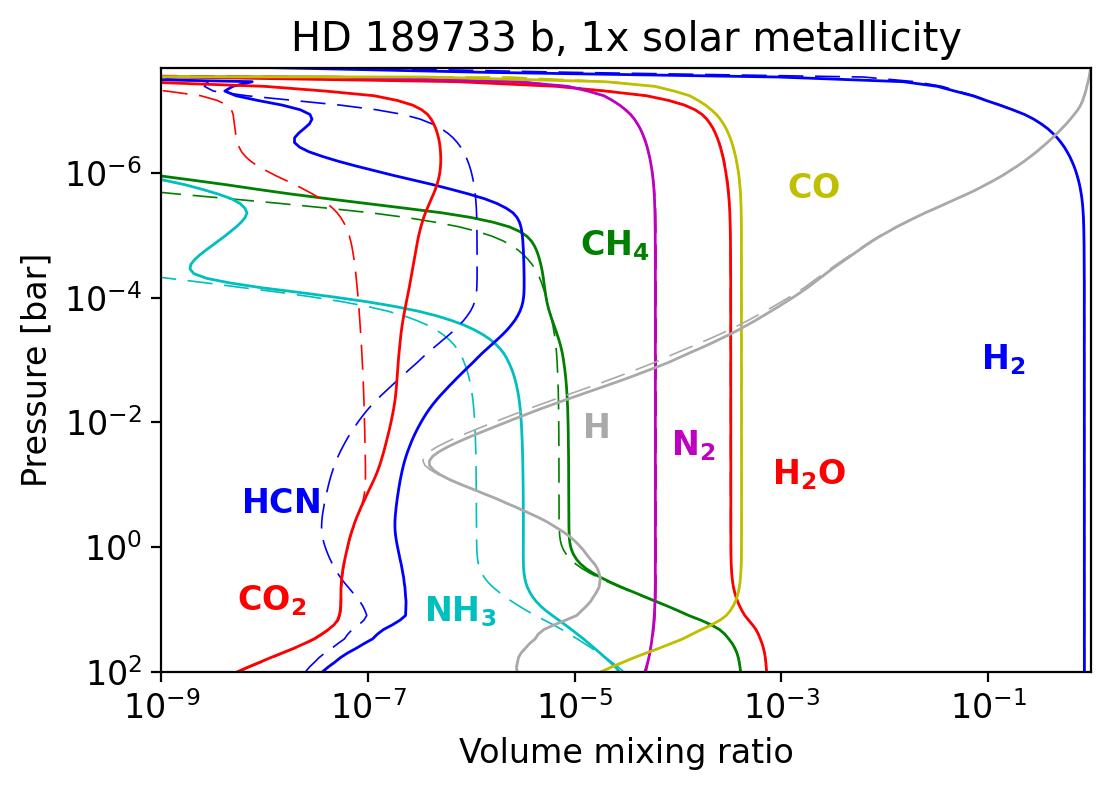}
                \caption{Abundance profiles of HD 189733 b for solar metallicity and a pressure-dependent eddy diffusion coefficient. Dashed lines are for V20, while solid lines are for V23.}
                \label{fig:abundance_HD189733b}
            \end{figure}

\begin{figure}[htbp]
                \centering
                \includegraphics[width=0.5\textwidth]{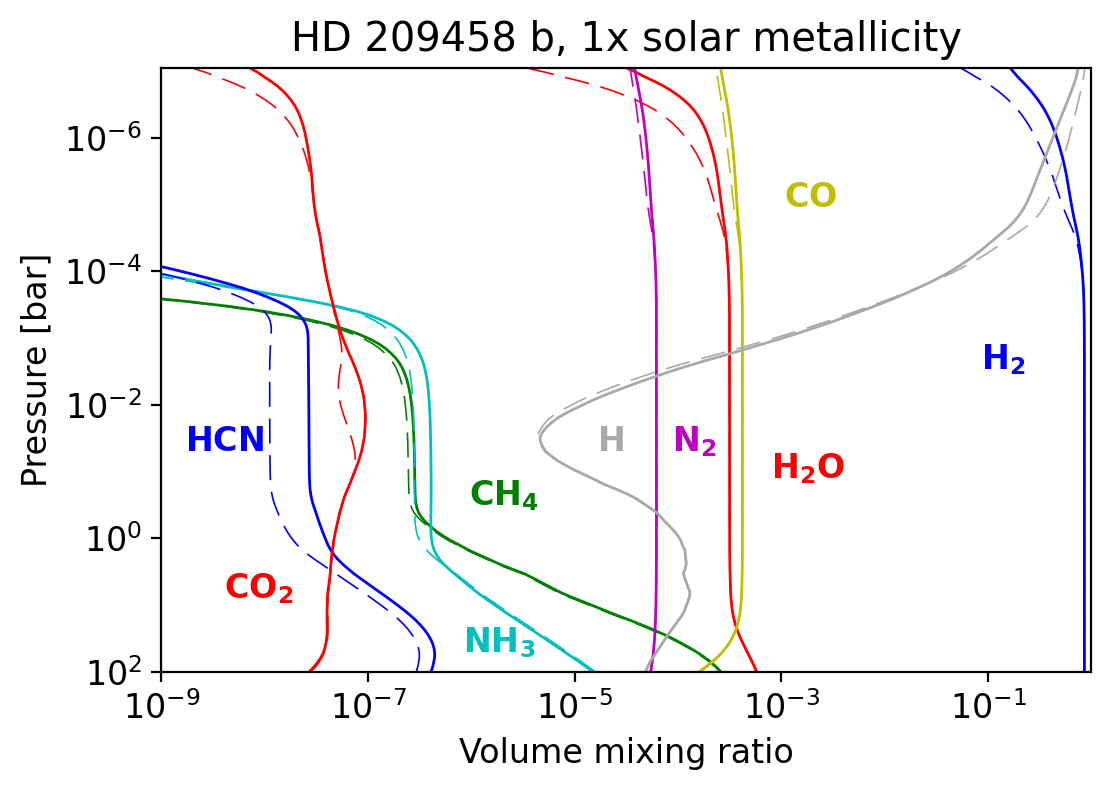}
                \caption{Abundance profiles of HD 209458 b for solar metallicity and a pressure-dependent eddy diffusion coefficient. Dashed lines are for V20, while solid lines are for V23.}
                \label{fig:abundance_HD209458b}
            \end{figure}

            While we observe that for warm Neptunes the most significant impacts of our new chemical scheme are found for low metallicity atmospheres, we examined the effect on two hot Jupiters: HD 189733 b and HD 209458 b.
            We used the same P-T profiles, UV fluxes, eddy diffusion coefficients, and metallicities as in \cite{venot2020}.
            The abundances profiles obtained using the two chemical schemes V20 and V23 are shown in Figs. \ref{fig:abundance_HD189733b} and \ref{fig:abundance_HD209458b}.


            For HD 189733 b, the only differences are for species lower than $10^{-5}$ abundance. \ce{CO2} is more abundant in V23 around $10^{-6}$ bar than in V20, due to the \ce{CO + OH -> CO2 + H} reaction (Sect. \ref{section:CO2_discussion}).
            %\ce{NH3} and HCN levels are also a bit less than one order of magnitude higher in V23.
            The abundances of \ce{NH3} and HCN are slightly less than one order of magnitude higher in V23 than in V20 between 10 and $10^{-3}$ bar.
            The larger \ce{NH3} difference between $10^{-4}$ and $10^{-6}$ bar is due to the reaction \ce{CH2NH2 + H -> ^1CH2 + NH3} which is analogous to the reaction \ce{CH2NH2 + H -> CH3 + NH2} discussed in Sect. \ref{section:HCN_discussion}.
            The differences between 10 and $10^{-3}$ bar are explained by two reactions : \ce{NH3 + NH2 -> N2H3 + H2} and \ce{NH2 + NH2 -> N2H2 + H2}. The first one is not included in V23, while in V20 it contributes to \ce{NH3} consumption. The second one is included in both networks, but the parameters used in the reactions are different, especially for the activation energy, which is 10 kcal/mol higher in V23 than in V20.
            When combined with the changes in the thermochemical data, the differences in these rate constants explain the differences between the HCN abundance profiles for pressures higher than $10^{-5}$ bar. For lower pressures, the reaction \ce{H2CN + H -> CH2NH} increases HCN consumption in V23 in comparison to V20 as previously discussed, leading to over one order of magnitude less HCN in V23 than in V20.
            A very detailed comparison between the chemical schemes of \cite{venot2012} and \cite{moses2011} (which we name V12 and M11 in the following, respectively) has been performed by \cite{moses2014}, taking  HD 189733 b as a case study, with the parameters used in \cite{venot2012}.
            Thus, it is interesting to evaluate how the results obtained with our new scheme compare with these two schemes.
            %This planet had also been subject to a comparison between Venot 2012 and Moses 2011 in \cite{moses2014}.
            The highlighted differences concerned the species \ce{NH3}, \ce{CH4}, and HCN.
            
                With M11, \ce{NH3} was one order of magnitude above V12 in the range 10 to $10^{-3}$ bar and one order of magnitude under V12 in the range $10^{-5}$ to $10^{-7}$ bar.
                The V23 \ce{NH3} abundance profile halves this gap between M11 and V12 in the range 10 to $10^{-3}$ bar, being nearly half an order of magnitude above V12 and half an order of magnitude under M11.
                However, in the pressure range $10^{-5}$ to $10^{-7}$ bar, the V23 \ce{NH3} abundance is much higher than in both chemical networks, reaching a difference of seven orders of magnitude around $10^{-5}$ bar.
    
                For \ce{CH4}, the differences mainly concerned the pressure range 1 to $10^{-3}$ bar, with the M11 profiles being around half an order of magnitude above V12.
                The corresponding V23 profile slightly approaches the M11 profiles, also halving the gap between M11 and V12 for \ce{CH4}.
    
                For HCN, the differences concerned the range 100 to $10^{-7}$ bar, the M11 profile being an order of magnitude above V12. The V23 profile comes closer to the M11 profile in comparison to both V12 and V20 for the range 100 to $10^{-5}$ bar, but the HCN abundance drops for V23 around $10^{-6}$ bar, resulting in a difference of three orders of magnitude with M11 at this pressure.

            Overall, these results bring the abundances a bit closer to M11, but enhances differences in the upper atmosphere where new \ce{CH2NH} chemical pathways begin to take effect.

\begin{figure}[htbp]
                \centering
                \includegraphics[width=0.5\textwidth]{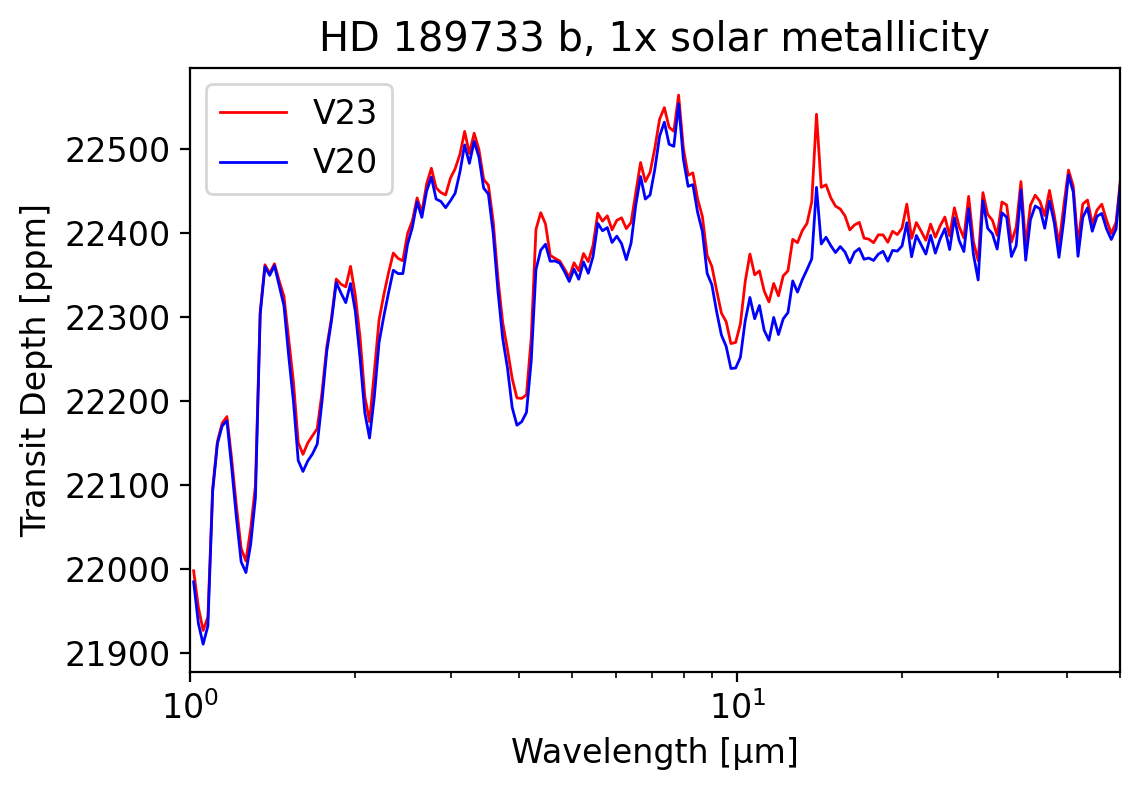}
                \caption{Synthetic transmission spectra of HD 189733 b, at a resolution of 50, corresponding to the atmospheric compositions calculated with V23 (in red) and V20 (in blue).}
                \label{fig:spectrum_HD189733b}
            \end{figure}

            %These slight differences manage to affect the spectrump resented in Fig. \ref{fig:spectrum_HD189733b}.
            We calculated the synthetic transmission spectrum corresponding to the abundances obtained with V23 and V20 and observed some differences (see Fig. \ref{fig:spectrum_HD189733b}).
            The main effect is an increase of the transit depth between 10 and 20 \si{\micro\metre}, with an amplitude of up to 50 ppm. No new feature is created in the spectrum, but the differences generated are above the instrumental precision. Thus, the change of chemical scheme could impact the interpretations of the transmission spectrum and the retrieval of \ce{NH3} abundance.

            For the hotter HD 209458 b, the differences are even smaller than what we observed for HD189733b, with an amplitude lower than one order of magnitude. The \ce{NH3} and \ce{HCN} abundances are still a bit higher in V23 as with all previous exoplanets, except for GJ 436 b with a solar metallicity. The reactions causing these differences are the same as for the case of HD 189733 b.
            Another minor difference is that \ce{CO2}, \ce{H2O} and \ce{H2} are more abundant in V23 for pressures lower than $10^{-6}$ bar, unlike H atoms.
            The formation pathways of these species at this pressure being the same in V20 and V23, the differences must come from slight differences in the parameters of the reactions.

\begin{figure}[htbp]
                \centering
                \includegraphics[width=0.5\textwidth]{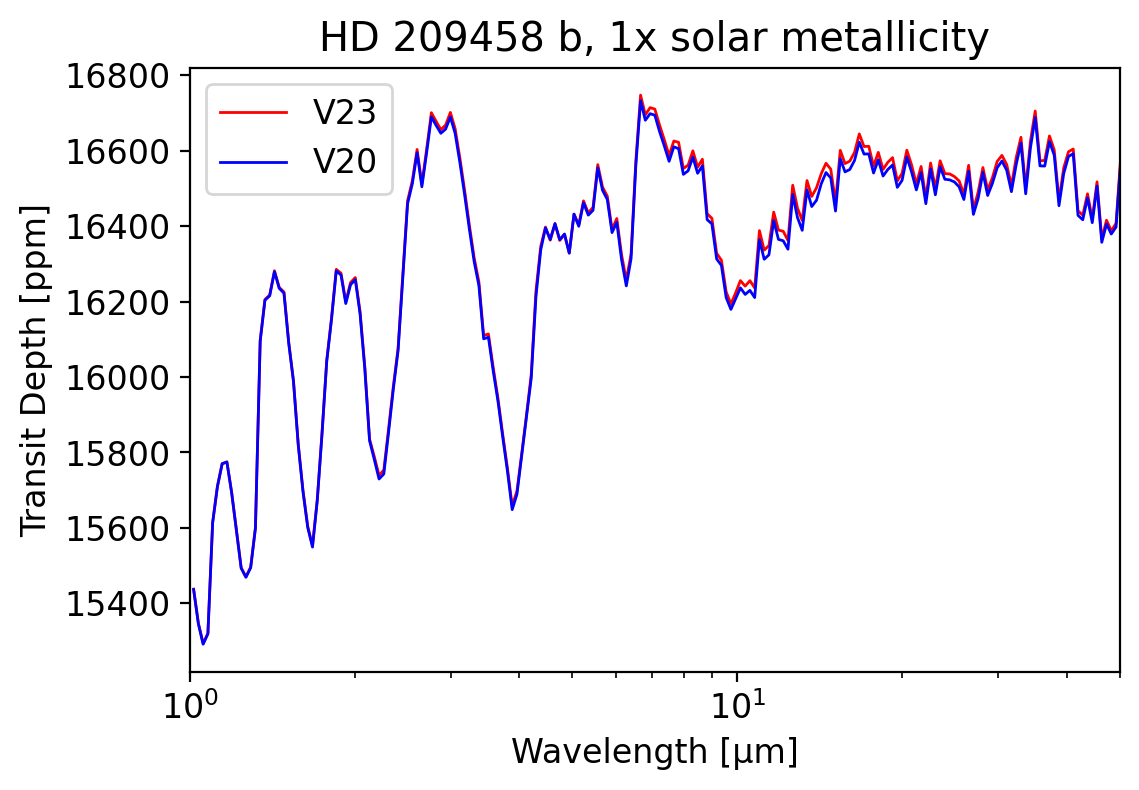}
                \caption{Synthetic transmission spectra of HD 209458 b, at a resolution of 50, corresponding to the atmospheric compositions calculated with V23 (in red) and V20 (in blue).}
                \label{fig:spectrum_HD209458b}
            \end{figure}

            Overall, the variations of abundances observed with V23 for this planet in comparison to V20 are very small, and concern mainly species with low abundances (<$10^{-6}$). As a consequence, the synthetic transmission spectra calculated for both networks (Fig. \ref{fig:spectrum_HD209458b}) are very similar.
            
    \section{Conclusion}
\label{section:conclusion}

    In this work, we developed a new \ce{C2} C/H/O/N detailed chemical scheme to model exoplanet disequilibrium chemistry. It was derived from experimentally tested combustion networks, extensively validated on 1618 experimental measurements on a wide range of conditions and compared to other chemical networks performances, such as V20 through a statistical study. Verifications, additions, and detailing of possibly missing radical reactions were also performed, resulting in a much more reliable network than Venot 2020.
    This network was then  used to model two warm Neptunes and two hot Jupiters using the kinetic model FRECKLL, and the results were compared to those obtained with our previous chemical scheme, V20.
    The chemistry differences were analyzed and new chemical pathways were found, such as the importance of \ce{CH2NH} and its formation through \ce{H2CN} to couple the nitrogen chemistry to \ce{CH3} radicals. This effect has been highlighted in a solar metallicity warm Neptune, and is such expected to mainly impact warm exoplanets with a low metallicity.
    Transmission spectra were also simulated using TauREx 3.1 and the resulting changes in abundances were found to significantly impact the spectrum for GJ 436 b, with differences around 100 ppm for \ce{CO2} at 4.2 µm and 200 ppm for HCN at 13 µm. The amplitude of these features is within the detection capabilities of JWST, which confirms that disequilibrium chemistry model accuracy is crucial to draw correct conclusions from observations.
    In the context of ongoing and future missions (e.g., JWST and Ariel), disequilibrium chemistry modelling will become increasingly important as we get access to higher precision observations.
    Improvements are still awaited in the development of experimentally validated sulfur chemical scheme: a compound such as \ce{SO2} has been recently been detected with the JWST and is typically a product of photochemistry \citep{tsai2022}. Expanding our network to sulfur species and their coupling to carbon and nitrogen species will be the next step toward a more complete chemical scheme addressing modern problematics in the exoplanet chemistry field.
    %tackle these issues.
    More in-depth insights on the critical reactions for each species through sensitivity analysis or other methods in a wide range of exoplanetary conditions could also help to further improve the reliability of these networks, enabling to identify the key reactions in the mechanism and could help to focus the community's efforts and reduce the associated uncertainty through more accurate but computationally intensive ab initio calculations such as VRC-TST for barrierless reactions and RRKM/ME for pressure dependence.
    As in situ experimentation with probes is impossible in the field of exoplanet chemistry, the use of chemical networks validated on combustion experiments remains the only way to validate our kinetic models to this day.
    
% ------------------------------------------------------------------------------
%  Acknowledgements
% ------------------------------------------------------------------------------

\begin{acknowledgements}
This project is fund by the ANR project `EXACT' (ANR-21-CE49-0008-01). In addition, O.V. acknowledges funding from the Centre National d'\'{E}tudes Spatiales (CNES), and from the CNRS/INSU Programme National de Plan\'etologie (PNP).
\end{acknowledgements}

% ------------------------------------------------------------------------------
%  Bibliography
% ------------------------------------------------------------------------------

\bibliographystyle{aa} % style aa.bst
\bibliography{bibliography.bib} % your references Yourfile.bib

% ------------------------------------------------------------------------------
%  Appendix
% ------------------------------------------------------------------------------

\begin{appendix}

\clearpage% Flush earlier floats (otherwise order might not be correct)
\thispagestyle{empty}% empty page style (?)
%\begin{sidewaystable}[htpb]
\newgeometry{margin=1cm}
\begin{landscape}
\section{Experimental data}
%\begin{Annexe}
%\begin{adjustwidth}{100pt}{100pt}
\begin{table}[hp]
\centering
\caption{\centering Experimental conditions used for comparison and simulation with each C/H/O/N mechanism\tablefootnote{\textbf{N°} is the condition reference number and $\mathbf{N_{points}}$ is the number of experimental measurements in the set.}.}%IDTs have no defined reaction time.}
\renewcommand{\arraystretch}{1.875}
        \scalebox{1}{
                %               \rotatebox{90}{%
                        %\begin{tabular}{|c|c|c|c|c|c|c|c|}
            \begin{tabular}{|c|c|c|c|c|c|c|c|}
                                \hline
                                \textbf{N°} & $\mathbf{N_{points}}$ & \textbf{Reaction conditions}                & \textbf{Measured quantity} & \textbf{Variable}    & \textbf{Temperature (K)} & \textbf{Pressure (bar)} & \textbf{Reaction time (s)} \\ \hline
                                1         & 15           & \ce{CH3OH} Combustion              & IDT               & Temperature & 1000 - 1500 & 10.18, 49.18   & N/A \\ %\hline
                                2         & 33           & \ce{CH3OH} Combustion              & Mole Fraction     & Time        & 823         & 0.26           & 26 \\ %\hline
                                3         & 112          & \ce{CH3OH} Combustion              & Mole Fraction     & Time        & 949         & 2.5            & 0.5 \\ %\hline
                                4         & 99           & \ce{CH3OH} Pyrolysis               & Mole Fraction     & Time        & 1266, 1368, 1458, 1567, 1610 & 2.5, 2.4, 2.3, 2.1, 2.2 & 0.0015 \\ %\hline
                                5         & 48           & \ce{CH3OH} Combustion              & Mole Fraction     & Temperature & 700 - 1200  & 20             & 1 \\ %\hline
                                6         & 54           & \ce{CH3OH} Combustion              & Mole Fraction     & Time        & 1000        & 1              & 0.27 \\ %\hline
                                7         & 64           & \ce{CH3OH} Combustion              & Mole Fraction     & Time        & 1031        & 1              & 0.125 \\ %\hline
                                8         & 68           & \ce{CH3OH} Combustion              & Mole Fraction     & Time        & 783         & 15             & 3.5 \\ %\hline
                                9         & 72           & \ce{CH4} Combustion                & Mole Fraction     & Temperature & 900 - 1200  & 10             & 1 \\ %\hline
                                10        & 24           & \ce{H2} Combustion                 & Mole Fraction     & Temperature & 800 - 1150  & 10             & 1 \\ %\hline
                                11        & 61           & \ce{H2} Combustion                 & Mole Fraction     & Time        & 934         & 3.02           & 0.7 \\ %\hline
                                12        & 16           & \ce{H2} Combustion in CO           & Mole Fraction     & Temperature & 850 - 1200  & 1              & 1 \\ %\hline
                                13        & 75           & CO + \ce{H2O}                      & Mole Fraction     & Time        & 1040        & 1, 2.4, 3.46, 6.5, 9.6 & 1.05 \\ %\hline
                                14        & 45           & \ce{H2} Combustion                 & IDT               & Temperature & 900 - 1700  & 1, 4, 16       & N/A \\ %\hline
                                15        & 72           & HCN Combustion                     & Mole Fraction     & Temperature & 900 - 1500  & 1              & 195 / T \\ %\hline
                                16        & 83           & \ce{CH4} Pyrolysis                 & Mole Fraction     & Temperature & 1250 - 1500 & 1              & 4550 / T \\ %\hline
                                17        & 60           & \ce{CH4} Pyrolysis                 & Mole Fraction     & Temperature & 1600 - 2400, 1300 - 2000 & 1.5, 30 & 0.003, 0.015 \\ %\hline
                                18        & 56           & Ethanol Pyrolysis                  & Mole Fraction     & Temperature & 1000 - 1300 & 0.74           & 0.006 \\ %\hline
                                19        & 399          & Ethanol Pyrolysis                  & Mole Fraction     & Time        & 950         & 3, 6, 9, 12    & 0.4, 0.8, 1.2, 1.4 \\ %\hline
                                20        & 87           & HCN + NO ( + \ce{NH3} ) Combustion & Mole Fraction     & Temperature & 900 - 1500  & 1              & 197 / T \\ %\hline
                                21        & 75           & \ce{H2} + \ce{N2O}                 & Mole Fraction     & Time        & 995         & 3              & 0.5 \\ \hline
                        \end{tabular}
                }%
                %       }
\label{tab:conditions}
\end{table}
%\end{sidewaystable}
%\end{Annexe}
%\end{adjustwidth}
\end{landscape}

\clearpage% Flush earlier floats (otherwise order might not be correct)
\thispagestyle{empty}% empty page style (?)
%\begin{sidewaystable}[htpb]
\begin{landscape}
%\begin{Annexe}
%\begin{adjustwidth}{5000pt}{5000pt}
\begin{table}[hp]
\ContinuedFloat  %%
\centering
\caption{\centering \textbf{(Continued)}. \textbf{Species data} refers to the species whose mole fraction has been measured.} %PSR refers to a Perfectly Stirred Reactor (usually a Jet-stirred reactor).}
\renewcommand{\arraystretch}{1.875}
        \scalebox{1}{
                %               \rotatebox{90}{%
                        \begin{tabular}{|c|c|c|c|c|c|}
                                \hline
                                \textbf{N°} & \textbf{Reactor}        & \textbf{Parameter study}   & \textbf{Reactants} & \textbf{Species data} & \textbf{Data Source} \\ \hline
                                1         & Shock Tube     & Pressure          & 5.7 \% \ce{CH3OH}, 8.55 \% \ce{O2}, Ar & N/A & \cite{curranmodel} \\ %\hline
                                2         & Static Reactor & N/A               & 5.89 \% \ce{CH3OH}, 8.84 \% \ce{O2}, \ce{N2}& \ce{H2}, CO, \ce{CO2}, \ce{CH3OH} and \ce{CH2O} & \cite{cathonnet1982} \\ %\hline
                                3         & Plug Flow      & N/A               & 0.333 \% \ce{CH3OH}, 0.601 \% \ce{O2}, \ce{N2} & \ce{CH3OH}, \ce{O2}, \ce{H2O}, CO, \ce{H2} and \ce{CO2} & \cite{held1994} \\ %\hline
                                4         & Shock Tube     & N/A               & 1 \% \ce{CH3OH}, Ar & \ce{CH3OH}, CO & \cite{ren2013} \\ %\hline
                                5         & PSR            & N/A               & 0.24 \% \ce{CH3OH}, 0.3 \% \ce{O2}, \ce{N2} & \ce{CH3OH}, \ce{O2}, CO and \ce{CO2} & \cite{curranmodel} \\ %\hline
                                6         & Plug Flow      & N/A               & 0.735 \% \ce{CH3OH}, 0.65 \% \ce{O2}, \ce{N2} & \ce{CH3OH}, \ce{CH2O}, CO & \cite{aronowitz1979} \\ %\hline
                                7         & Plug Flow      & N/A               & 0.93 \% \ce{CH3OH}, 1.18 \% \ce{O2}, \ce{N2} & \ce{O2}, \ce{CH3OH}, CO, \ce{CO2} and \ce{CH2O} & \cite{norton1989} \\ %\hline
                                8         & Plug Flow      & N/A               & 0.415 \% \ce{CH3OH}, 0.6 \% \ce{O2}, \ce{N2} & \ce{O2}, \ce{CH3OH}, CO and \ce{CH2O} & \cite{held1994} \\ %\hline
                                9         & PSR            & N/A               & 0.3 \% \ce{CH4}, 0.00055 \% \ce{C2H6}, 1.2 \% \ce{O2}, \ce{N2} & CO, \ce{CO2}, \ce{CH4}, \ce{C2H6}, \ce{C2H4}, \ce{C2H2} & \cite{dagaut1990} \\ %\hline
                                10        & PSR            & N/A               & 0.93 \% \ce{H2}, 5 \% \ce{O2}, \ce{N2} & \ce{H2}, \ce{H2O}  & \cite{lecong2009} \\ %\hline
                                11        & Plug Flow      & N/A               & 0.95 \% \ce{H2}, 0.495 \% \ce{O2}, \ce{N2}  & \ce{H2}, \ce{O2} and \ce{H2O} & \cite{mueller1999} \\ %\hline
                                12        & PSR            & N/A               & 0.2 \% \ce{H2}, 0.2 \% CO, 2 \% \ce{O2}, \ce{N2} & \ce{H2}, \ce{CO2} & \cite{lecong2009} \\ %\hline
                                13        & Plug Flow      & Pressure          & 1 \% CO, 0.65 \% \ce{H2O}, 0.5 \% \ce{O2}, \ce{N2} & CO & \cite{kim1994} \\ %\hline
                                14        & Shock Tube     & Pressure          & 0.81 \% \ce{H2}, 4.03 \% \ce{O2}, Ar & N/A & \cite{keromnes2013} \\ %\hline
                                15        & Plug Flow      & Equivalence Ratio & 0.1 \% HCN, \ce{O2}, 0.7 \% \ce{H2O}, 25 \% Ar, \ce{CO2} & HCN, CO & \cite{gimenez-lopez2010} \\ %\hline
                                16        & Plug Flow      & Mixtures          & \ce{CH4}, \ce{CO2}, \ce{C2H6}, \ce{N2} & \ce{CH4}, \ce{H2}, \ce{C2H2}, \ce{C2H4}, \ce{C2H6}, CO, \ce{CO2} & \cite{keromnes2013} \\ %\hline
                                17        & Shock Tube     & Pressure          & 10 \% \ce{CH4}, 90 \% Ar & \ce{CH4}, \ce{C2H2}, \ce{C2H4} & \cite{nativel2019} \\ %\hline
                                18        & Plug Flow      & N/A               & 2.8 \% \ce{C2H5OH}, 97.2 \% Ar & \ce{C2H5OH}, \ce{C2H4}, \ce{CH3CHO}, \ce{H2}, \ce{CH4}& \cite{rotzoll1985} \\ %\hline
                                19        & Plug Flow      & Pressure          & 0.3 \% \ce{C2H5OH}, 0.0035 \% \ce{O2}, Ar & \ce{C2H5OH}, \ce{H2O}, \ce{C2H4}, \ce{CH3CHO}, CO, \ce{CH2O}, \ce{CH4} & \cite{li2001} \\ %\hline
                                20        & Plug Flow      & Mixtures          & HCN, \ce{O2}, NO, \ce{H2O}, CO, \ce{NH3} & NO, CO, HCN, \ce{N2O}, \ce{NH3} & \cite{dagaut2008} \\ %\hline
                                21        & Plug Flow      & N/A               & 0.53 \% \ce{H2}, 1.1 \% \ce{N2O}, \ce{N2} & \ce{N2O}, \ce{H2O}, \ce{H2}, NO, \ce{NH3} & \cite{allen1998} \\ \hline
                        \end{tabular}
                }%
                %       }
\label{tab:conditions2}
\end{table}
%\end{adjustwidth}
%\end{sidewaystable}
%\end{Annexe}
\end{landscape}

\section{Photodissociation data}
\renewcommand{\arraystretch}{1.15}
\begin{table*}[htbp]
\centering\caption{Photodissociation pathways, cross-sections, and quantum yields used in this study.}
\scalebox{0.9}{
\centering\begin{tabular}{|c|c|c|c|}
\hline
\textbf{Species} & \textbf{Products} & \textbf{Cross sections} & \textbf{Quantum yields} \\ \hline
\ce{H2O}     & H + OH & \cite{chan1993b, fillion2004}; & \cite{heays2017} \\
 & \ce{H2} + \ce{O(^1D)} & \cite{mota2005, ranjan2020} &    \\
 & H + H + \ce{O(^3P)} & &    \\
 
\ce{CO2}     & CO + \ce{O(^1D)} & \cite{stark2007, huestis2010}; & \cite{heays2017} \\
 & CO + \ce{O(^3P)} & \cite{venot2018} &    \\
 
\ce{H2CO}    & \ce{H2} + CO & \cite{cooper1996,meller2000} & \cite{heays2017} \\
 & H + HCO &  & \\
 & H + H + CO &  &    \\

OH           & \ce{O(^1D)} + H & \cite{huebner1992} & \cite{heays2017} \\
 & \ce{O(^3P)} + H &  &    \\
 
OOH          & OH + \ce{O(^3P)} & \cite{heays2017} &  \\

CO           & C + \ce{O(^3P)} & \cite{olney1997} & \cite{heays2017} \\
 & C + \ce{O(^1D)} & & \\

\ce{H2}      & H + H & \cite{samson1994}; & \cite{heays2017} \\
 &  & \cite{chan1992, olney1997} &    \\

HCO          & H + CO & \cite{loison1991, hochanadel1980} &    \\

\ce{CH3OH}   & \ce{H2CO} + \ce{H2} & \cite{burton1992, cheng2002} & \cite{heays2017} \\
 & \ce{CH3} + OH & & \\
 
\ce{CH3OOH}  & \ce{CH3O} + OH & \cite{vaghjiani1989, matthews2005} &    \\

%CHCO         &     &  &    \\
\ce{CH2CO}   & \ce{^3CH2} + CO & \cite{laufer1971} &    \\

\ce{CH3CHO}  & \ce{CH4} + CO & \cite{limao2003, burkholder2020} & \cite{heays2017} \\
 & \ce{CH3} + HCO & & \\
 
\ce{CH3}     & \ce{^1CH2} + H & \cite{heays2017} & \cite{heays2017} \\

\ce{CH4}     & \ce{CH3} + H & \cite{Au1993,lee2001}; & \cite{peng2014} \\
& \ce{^1CH2} + \ce{H2} & \cite{kameta2002,chen2004} & \\
& \ce{^1CH2} + H + H &  & \\
& \ce{^3CH2} + H + H & & \\
& CH + \ce{H2} + H & & \\

\ce{C2H}     & C + C + H & \cite{heays2017, fahr2003} &  \\

\ce{C2H2}    & \ce{C2H} + H & \cite{cooper1995a,wu2001} & \cite{heays2017, hebrard2013} \\

\ce{C2H3}    & \ce{C2H2} + H & \cite{fahr1998} &  \\

\ce{C2H4}    & \ce{C2H2} + \ce{H2} & \cite{cooper1995b,orkin1997}; & \cite{heays2017} \\
& \ce{C2H2} + H + H & \cite{wu2004} & \\

\ce{C2H6}    & \ce{C2H4} + \ce{H2} & \cite{Au1993,kameta1996}; & \cite{heays2017, hebrard2006} \\
& \ce{C2H4} + H + H & \cite{lee2001,chen2004} & \\
& \ce{C2H2} + \ce{H2} + \ce{H2} &  & \\
& \ce{CH4} + \ce{^1CH2} & & \\
& \ce{CH3} + \ce{CH3} & & \\

\ce{N2}      & \ce{N(^2D)} + \ce{N(^4S)} & \cite{samson1964,huffman1969}; &    \\
& & \cite{stark1992,chan1993c} & \\

\ce{NH2}     & NH + H & \cite{heays2017} & \cite{heays2017} \\

\ce{NH3}     & \ce{NH2} + H & \cite{burton1993,chen1998}; & \cite{heays2017} \\
& NH + \ce{H2} & \cite{cheng2006} & \\
& NH + H + H & & \\

\ce{N2H4}    & \ce{N2H3} + H & \cite{vaghjiani1993, biehl1991} &    \\

HCN, HNC          & CN + H & \cite{nuth1982};Bénilan et al. (in prep.); & \cite{heays2017} \\
& & Venot et al. (in prep.) & \\

%HNC          & CN + H & \cite{venot2020, nuth1982} & \cite{heays2017} \\
%& & Bénilan et al. (in prep.); Venot et al. (in prep.) & \\

\ce{H2CN}    & HCN + H & \cite{nizamov2003, teslja2006} & \\

\ce{C2N2}    & CN + CN & Bénilan et al. (in prep.) &    \\

NO           & \ce{N(^4S)} + \ce{O(^3P)} & \cite{iida1986,chan1993a} & \cite{heays2017} \\

\ce{NO2}     & NO + \ce{O(^3P)} & \cite{au1997, vandaele2002} & \cite{heays2017} \\
& NO + \ce{O(^1D)} & & \\

\ce{NO3}     & \ce{NO2} + \ce{O(^3P)} & \cite{orphal2003, sander1986} & \cite{heays2017} \\
& NO + \ce{O2} & \cite{yokelson1994} & \\

\ce{HNO2}    & NO + OH & \cite{burkholder2020} &    \\

\ce{HNO3}    & \ce{NO2} + OH & \cite{burkholder2020} &    \\
\ce{N2O}     & \ce{N2} + \ce{O(^1D)} & \cite{burkholder2020, hubrich1980} & \cite{heays2017} \\

%\ce{N2O4}    &     &       &    \\

%\ce{N2O3}    &     &       &    \\

%\ce{H2S}     &     &       &    \\
%\ce{SO2}     &     &       &    \\
%\ce{CH3SH}   &     &       &    \\
%SH           &     &       &    \\
%SO           &     &       &    \\
%\ce{H2SO}    &     &       &    \\
%HSOH         &     &       &    \\
%OCS          &     &       &    \\
\ce{O2}      & \ce{O(^3P)} + \ce{O(^3P)} & \cite{brion1979, yoshino2005} & \cite{heays2017} \\
 & \ce{O(^1D)} + \ce{O(^3P)} & \cite{fally2000, chan1993d} & \\
& \ce{O(^1D)} + \ce{O(^1D)} & & \\
%\ce{SO3}     &     &       &    \\
%\ce{S2}      &     &       &    \\
%\ce{S2O}     &     &       &    \\
\hline
\end{tabular}}
%\caption{Table of the different considered mechanisms for the C/H/O/N base and their characteristics.}
\label{tab:ref_sections}
\end{table*}

% Photolysis differences
\begin{figure*}[htbp]
     \begin{subfigure}{0.497\textwidth}
         \includegraphics[width=1.0\textwidth]{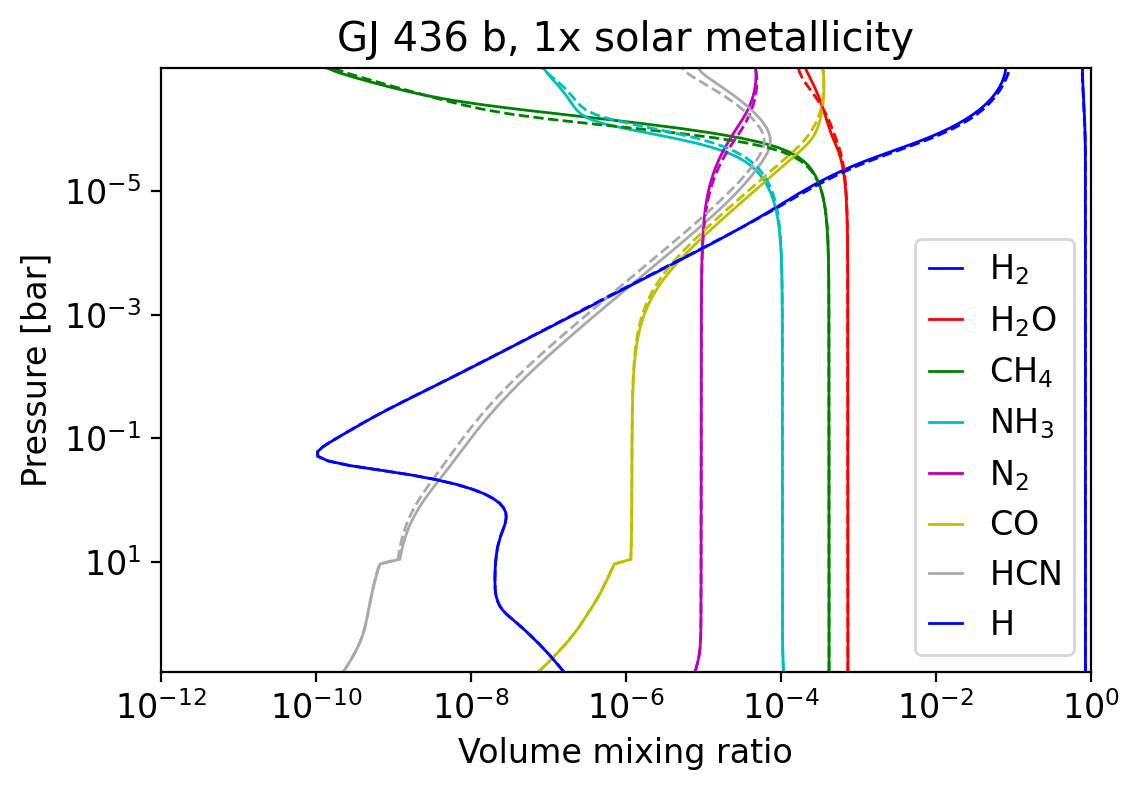}
     \end{subfigure}
     \hfill
     \begin{subfigure}{0.497\textwidth}
         \includegraphics[width=1.0\textwidth]{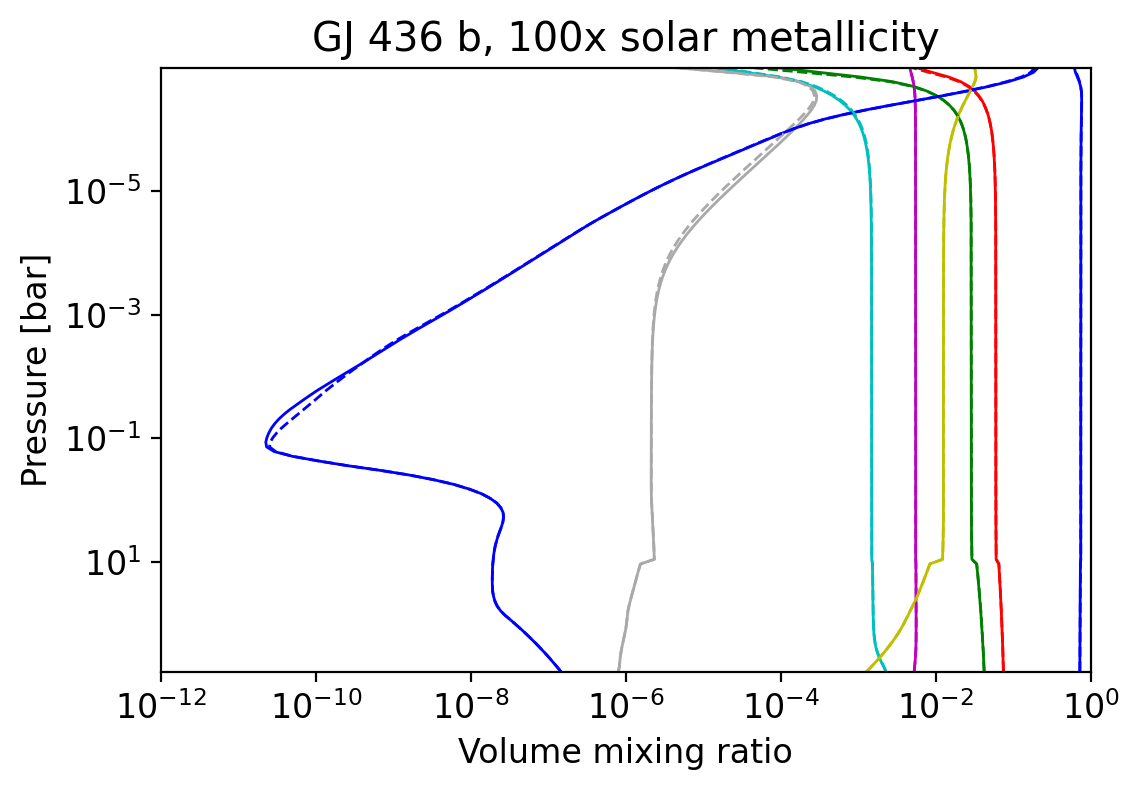}
     \end{subfigure}
     \begin{subfigure}{0.497\textwidth}
         \includegraphics[width=1.0\textwidth]{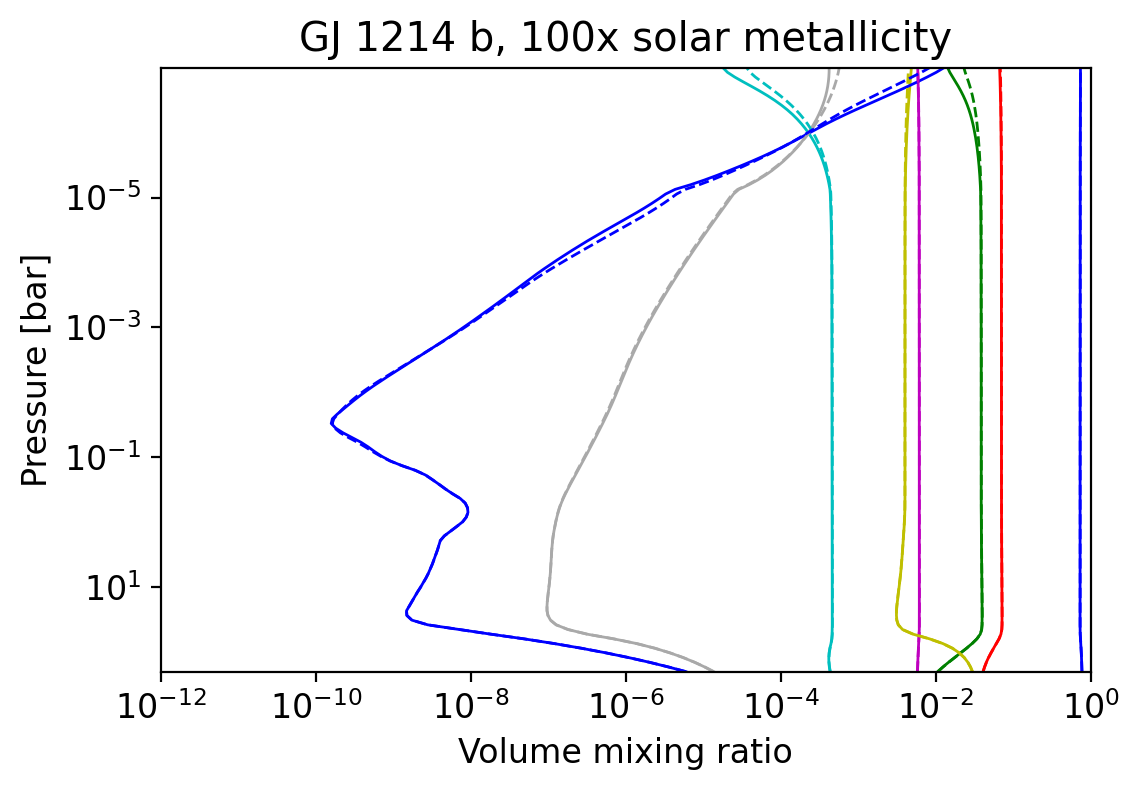}
     \end{subfigure}
     \hfill
     \begin{subfigure}{0.497\textwidth}
         \includegraphics[width=1.0\textwidth]{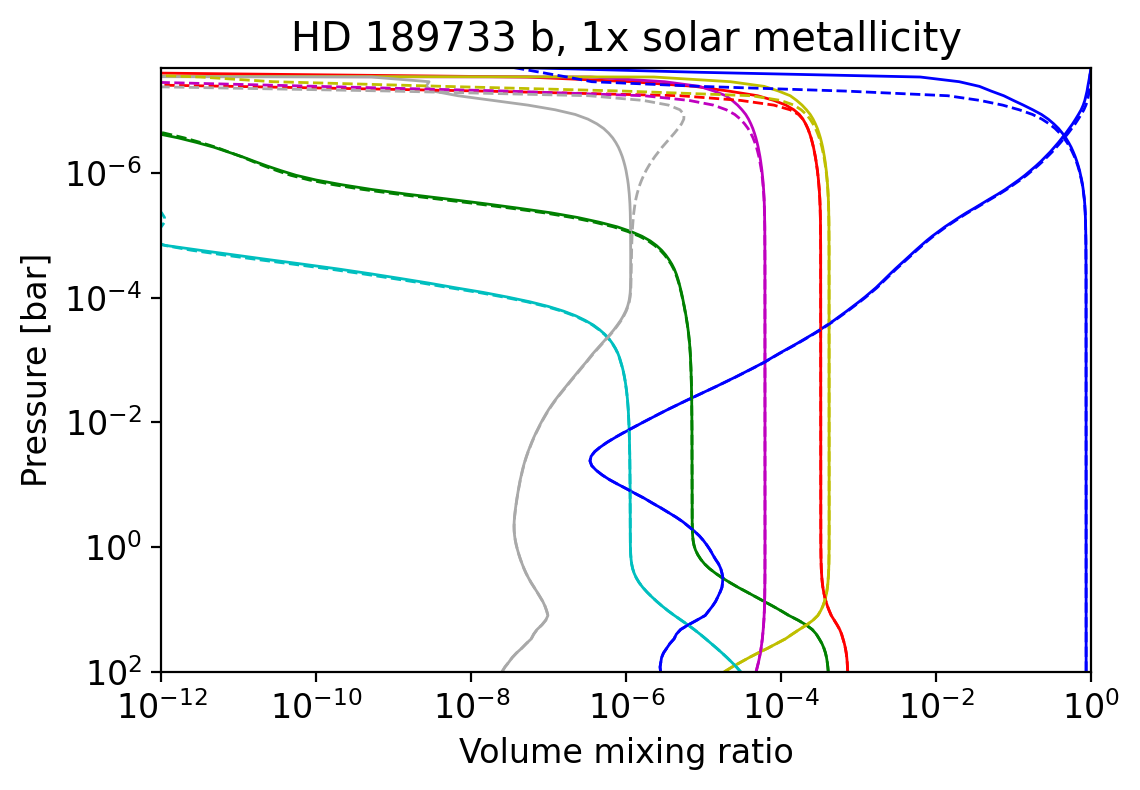}
     \end{subfigure}
    \centering
     \begin{subfigure}{0.497\textwidth}
         \includegraphics[width=1.0\textwidth]{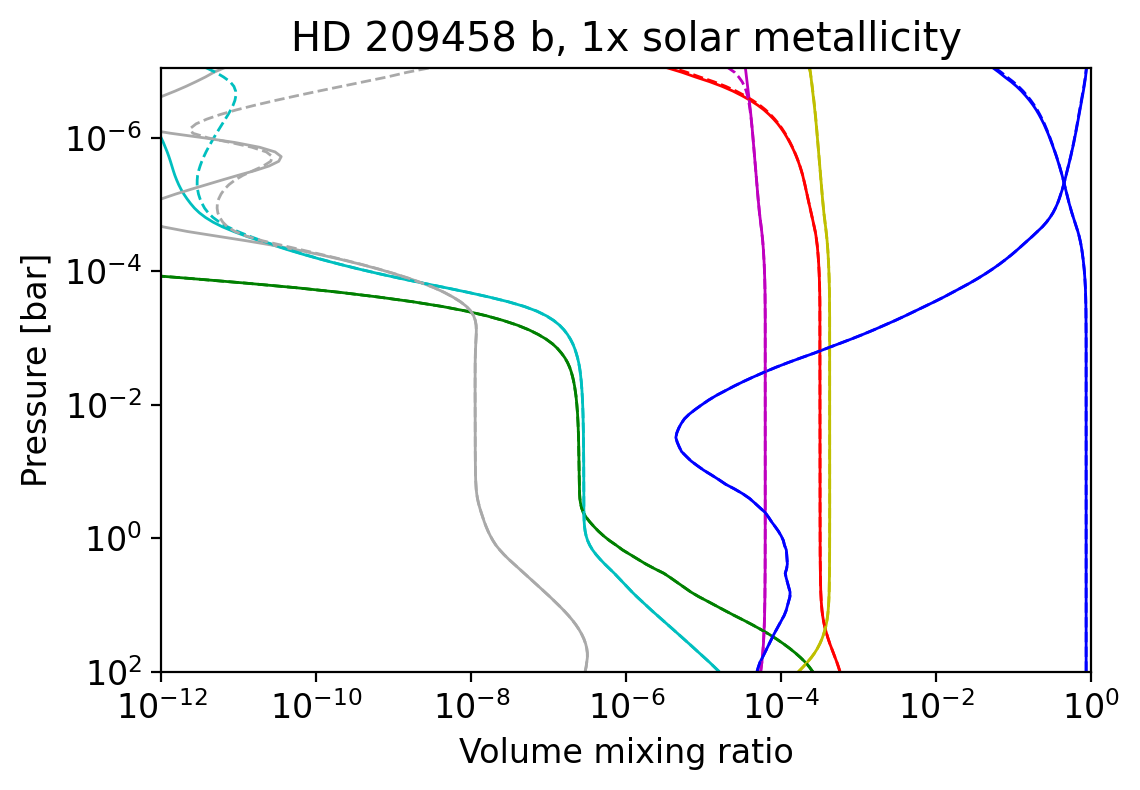}
     \end{subfigure}
        \caption{Abundance profiles of the main species for all exoplanet cases in Table \ref{tab:planets}, computed with V20. Dashed lines are abundances based on old photolysis data and solid lines are based on updated photolysis data.}
        \label{fig:photolysis_changes}
\end{figure*}

\clearpage
\section{Statistical distributions of network errors}

% Combustion Alcools
\begin{figure*}[htbp]
     \begin{subfigure}{0.42\textwidth}
         \includegraphics[width=1.0\textwidth]{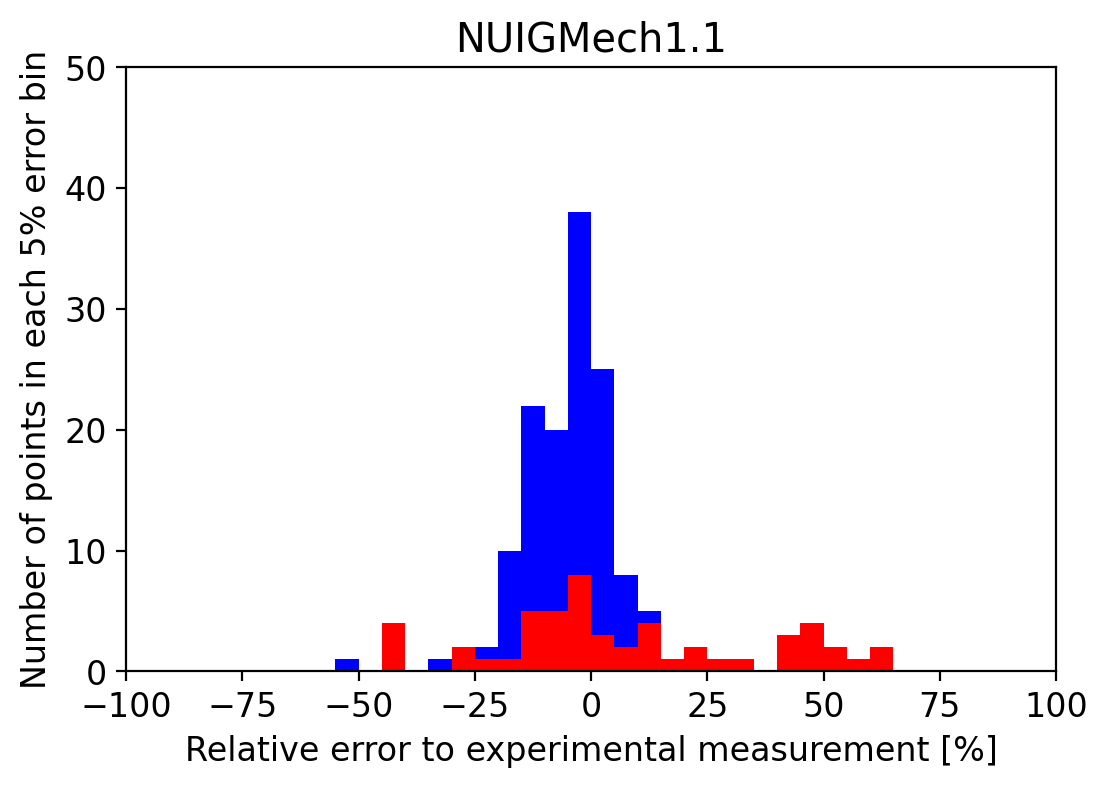}
     \end{subfigure}
     \hfill
     \begin{subfigure}{0.42\textwidth}
         \includegraphics[width=1.0\textwidth]{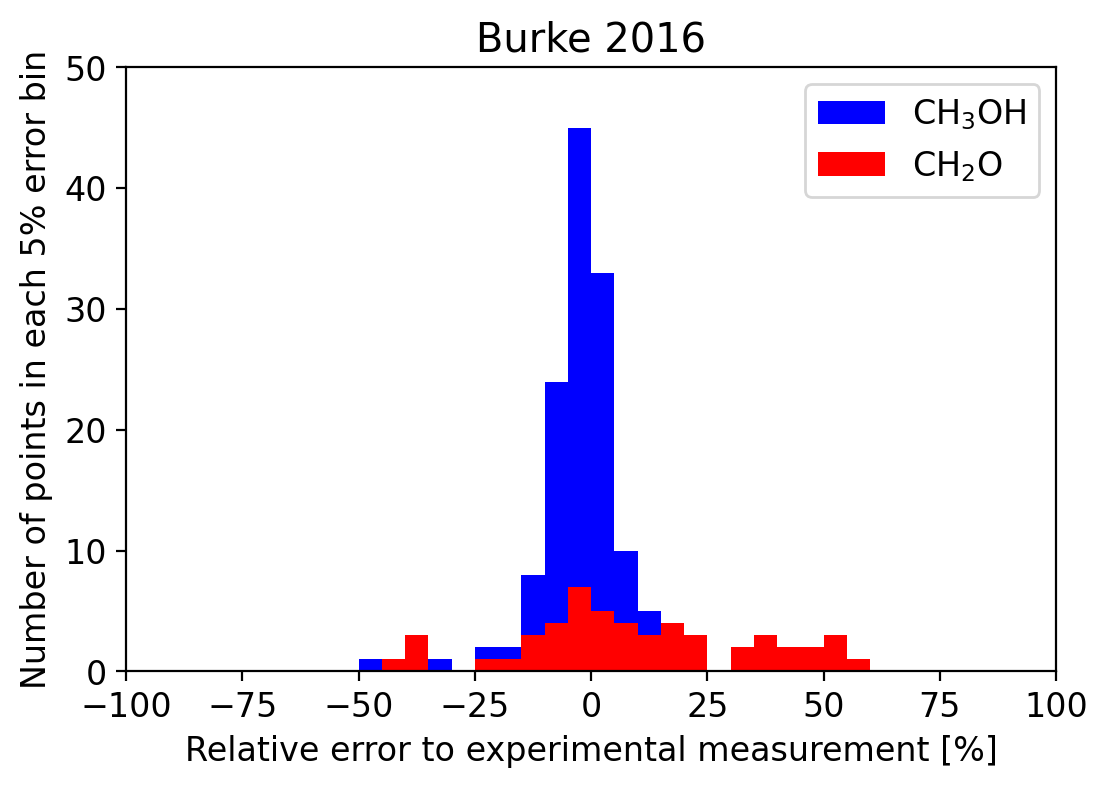}
     \end{subfigure}
     \begin{subfigure}{0.42\textwidth}
         \includegraphics[width=1.0\textwidth]{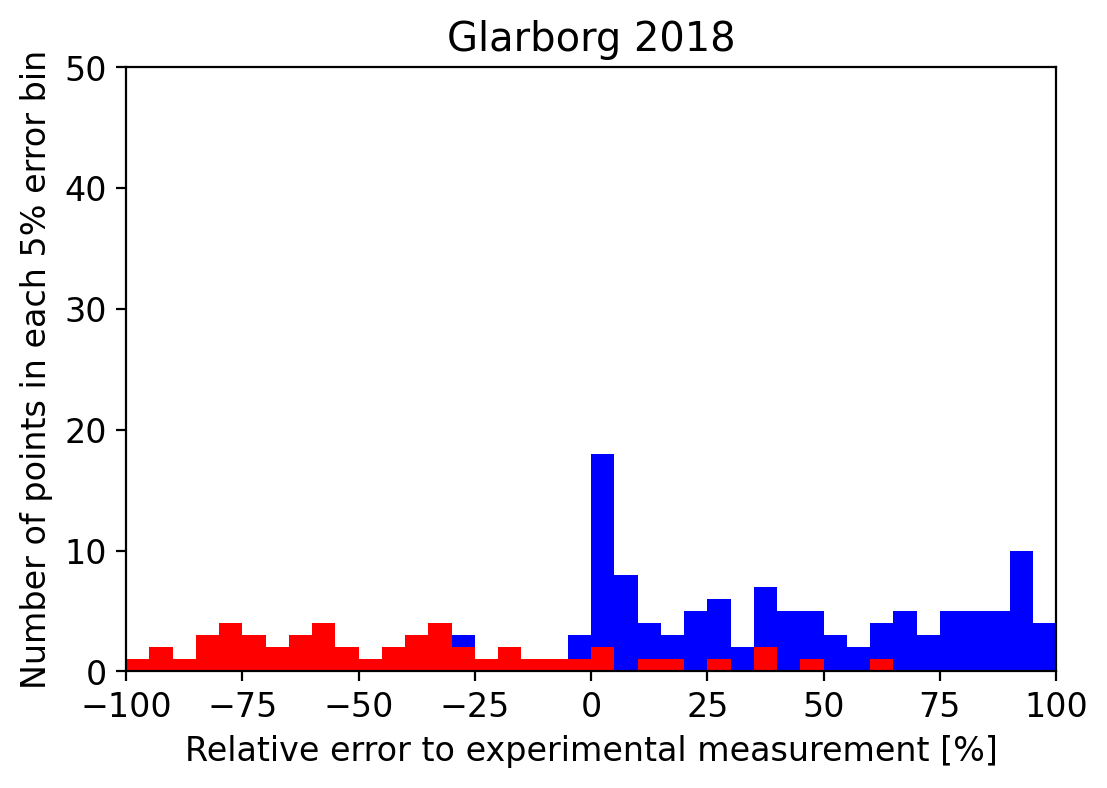}
     \end{subfigure}
     \hfill
     \begin{subfigure}{0.42\textwidth}
         \includegraphics[width=1.0\textwidth]{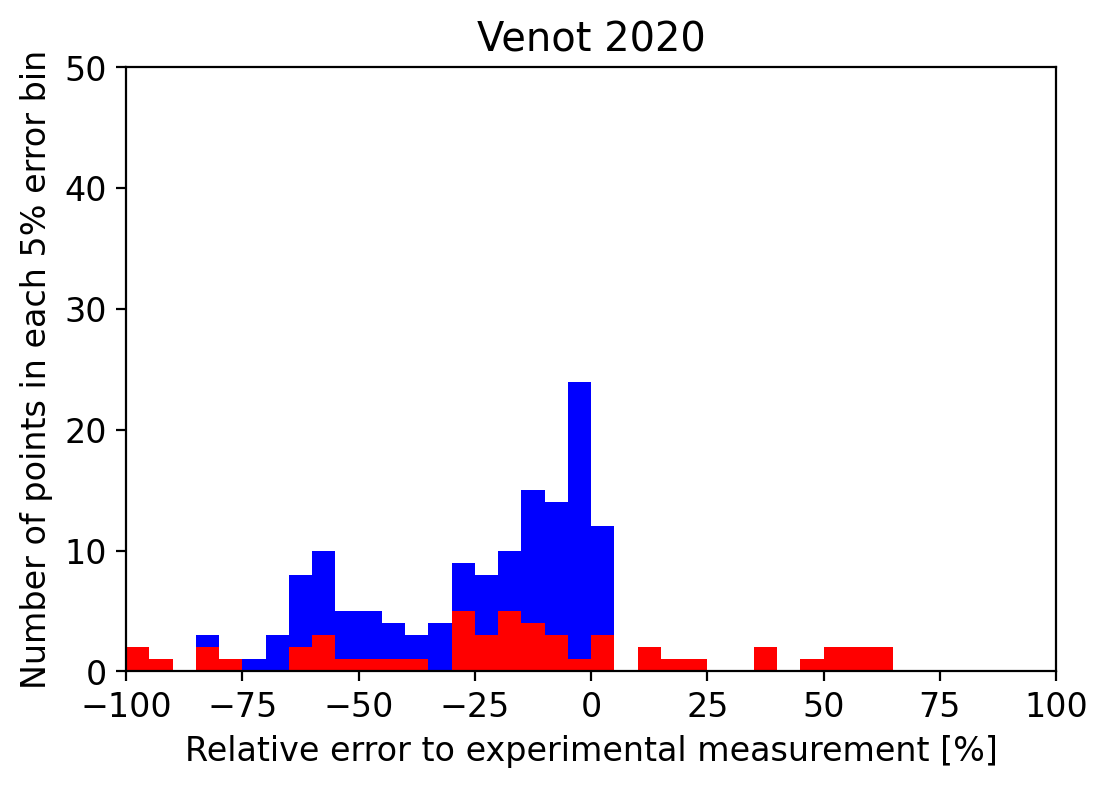}
     \end{subfigure}
        \caption{Statistical distribution of the relative error over the 155 experimental data points corresponding to combustion conditions of Fig. \ref{fig:species4_stats}. Each color gathers all molar fraction measurements of the corresponding species.}
        \label{fig:combustion_alcools}
\end{figure*}

% Pyrolysis Alcools
\begin{figure*}[htbp]
     \begin{subfigure}{0.42\textwidth}
         \includegraphics[width=1.0\textwidth]{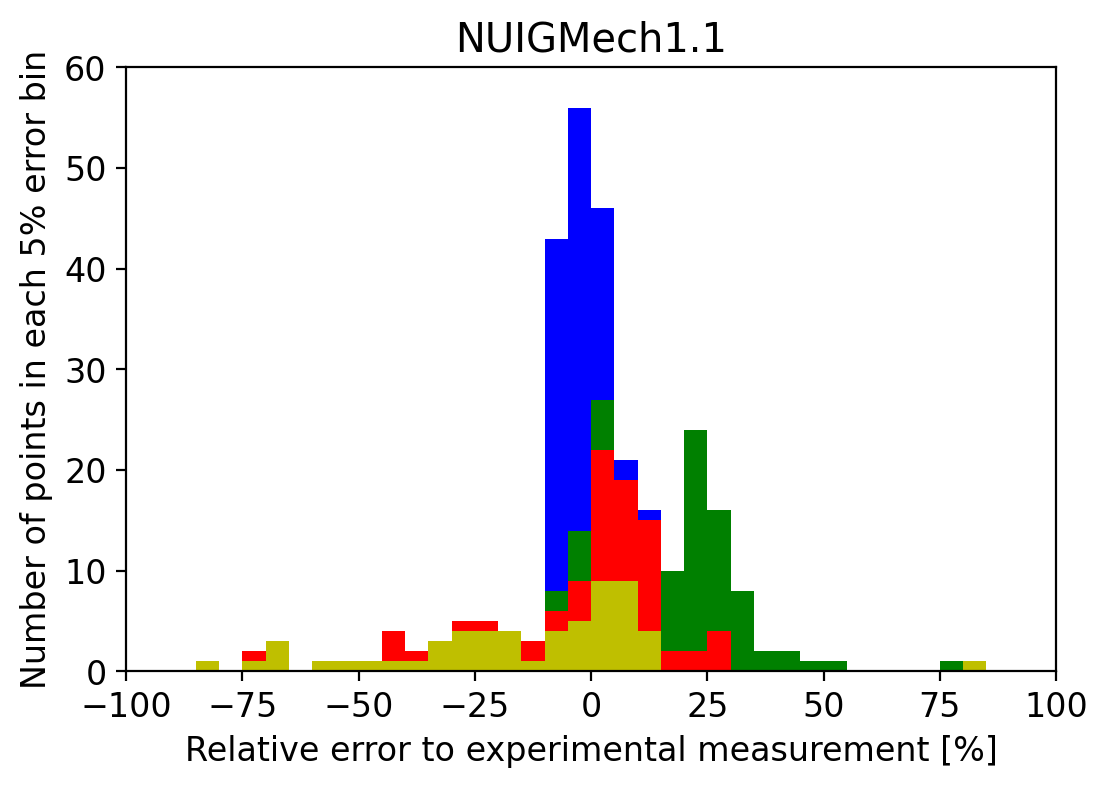}
     \end{subfigure}
     \hfill
     \begin{subfigure}{0.42\textwidth}
         \includegraphics[width=1.0\textwidth]{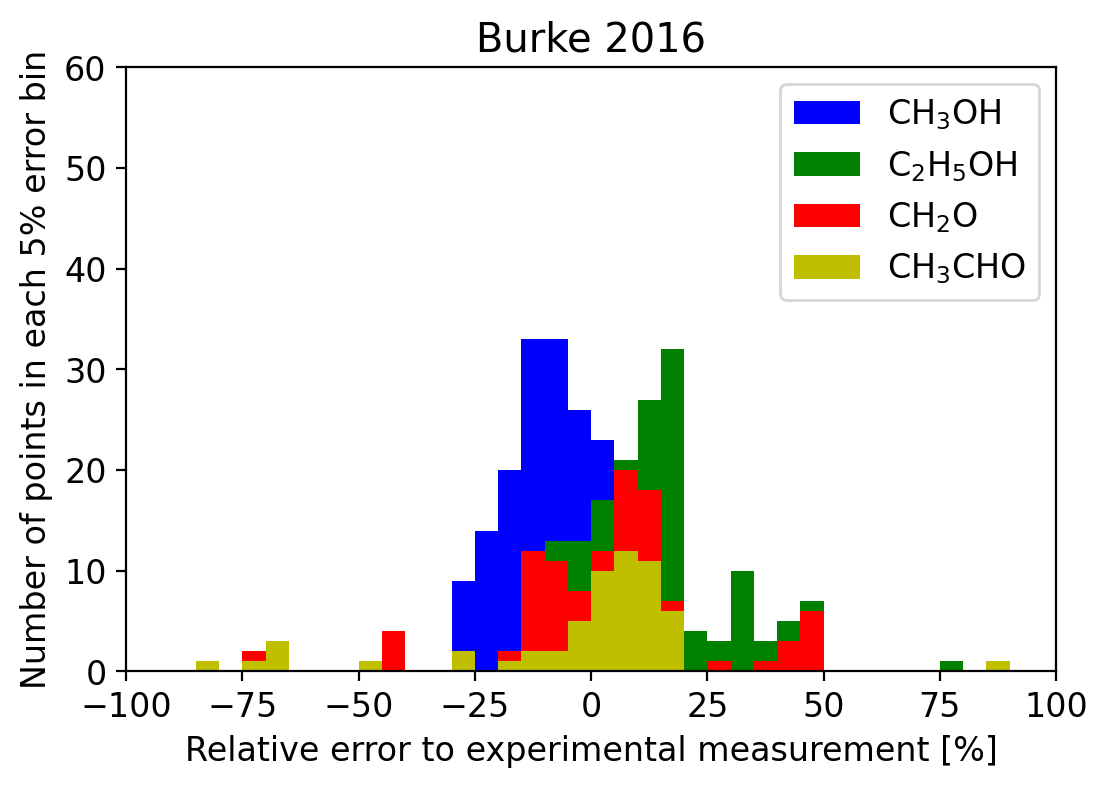}
     \end{subfigure}
     \begin{subfigure}{0.42\textwidth}
         \includegraphics[width=1.0\textwidth]{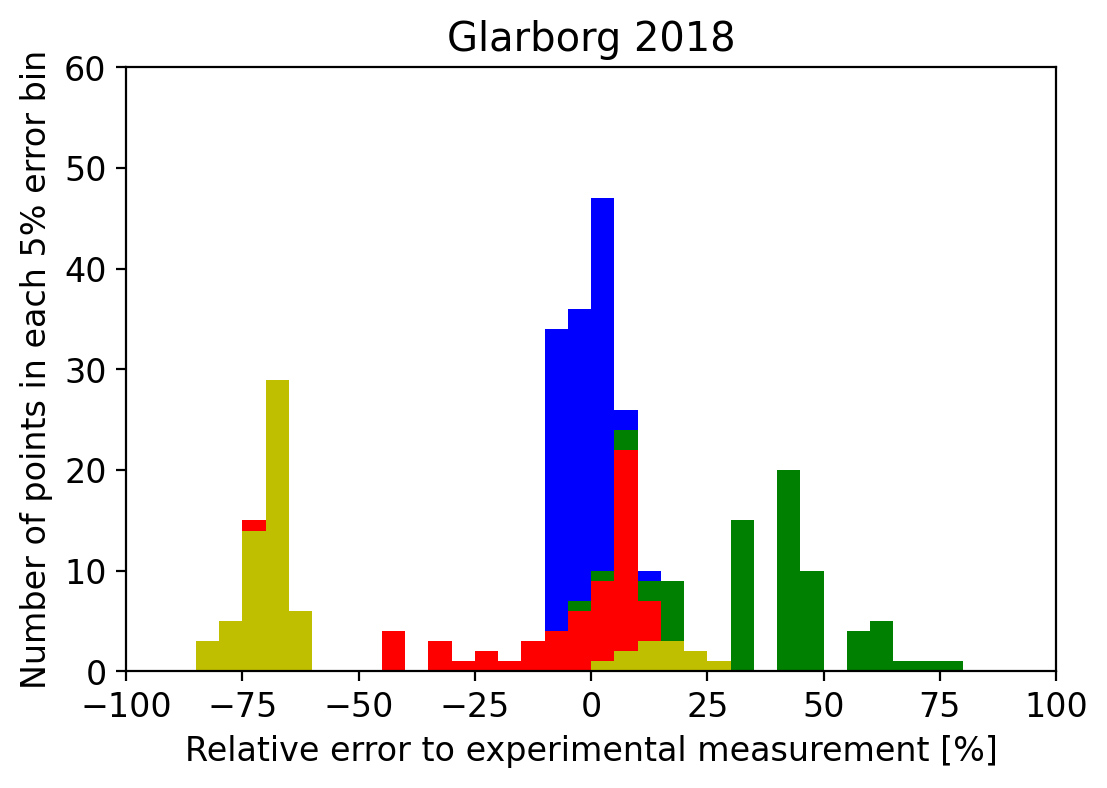}
     \end{subfigure}
     \hfill
     \begin{subfigure}{0.42\textwidth}
         \includegraphics[width=1.0\textwidth]{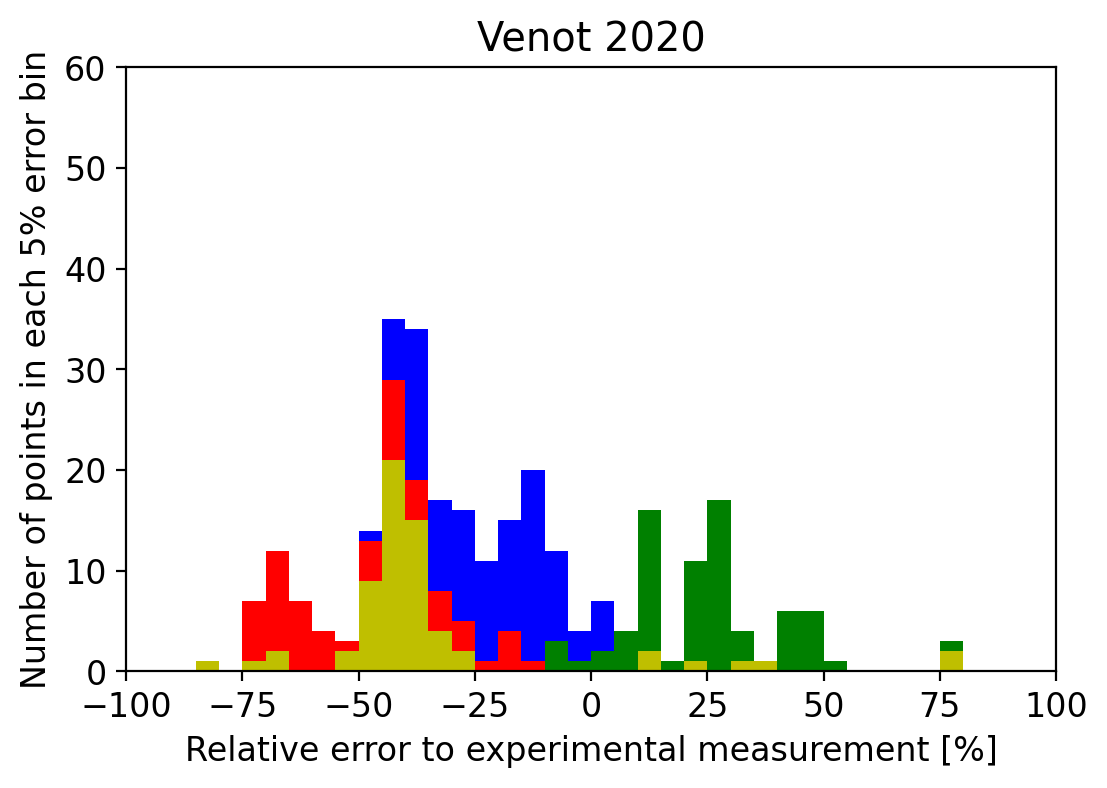}
     \end{subfigure}
        \caption{Statistical distribution of the relative error over the 294 experimental data points corresponding to pyrolysis conditions of Fig. \ref{fig:species4_stats}. Each color gathers all molar fraction measurements of the corresponding species.}
        \label{fig:pyrolysis_alcools}
\end{figure*}

% Combustion Main Species
\begin{figure*}[htbp]
     \begin{subfigure}{0.46\textwidth}
         \includegraphics[width=1.0\textwidth]{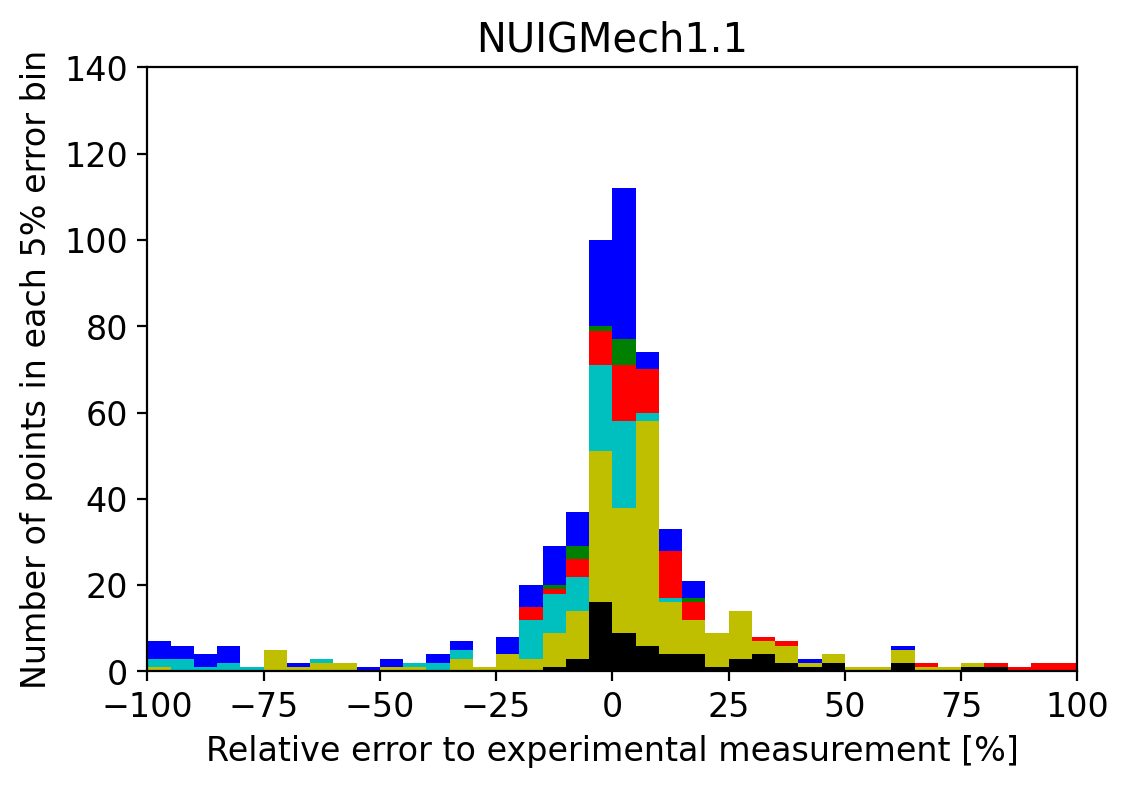}
     \end{subfigure}
     \hfill
     \begin{subfigure}{0.46\textwidth}
         \includegraphics[width=1.0\textwidth]{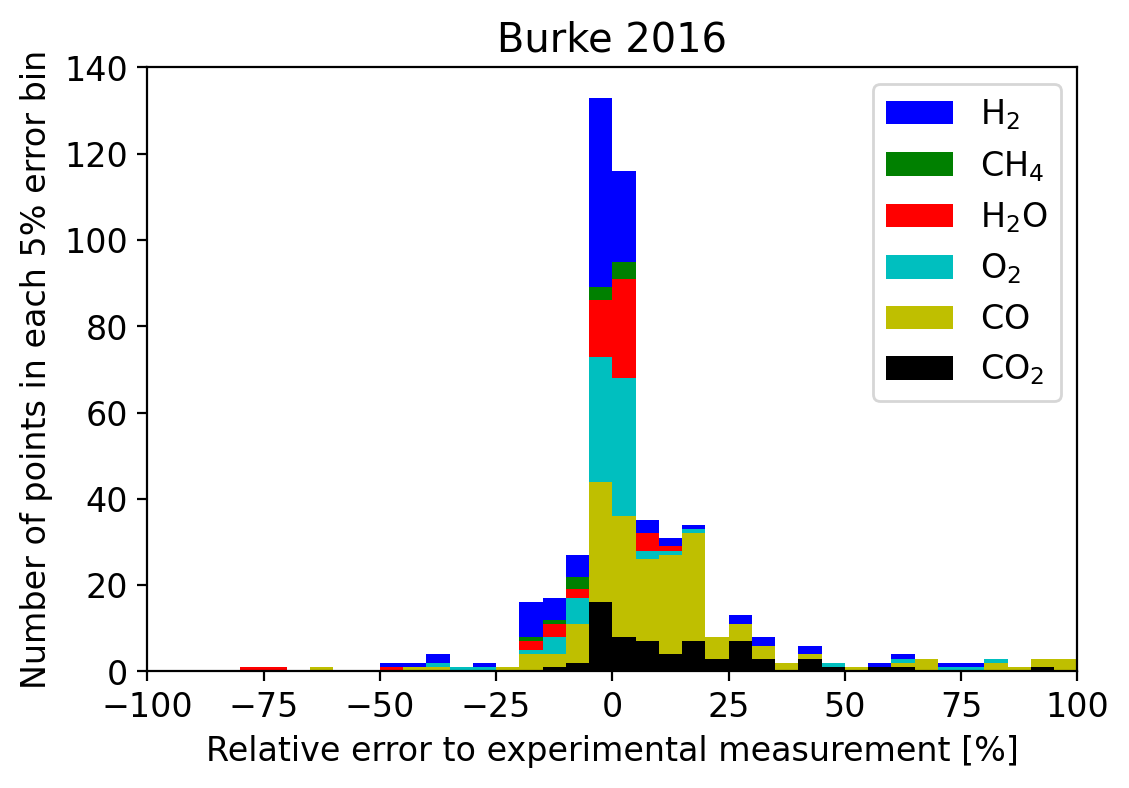}
     \end{subfigure}
     \begin{subfigure}{0.46\textwidth}
         \includegraphics[width=1.0\textwidth]{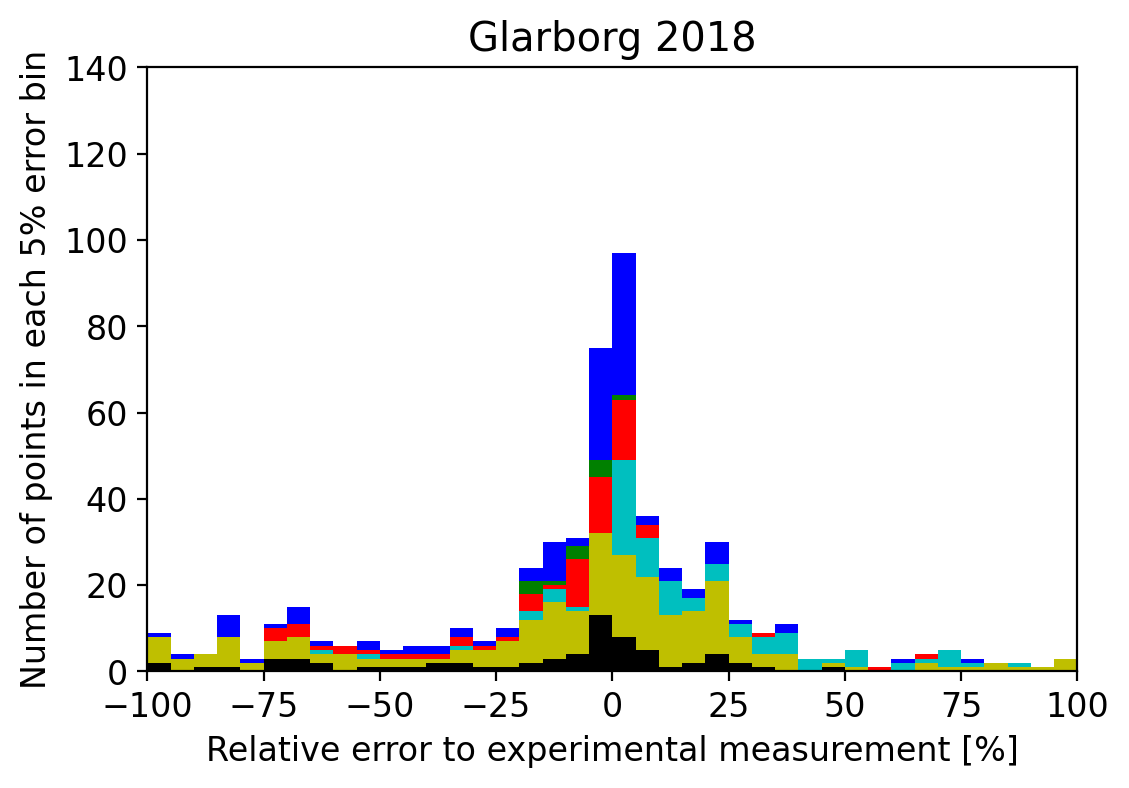}
     \end{subfigure}
     \hfill
     \begin{subfigure}{0.46\textwidth}
         \includegraphics[width=1.0\textwidth]{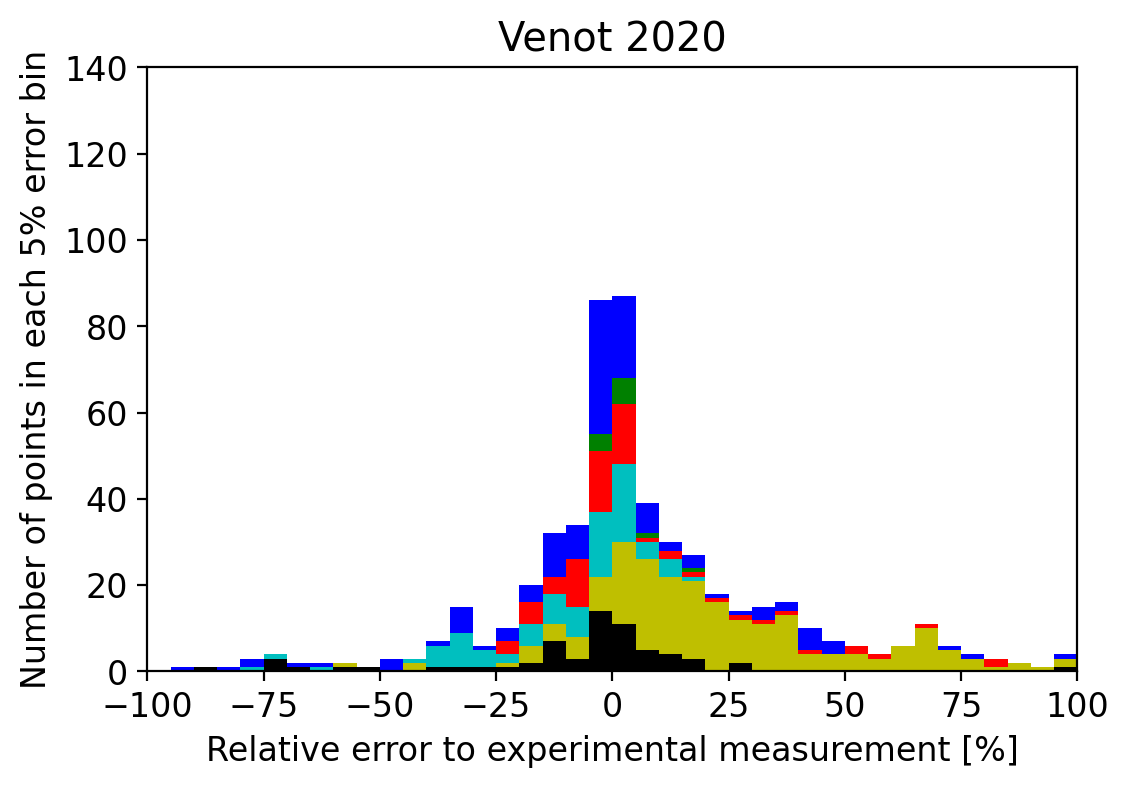}
     \end{subfigure}
        \caption{Statistical distribution of the relative error over the 571 experimental data points corresponding to combustion conditions of Fig. \ref{fig:species1_stats}. Each color gathers all molar fraction measurements of the corresponding species.}
        \label{fig:combustion_species}
\end{figure*}

% Pyrolysis Main Species
\begin{figure*}[htbp]
     \begin{subfigure}{0.46\textwidth}
         \includegraphics[width=1.0\textwidth]{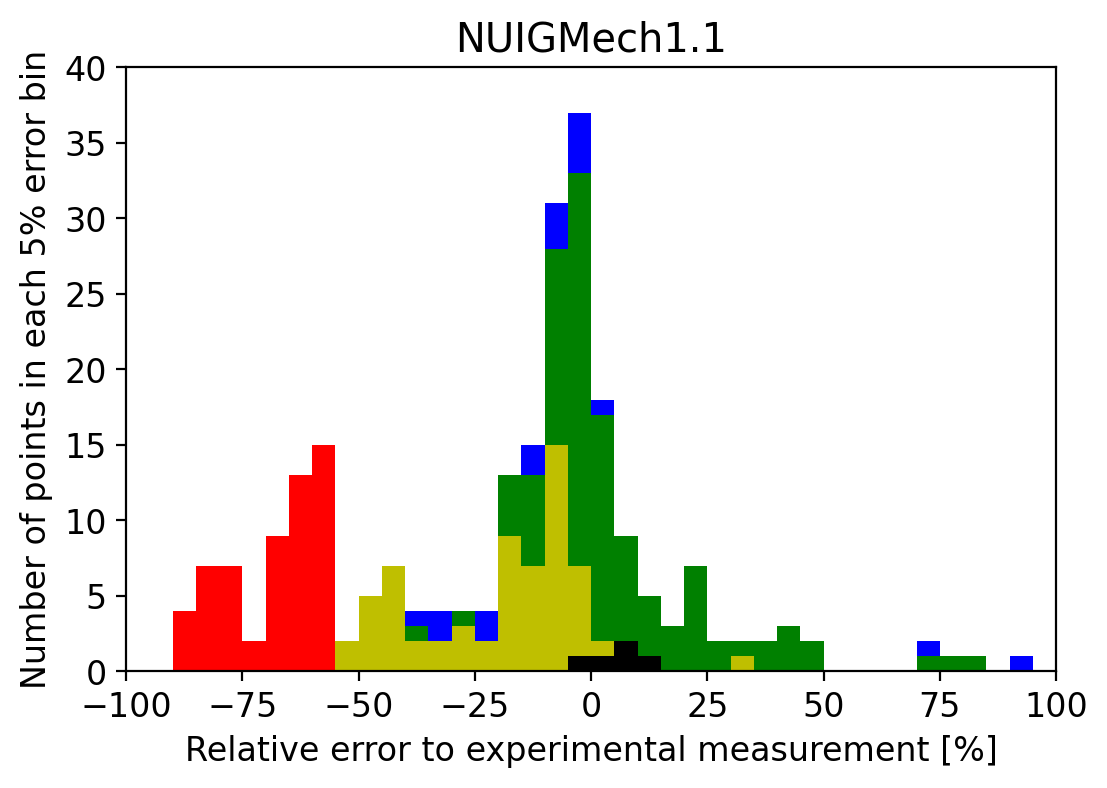}
     \end{subfigure}
     \hfill
     \begin{subfigure}{0.46\textwidth}
         \includegraphics[width=1.0\textwidth]{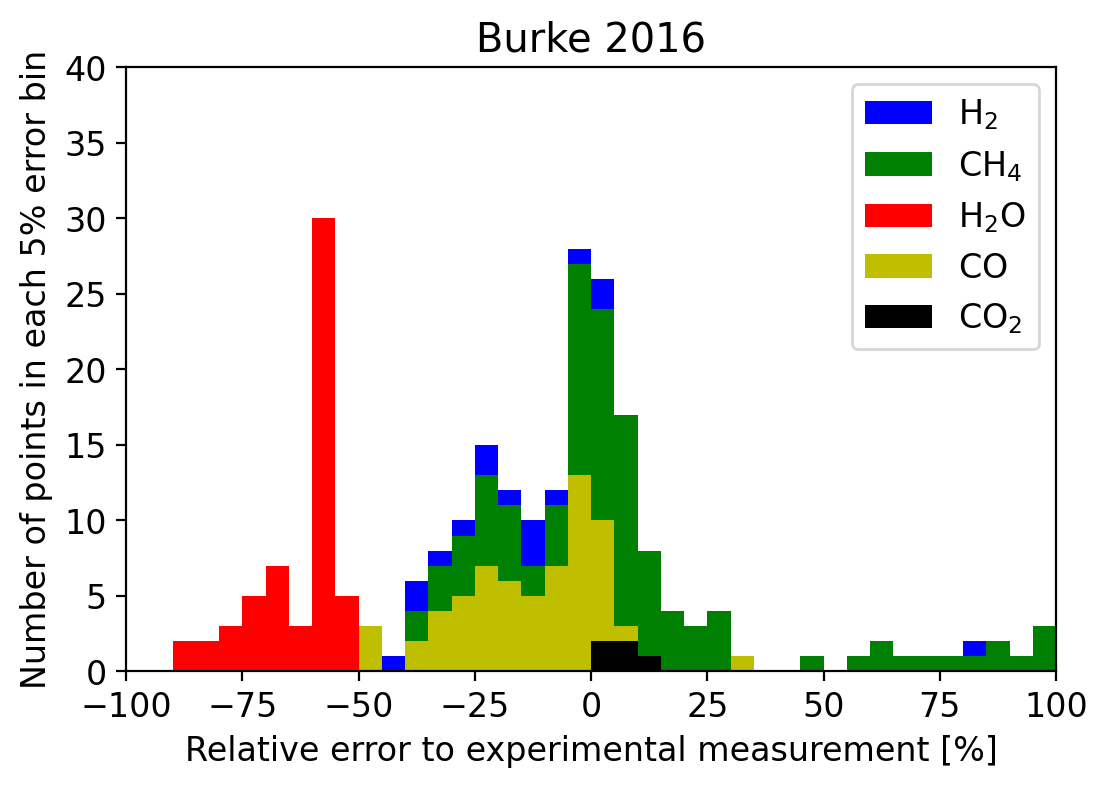}
     \end{subfigure}
     \begin{subfigure}{0.46\textwidth}
         \includegraphics[width=1.0\textwidth]{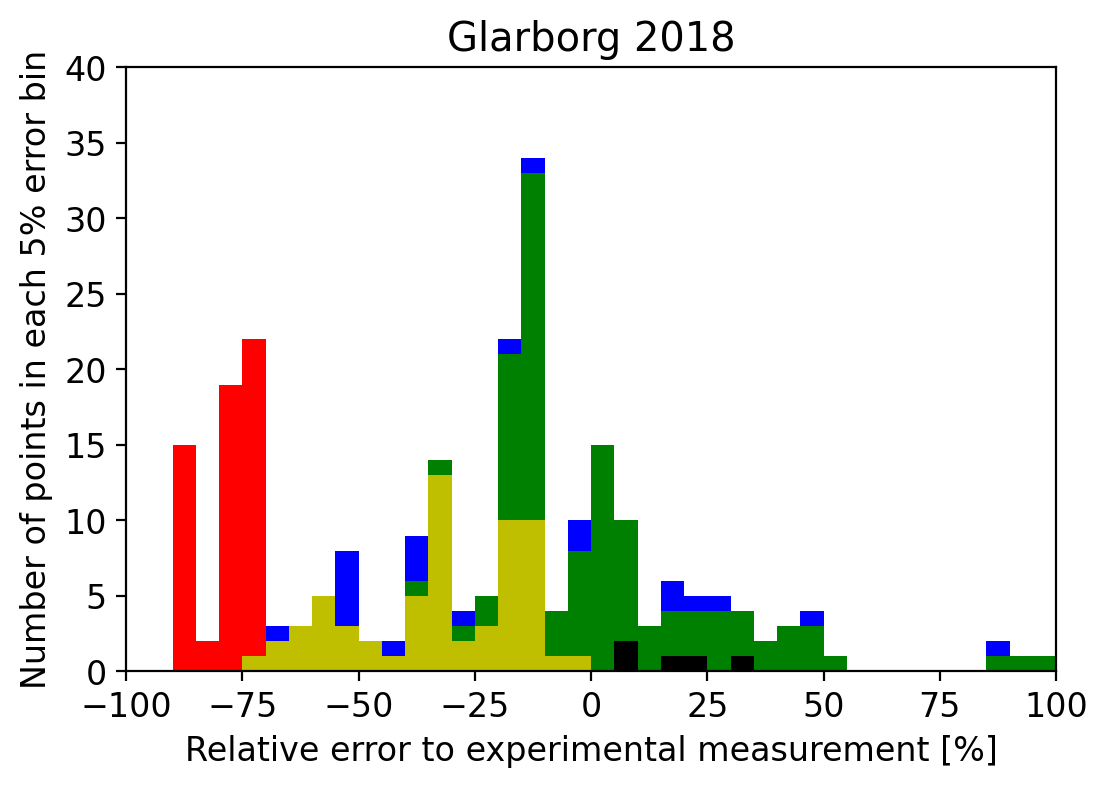}
     \end{subfigure}
     \hfill
     \begin{subfigure}{0.46\textwidth}
         \includegraphics[width=1.0\textwidth]{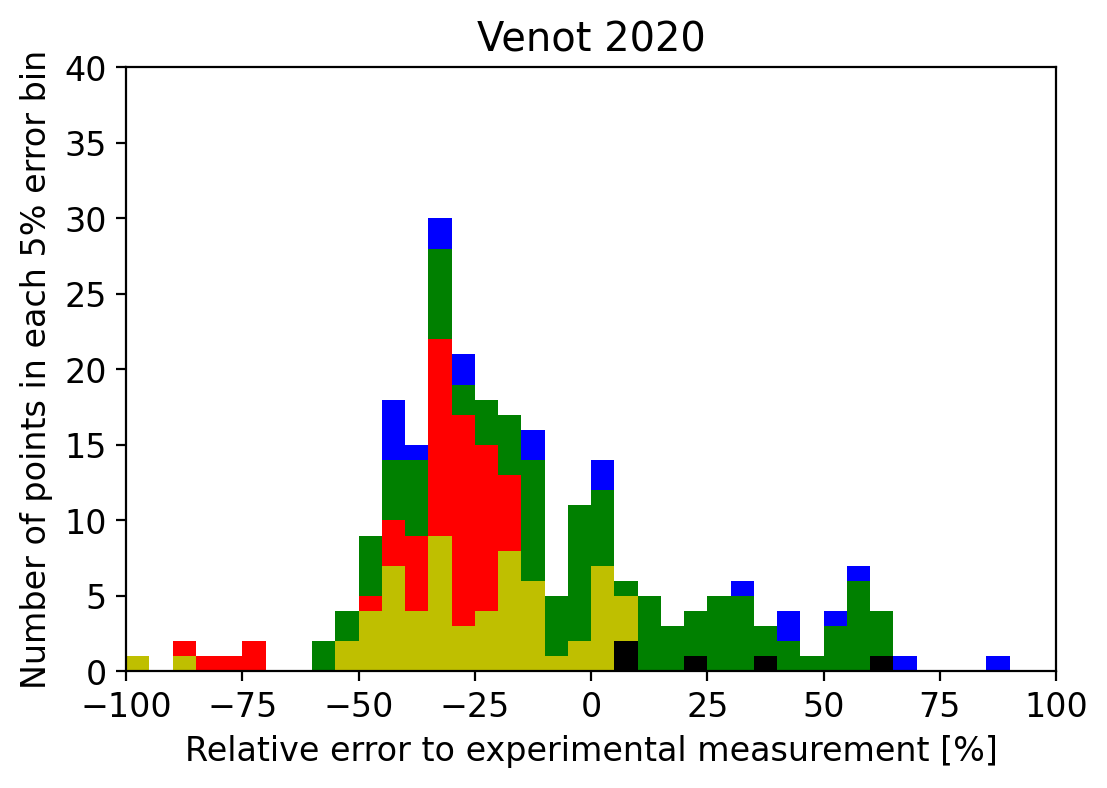}
     \end{subfigure}
        \caption{Statistical distribution of the relative error over the 252 experimental data points corresponding to pyrolysis conditions of Fig. \ref{fig:species1_stats}. Each color gathers all molar fraction measurements of the corresponding species.}
        \label{fig:pyrolysis_species}
\end{figure*}

\newpage
\section{Thermochemical data}

% Thermo differences
\begin{figure*}[htbp]
     \begin{subfigure}{0.497\textwidth}
         \includegraphics[width=1.0\textwidth]{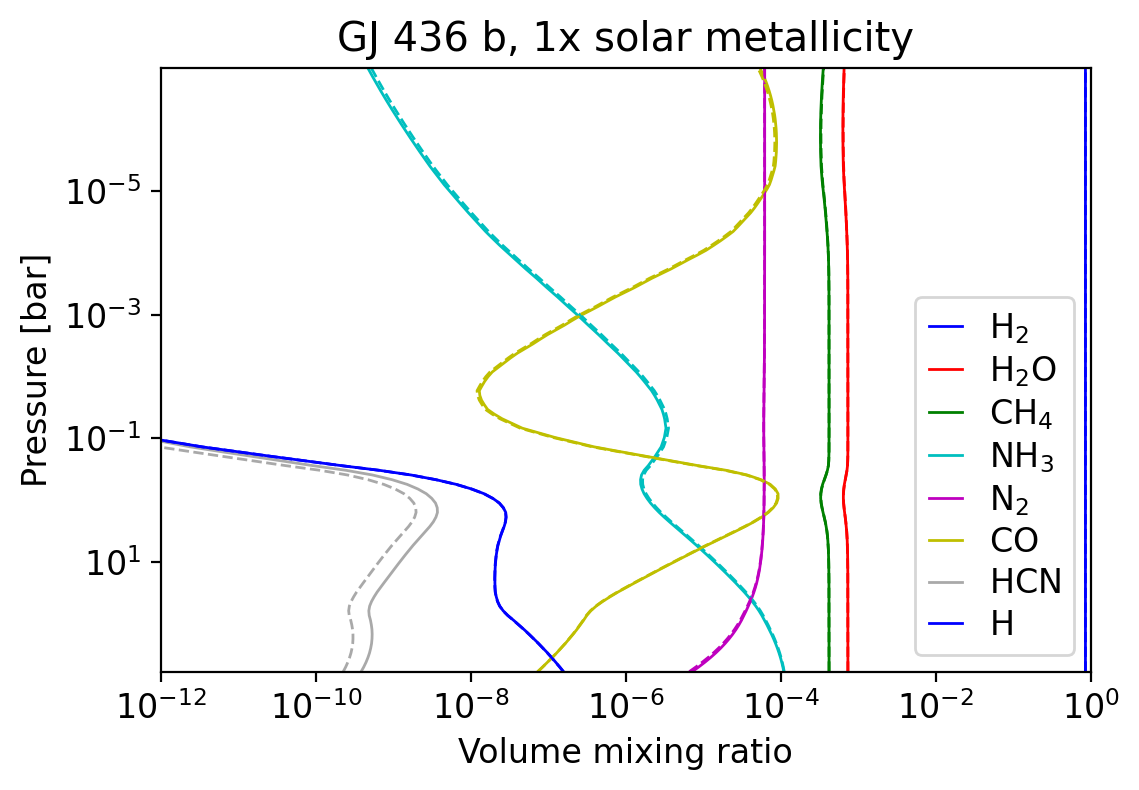}
     \end{subfigure}
     \hfill
     \begin{subfigure}{0.497\textwidth}
         \includegraphics[width=1.0\textwidth]{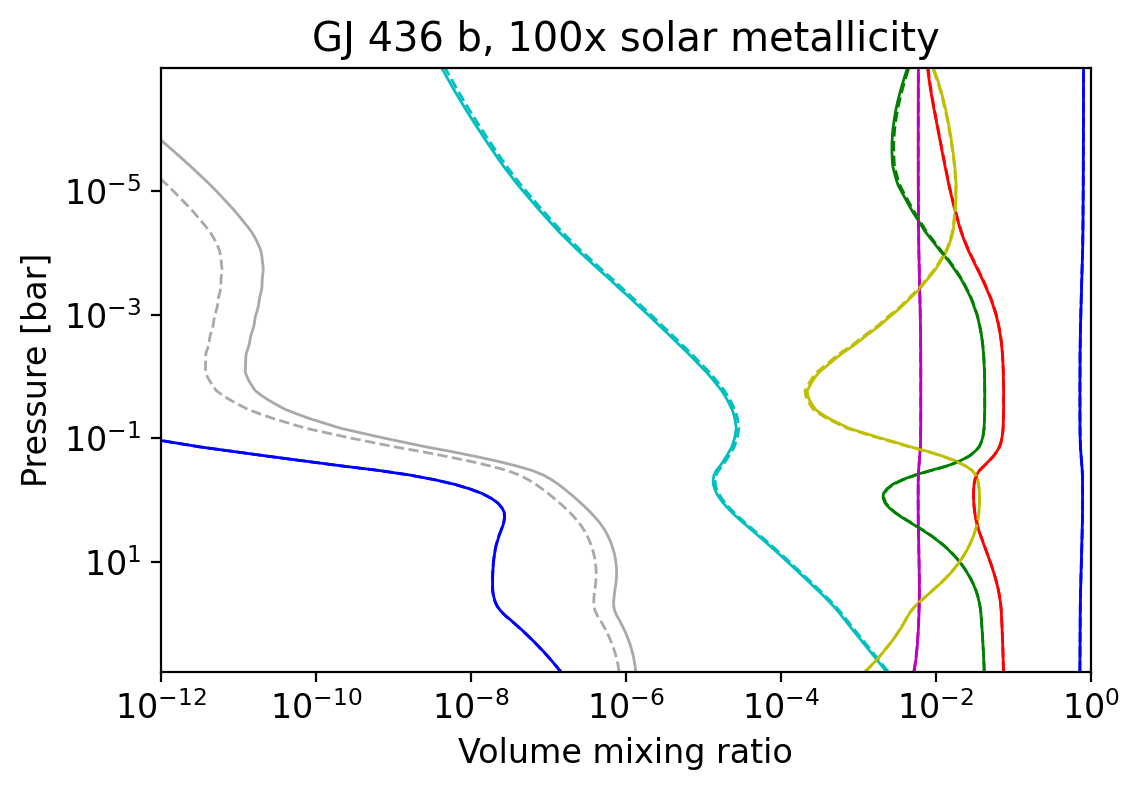}
     \end{subfigure}
     \begin{subfigure}{0.497\textwidth}
         \includegraphics[width=1.0\textwidth]{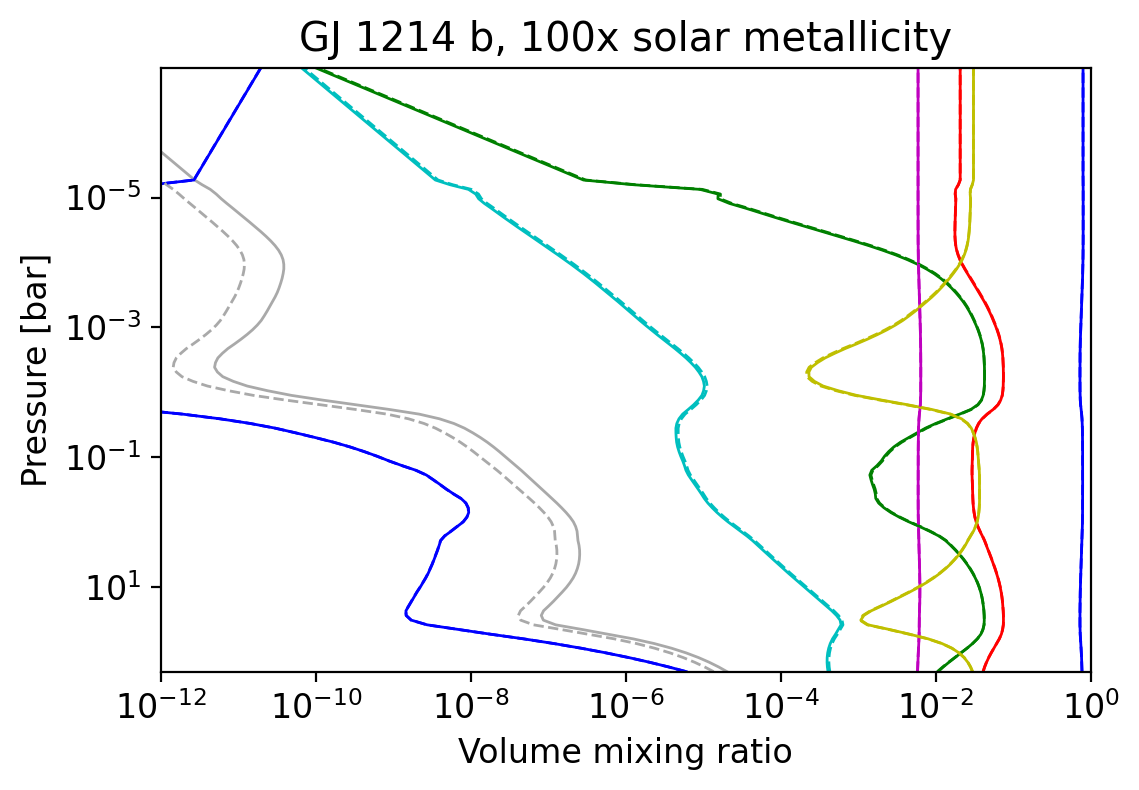}
     \end{subfigure}
     \hfill
     \begin{subfigure}{0.497\textwidth}
         \includegraphics[width=1.0\textwidth]{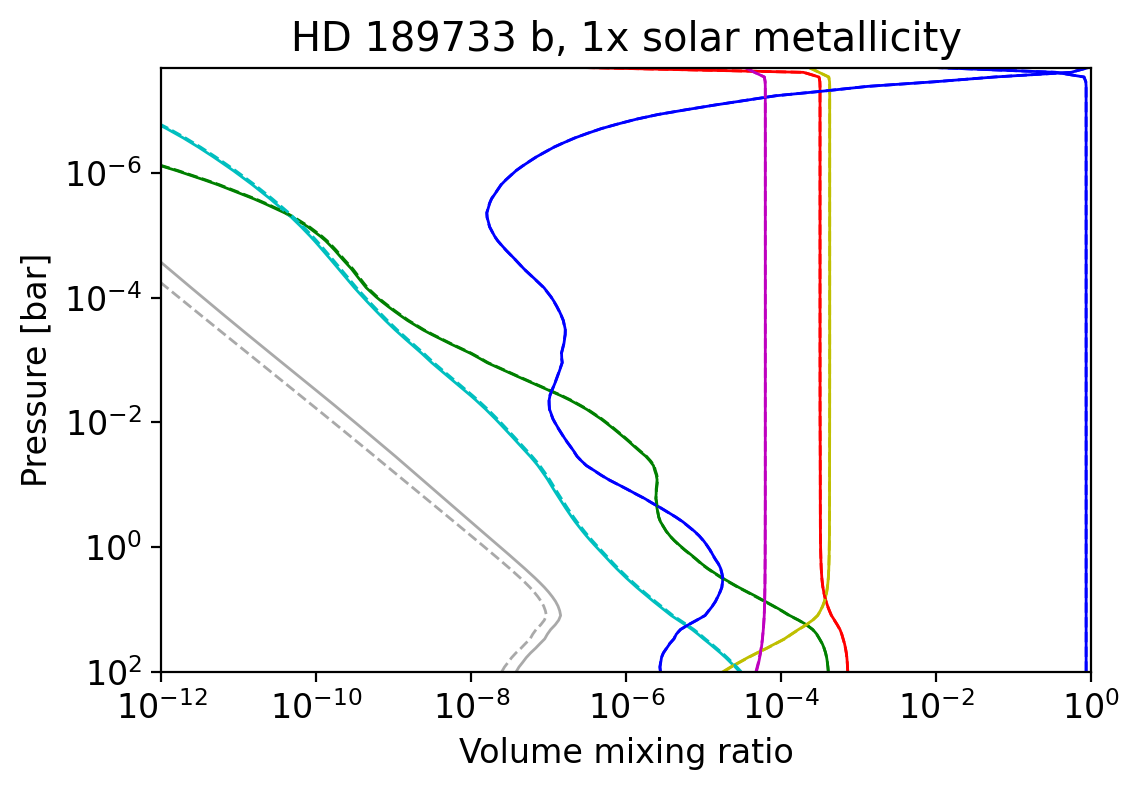}
     \end{subfigure}
     \centering
     \begin{subfigure}{0.497\textwidth}
         \includegraphics[width=1.0\textwidth]{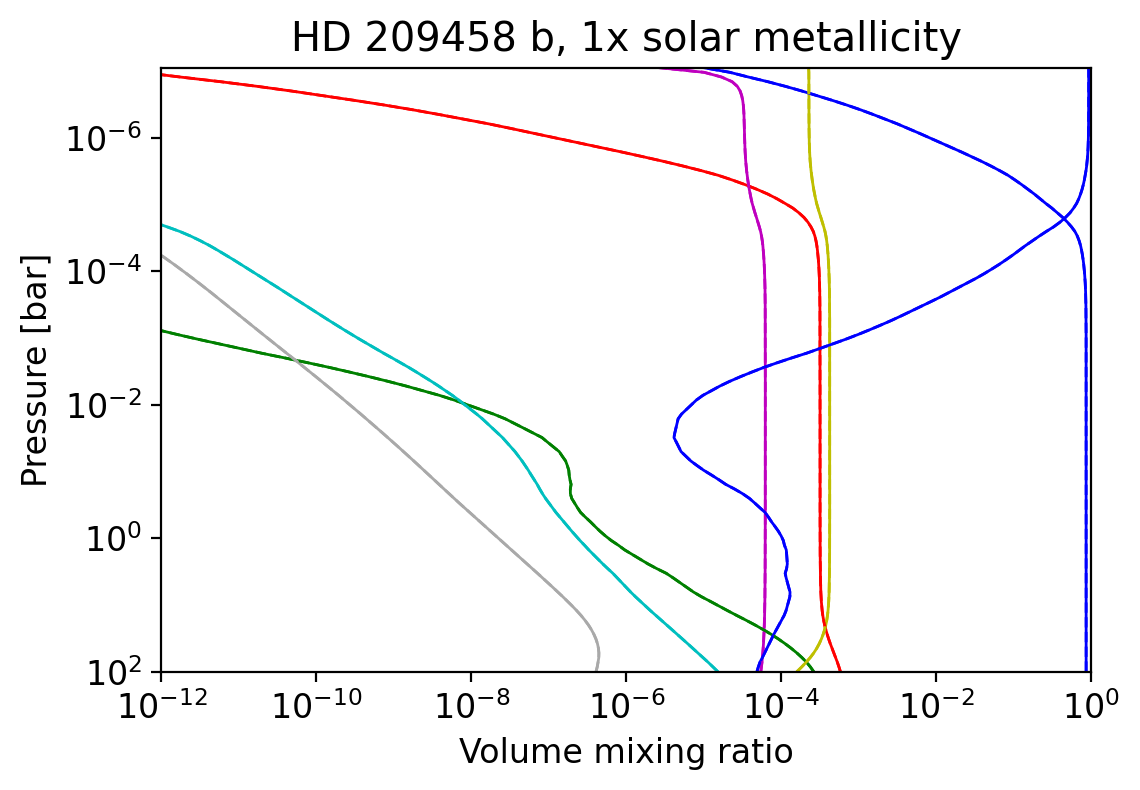}
     \end{subfigure}
        \caption{Thermodynamic equilibrium abundance profiles of the main species for all exoplanets cases in Table \ref{tab:planets}. Dashed lines are computed with the V20 thermodynamic data and solid lines are based on V23 thermodynamic data. In the last panel, all these lines overlap perfectly, indicating no difference for this PT profile.}
        \label{fig:thermo_changes}
\end{figure*}

% Spectra contrib
\newpage
\section{Transmission spectrum contributions}

\begin{figure*}[htbp]
     \begin{subfigure}{0.497\textwidth}
         \includegraphics[width=1.0\textwidth]{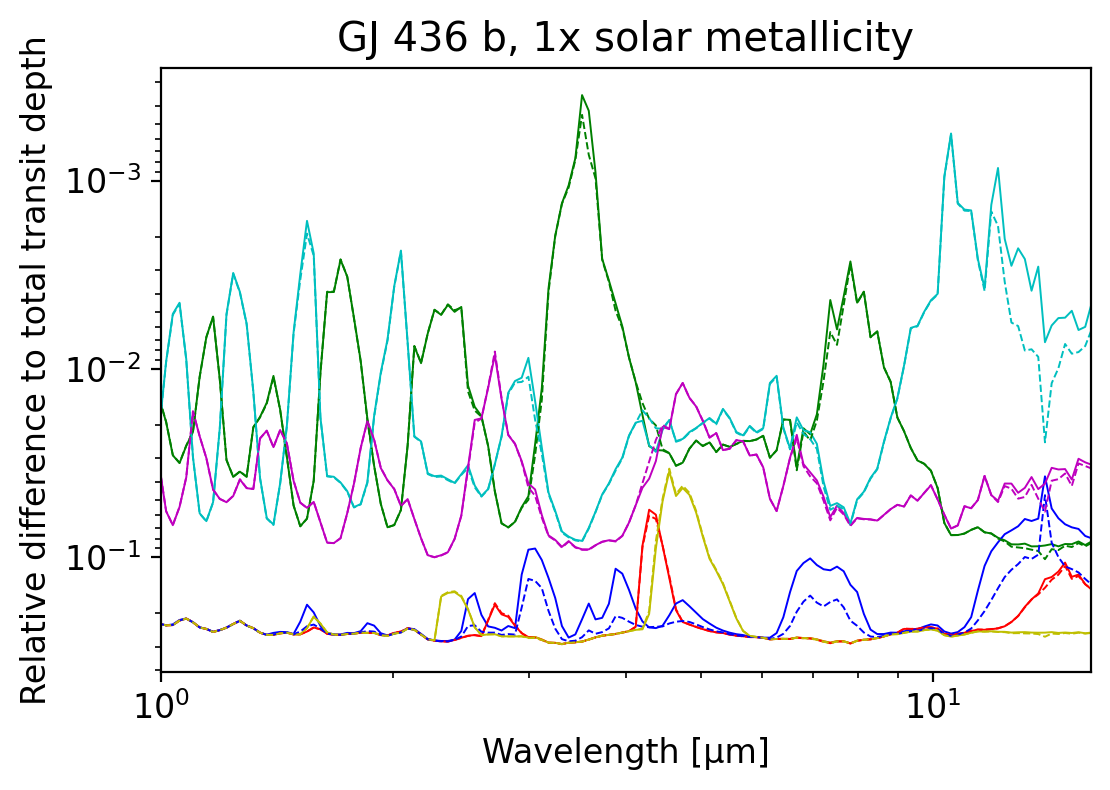}
     \end{subfigure}
     \hfill
     \begin{subfigure}{0.497\textwidth}
         \includegraphics[width=1.0\textwidth]{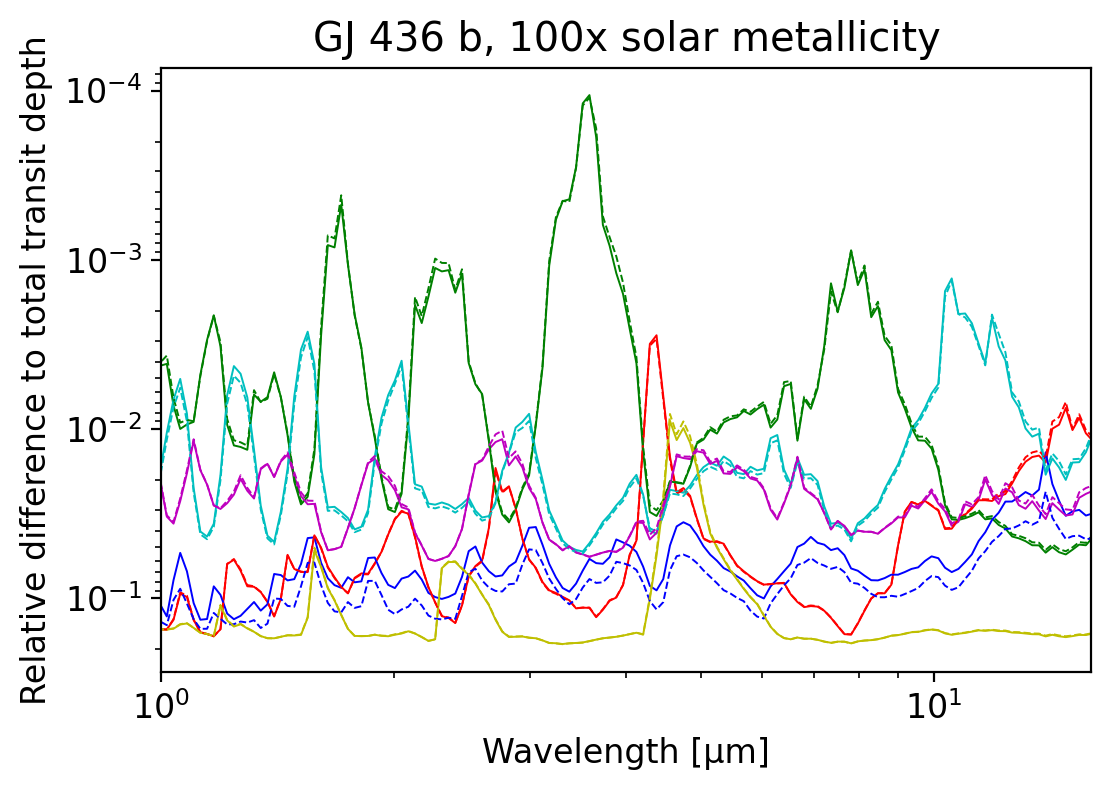}
     \end{subfigure}
     \begin{subfigure}{0.497\textwidth}
         \includegraphics[width=1.0\textwidth]{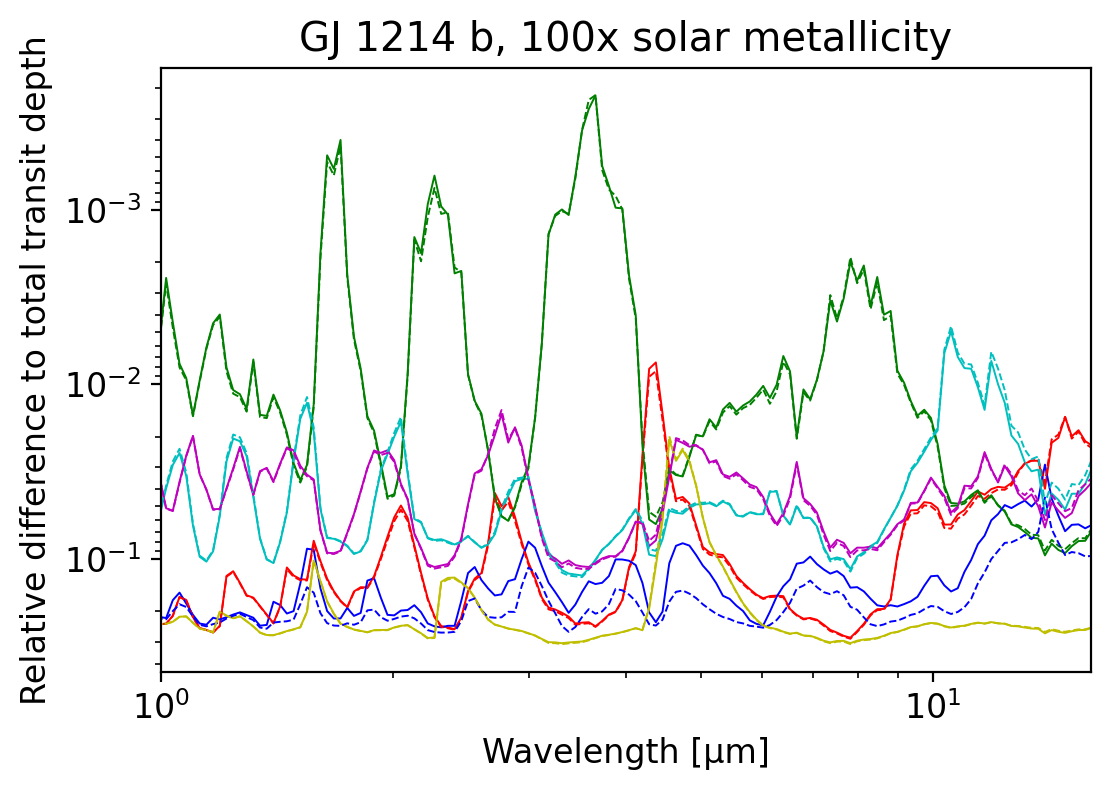}
     \end{subfigure}
     \hfill
     \begin{subfigure}{0.497\textwidth}
         \includegraphics[width=1.0\textwidth]{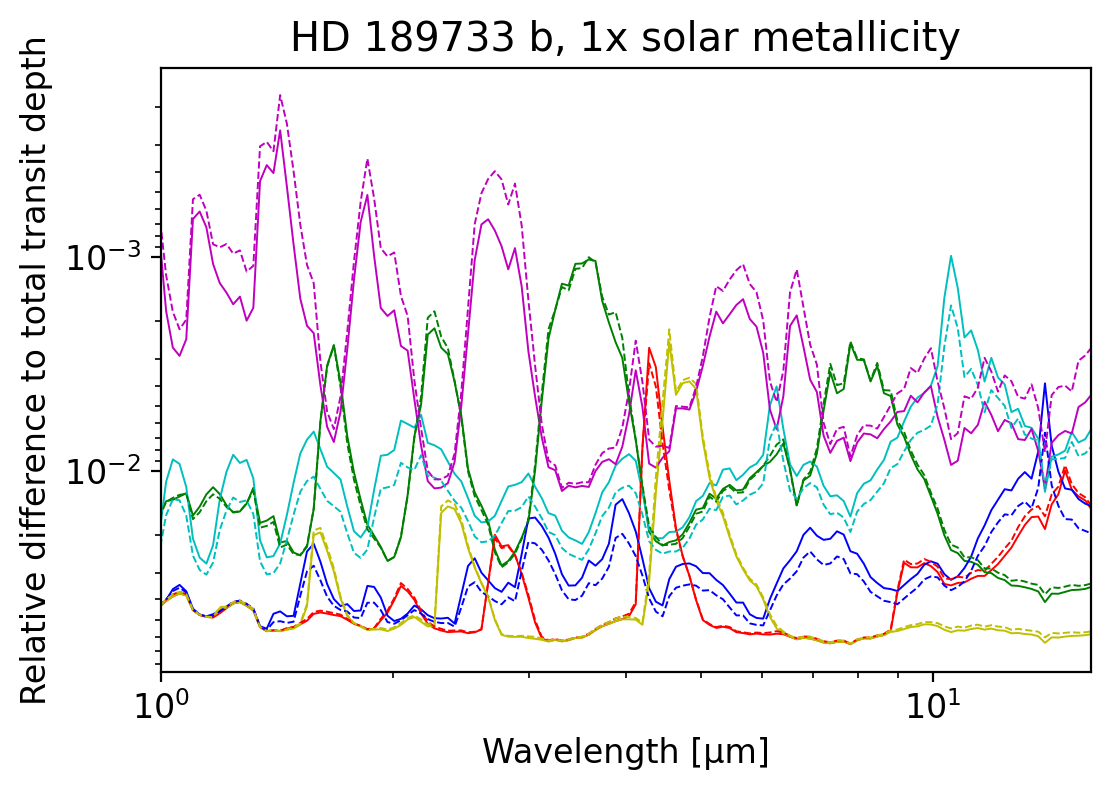}
     \end{subfigure}
     \centering
     \begin{subfigure}{0.497\textwidth}
         \includegraphics[width=1.0\textwidth]{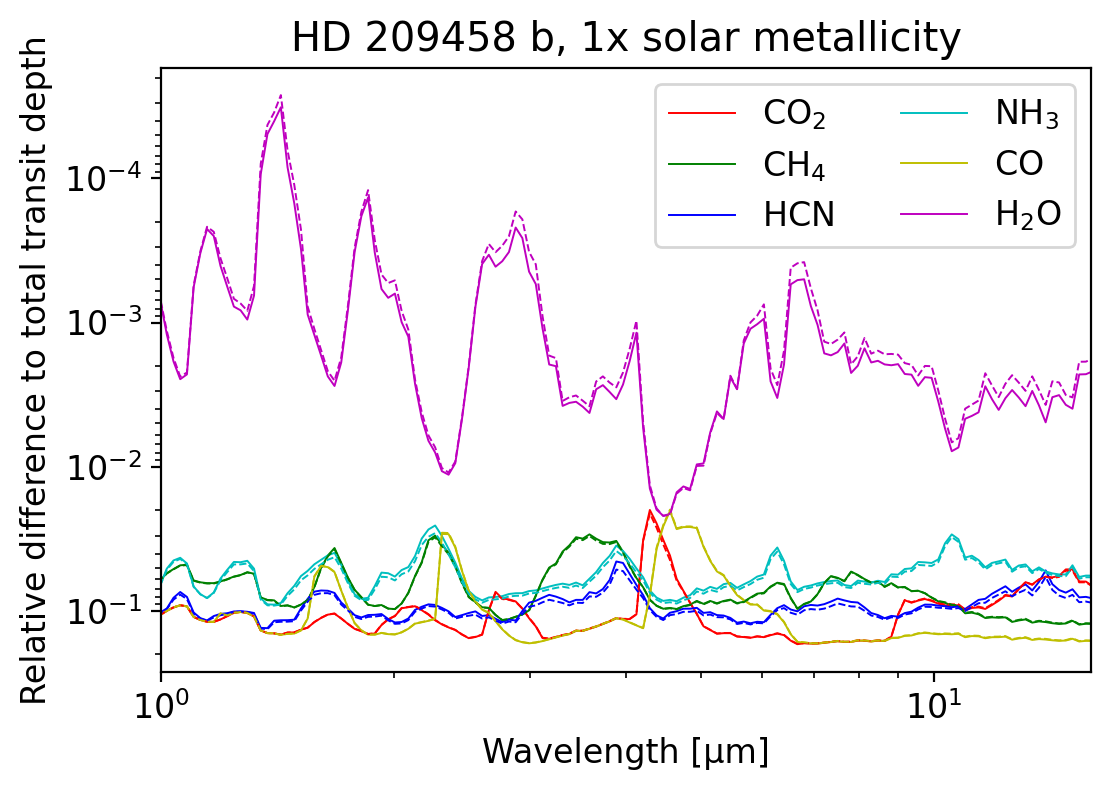}
     \end{subfigure}
        \caption{Contribution to the synthetic transmission spectra for all considered absorbing species. Dashed lines are computed from abundances obtained with V20, and solid lines from abundances obtained with V23. The plotted quantity is the relative difference to the total transit depth, $\frac{D_{tot}-D_{spec}}{D_{tot}}$, with $D_{tot}$ being the total transit depth and $D_{spec}$ being the contribution of the species to the total transit depth. For each wavelength, only the uppermost and close lines have significant impact on the spectrum.}
        \label{fig:spectra_contrib}
\end{figure*}

\end{appendix}

\end{document}